\documentclass[prd,preprint,showpacs,nofootinbib,preprintnumbers]{revtex4}
 \usepackage{graphicx, epsfig,color, amsmath, bm} 

\def\ba{\begin{array}} 
\def\ea{\end{array}}
\def\to{\rightarrow}

\def\bi{\begin{itemize}}
\def\ei{\end{itemize}}

\def\tu{\tilde u}

\def\tf{\tilde f}

\def\tg{\tilde g}

\def\tq{\tilde q} 
 
\def\tz{\widetilde Z}

\def\be{\begin{equation}}
\def\ee{\end{equation}} 
\def\sn{s} 
\def\cs{c}
\def\snp{s'} 
\def\csp{c'}
\def\dblone{\hbox{$\bm{1}\hskip -1.2pt\vrule depth 0pt height 1.6ex width 0.7pt
                  \vrule depth 0pt height 0.3pt width 0.12em$}}

\newcommand{\hepph}[1]{hep-ph/#1}

\newcommand{\by}{\mathbf{Y}}
\newcommand{\bu}{\mathbf{U}}
\newcommand{\bv}{\mathbf{V}}
\newcommand{\bw}{\mathbf{W}}
\newcommand{\bx}{\mathbf{X}}
\newcommand{\bd}{\mathbf{D}}
\newcommand{\bE}{\mathbf{E}}
\newcommand{\bdf}{\mathbf{f}}
\newcommand{\bdfm}{\mathbf{f}^{\rm diag}}
\newcommand{\bdlm}{\bm{\lambda}^{\rm diag}}
\newcommand{\bft}{\tilde{\mathbf{f}}}
\newcommand{\bgt}{\tilde{\mathbf{g}}}
\newcommand{\bbet}{\bm{\beta}}
\newcommand{\bftuq}{(\tilde{\mathbf{f}}^Q_u)}
\newcommand{\bftuqnb}{\tilde{\mathbf{f}}^Q_u}
\newcommand{\bftur}{(\tilde{\mathbf{f}}^{u_R}_u)}
\newcommand{\bfturnb}{\tilde{\mathbf{f}}^{u_R}_u}
\newcommand{\bftdq}{(\tilde{\mathbf{f}}^Q_d)}
\newcommand{\bftdr}{(\tilde{\mathbf{f}}^{d_R}_d)}
\newcommand{\bftel}{(\tilde{\mathbf{f}}^L_e)}
\newcommand{\bfter}{(\tilde{\mathbf{f}}^{e_R}_e)}
\newcommand{\bgtpq}{(\tilde{\mathbf{g}}'^Q)}
\newcommand{\bgtpqnb}{\tilde{\mathbf{g}}'^Q}
\newcommand{\bgtpl}{(\tilde{\mathbf{g}}'^L)}
\newcommand{\bgtpur}{(\tilde{\mathbf{g}}'^{u_R})}
\newcommand{\bgtpurnb}{\tilde{\mathbf{g}}'^{u_R}}
\newcommand{\bgtpdr}{(\tilde{\mathbf{g}}'^{d_R})}
\newcommand{\bgtper}{(\tilde{\mathbf{g}}'^{e_R})}
\newcommand{\bgtq}{(\tilde{\mathbf{g}}^Q)}
\newcommand{\bgtqnb}{\tilde{\mathbf{g}}^Q}
\newcommand{\bgtl}{(\tilde{\mathbf{g}}^L)}
\newcommand{\bgtsq}{(\tilde{\mathbf{g}}_s^{Q})}
\newcommand{\bgtsqnb}{\tilde{\mathbf{g}}_s^{Q}}
\newcommand{\bgtsur}{(\tilde{\mathbf{g}}_s^{u_R})}
\newcommand{\bgtsurnb}{\tilde{\mathbf{g}}_s^{u_R}}
\newcommand{\bgtsdr}{(\tilde{\mathbf{g}}_s^{d_R})}
\newcommand{\bgtsdrnb}{\tilde{\mathbf{g}}_s^{d_R}}

\newcommand{\gtphu}{\tilde{g}'^{h_u}}
\newcommand{\gtphus}{(\tilde{g}'^{h_u})^{*}}
\newcommand{\gtphd}{\tilde{g}'^{h_d}}
\newcommand{\gtphds}{(\tilde{g}'^{h_d})^{*}}

\newcommand{\gthu}{\tilde{g}^{h_u}}
\newcommand{\gthd}{\tilde{g}^{h_d}}
\newcommand{\gthus}{(\tilde{g}^{h_u})^{*}}
\newcommand{\gthds}{(\tilde{g}^{h_d})^{*}}

\newcommand{\h}{\theta_h}
\newcommand{\Hh}{\theta_H}
\newcommand{\Go}{\theta_{G^0}}

\newcommand{\Gp}{\theta_{G^+}}
\newcommand{\A}{\theta_{A}}

\newcommand{\Hp}{\theta_{H^+}}
\newcommand{\sh}{\theta_{\tilde{h}}}
\newcommand{\shram}{\theta_{\tilde{h}^{0}_{1}}}
\newcommand{\sH}{\theta_{\tilde{h}^{0}_{2}}}
\newcommand{\shc}{\theta_{\tilde{h}^{\pm}_1}}
\newcommand{\sHc}{\theta_{\tilde{h}^{\pm}_2}}
\newcommand{\sq}{\theta_{{\tilde{Q}_j}}}
\newcommand{\sqk}{\theta_{{\tilde{Q}_k}}}
\newcommand{\sql}{\theta_{{\tilde{Q}_l}}}
\newcommand{\su}{\theta_{{\tilde{u}_i}}}
\newcommand{\suk}{\theta_{{\tilde{u}_k}}}
\newcommand{\sul}{\theta_{{\tilde{u}_l}}}

\newcommand{\sdk}{\theta_{{\tilde{d}_k}}}
\newcommand{\sdl}{\theta_{{\tilde{d}_l}}}

\newcommand{\sll}{\theta_{{\tilde{L}_l}}}
\newcommand{\slk}{\theta_{{\tilde{L}_k}}}

\newcommand{\sek}{\theta_{{\tilde{e}_k}}}
\newcommand{\sel}{\theta_{{\tilde{e}_l}}}
\newcommand{\sbi}{\theta_{\tilde{B}}}
\newcommand{\swi}{\theta_{\tilde{W}}}
\newcommand{\sgl}{\theta_{\tilde{g}}}
\newcommand{\mgtphusq}{\left|\gtphu\right|^{2}}
\newcommand{\mgtphdsq}{\left|\gtphd\right|^{2}}
\newcommand{\mgthusq}{\left|\gthu\right|^{2}}
\newcommand{\mgthdsq}{\left|\gthd\right|^{2}}

\begin{document}

\preprint{UH-511-1116-07}

\title{Threshold and Flavour Effects in the Renormalization Group Equations of the MSSM I: Dimensionless Couplings}

\author{Andrew D. Box}
 \email{abox@phys.hawaii.edu}
\author{Xerxes Tata}%
 \email{tata@phys.hawaii.edu}
\affiliation{%
\mbox{Dept. of Physics and Astronomy}, University of Hawaii, Honolulu, HI 96822, U.S.A.\\
}%

\begin{abstract}
\noindent In a theory with broken supersymmetry, gaugino couplings renormalize differently from gauge couplings, as do higgsino couplings from Higgs boson couplings. As a result, we expect the gauge (Higgs boson) couplings and the corresponding gaugino (higgsino) couplings to evolve to different values under renormalization group evolution. We re-examine the renormalization group equations (RGEs) for these couplings in the Minimal Supersymmetric Standard Model (MSSM). To include threshold effects, we calculate the $\beta$-functions using a sequence of (non-supersymmetric) effective theories with heavy particles decoupled at the scale of their mass. We find that the difference between the SM couplings and their SUSY cousins that is ignored in the literature may be larger than two-loop effects which are included, and further that renormalization group evolution induces a non-trivial flavour structure in gaugino interactions. We present here the coupled set of RGEs for these dimensionless gauge and ``Yukawa''-type couplings. The RGEs for the dimensionful SSB parameters of the MSSM will be presented in a companion paper.
\end{abstract}



\maketitle

\section{Introduction}\label{intro}

Renormalization group equations (RGEs) have played an important role in
extracting phenomenological predictions of theories valid at very high
energy scales. One of the best known examples of this is
the prediction of the weak mixing angle from the simplest SUSY $SU(5)$ grand
unified theories (GUTs): the equality of the gauge couplings at 
$Q=M_{\rm GUT}$, which implies $\sin^{2}{\theta_{W}}(M_{\rm GUT})=3/8$, leads 
to the measured value of $\sin^{2}{\theta_{W}}$ only when these couplings are 
evolved to the scale $Q\sim100\ \textrm{GeV}$ relevant to experiments
\cite{gutbounds}. Since at least the 1980's, RGEs have played a role in 
the analysis of the implications of widely different models of new physics, 
including supersymmetric models, the subject of this paper 
\cite{earlyrges}. 

Supersymmetry (SUSY) is a novel symmetry that links bosons and fermions,
providing a level of synthesis never before attained 
\cite{wss,drees,binetruy,primer}. Since
supersymmetry implies a super-partner for every chirality state of the
Standard Model (SM) with the same internal quantum numbers, it implies
an approximate doubling of the degrees of freedom. Supersymmetry, even if 
softly broken, fixes
the dimensionless couplings that determine the tree-level interactions of the
superpartners with SM particles and with one another ({\it i.e.}
interactions via mass dimension four), so that in this sense the theory
is completely predictive. The interest in SUSY phenomenology was sparked
once it was understood that the SUSY non-renormalization theorem implied
the stability of the hierarchy between the weak and GUT or Planck scales
to radiative corrections \cite{dimgeor}. Since SUSY is clearly not manifested in 
the observed spectrum of elementary particles, SUSY must be (explicitly or
spontaneously) broken. Moreover, the mass scale of SM superpartners --
at least of those that couple sizably to the Higgs sector -- must be
smaller than ${\cal O}(1)$~TeV, in order for SUSY breaking not to
destabilize the hard-won hierarchy between the weak and GUT scale just
mentioned. In view of our complete ignorance of the mechanism of SUSY
breaking, it is expedient to parametrize SUSY breaking by introducing
soft-supersymmetry-breaking (SSB) operators with dimension $\le 3$ that
respect Lorentz, gauge and any other symmetries into the effective
Lagrangian of the theory. These include gaugino mass parameters, 
the SSB Higgs mass parameter $b$ often written as $B\mu$,
sfermion mass matrices, and trilinear scalar coupling parameters,
usually denoted by $A_{ijk}$ \cite{SSB}. 

We emphasize that the proliferation of SSB parameters is a result of our
ignorance of how SUSY breaking is communicated to MSSM
super-partners. Until this is understood, we are led to resort to a
variety of models \cite{mSUGRA,GMSB,AMSB,gaugino}. Typically these models, 
which are based on
differing assumptions about {\it high scale} physics, lead to different
ans\"atze for the pattern of SSB parameters, in a Lagrangian
renormalized at this high scale $Q_{\rm high}$. This
Lagrangian cannot directly be used to perturbatively extract
phenomenological predictions at the $E \sim 100$~GeV scale of our
experiments, because large logarithms $\log \left(Q_{\rm high}/E
\right)$ would invalidate (fixed order) perturbation theory. Instead,
RGEs are used to evolve all the Lagrangian parameters to the scale $E$,
and this resulting ``low energy'' Lagrangian is used to evaluate
its implications \cite{earlyrges}. 

Since supersymmetric models automatically contain many scalars with the
same gauge quantum numbers, there are potential new sources of flavour
violation \cite{dimgeor,flavour}. Within the parametrization of the SSB
terms described above, these are encapsulated in the sfermion mass
matrices and in the $A$-parameters (which may be regarded as matrices in
the flavour space). Our ultimate goal is to develop a program that will
enable a general study of flavour violation, for arbitrary values of SSB
parameters at $Q=Q_{\rm high}$. For the present, we confine ourselves to
quark flavour violation.\footnote{The analysis can readily be extended
to lepton flavour violation. Although lepton flavour violation may be
more striking and experimentally easier to access than quark flavour
violation, it may depend on yet other unknowns such as the structure of
any singlet neutrino mass matrix.}

Toward this end, we need to know the RGEs for all dimensionless and
dimensionful parameters. These are already known, both in the SM
\cite{mv} and in the MSSM \cite{bbo,yamada,martv,jackjones} at the
two-loop level. Working to two loop accuracy means that we must include
threshold effects in the one-loop RGEs, since these can
numerically be as large as the two loop terms, and perhaps even larger
if the sparticle spectrum is very split. It is in this connection that
we extend the existing literature \cite{ramond,sakis} as described
below.  As in these studies, we implement SUSY and Higgs particle
thresholds as step functions in our evaluation of the
$\beta$-functions. We consider first the evolution of the dimensionless
couplings. We assume that for any particle with mass $M_i$ between the
weak scale and the scale $Q_{\rm high}$ introduced above, the effective
theory used to calculate the $\beta$-function includes this particle if
$Q>M_i$, but not otherwise. Thus, at a high scale $Q$ above the masses
of all sparticles and Higgs bosons, the $\beta$-functions are those of
the MSSM. As we come down in scale, the SUSY particles and the heavy
Higgs bosons decouple one-by-one until we are left with just the SM
particles, and the $\beta$-functions for the dimensionless couplings of
the SM. Although this sounds quite straightforward in principle, there
are several of issues that need to be confronted to implement this
program.
\begin{itemize}
\item The step-function decoupling of particles at the scale of their 
  mass makes sense for couplings of fields only in their (at least, 
  approximate) mass basis. For instance, in the MSSM Higgs sector, it is 
  quite clear where the spin zero mass eigenstates $h$ or $H$ (or $A$ 
  and $H^{\pm}$) decouple, but 
  the decoupling point for the $h_{u}$ or $h_{d}$ fields (or of the 
  higgsinos) is ambiguous, because the mass eigenstates are 
  combinations of the $h_u$ and $h_d$
  type fields (or the corresponding higgsinos). Moreover, since the 
  Higgs bosons and higgsinos acquire mass from different origins, these
  are not ``diagonalized'' by a common rotation, so that in order to
  implement the decoupling we must in principle work with Higgs
  boson/higgsino bases that are not SUSY transforms of one another.
  See, however, Sec.~\ref{subsec:massbasis} for a simplification.

\item Since we decouple the superpartners but not the lighter SM
  particles, the effective theory below the scale of the heaviest
  sparticle ``knows'' about SUSY breaking, and necessarily includes new
  dimensionless couplings. For instance, while in the SUSY limit, the
  coupling of the gauge boson to quarks is exactly the same for the
  corresponding gaugino-quark-squark coupling, below the scale where we
  begin to decouple heavy particles, these couplings no longer run
  together. For every gauge coupling $g_i$, the low energy theory may
  include additional distinct gaugino-particle-sparticle couplings
  ${\tilde{g}}^{\Phi}_{i}$ (labelled by the scalar $\Phi$) that evolve
  differently from $g_i$. For instance, if $m_{\tilde{q}}\ll
  m_{\tilde{g}}$, below $Q=m_{\tilde{g}}$ the bino coupling
  $\tilde{g}'^{Q}$ evolves differently from $g'$. Moreover,
  $\tilde{g}'^{Q}$ and $\tilde{g}'^L$ (as well as other 
  ${\tilde{g}}'^{\Phi}$ couplings) also do not evolve in the same way. 
  Indeed, as we will see later,
  these couplings also develop flavour-violating components and so
  become matrices in the flavour-space, which we will denote by
  $\tilde{\mathbf{g}}_i^\Phi$, once again labelled by the scalar $\Phi$.
  

\item The proliferation of these new fermion-fermion-scalar couplings
  also extends to the Higgs sector. For instance, if the heavier Higgs
  bosons $H^{\pm}$, $A$ and $H$ all decouple at the scale $Q\simeq m_H$
  that is hierarchically larger than $|\mu|$, and squarks are light, the
  effective theory in the range $|\mu| < Q <m_H$ includes both higgsino
  doublets ${\tilde{h}}_u$ and ${\tilde{h}}_d$ coupling to various
  flavours of quarks and squarks, but with quark (and squark) pairs
  coupling only to the light SM-like Higgs boson $h$. Thus, we must also
  have independent higgsino coupling matrices that we denote by
  ${\tilde{\bf{f}}}_{u,d,e}^\Phi$, once again, labelled by the scalar.

\end{itemize}

We see that even for the dimensionless couplings, there are new
complications which (to our knowledge) have not been included in
previous studies, but must be taken into account for a proper
implementation of threshold effects into the (one-loop) RGEs.
Motivated by the fact that our goal is to study flavour violation in
sparticle decays in a general way, we derive the
RGEs for all the relevant dimensionless couplings of the MSSM, including
flavour-violating superpotential interactions, in this paper. The number
of RGEs for the dimensionless couplings now expands from those for the
three gauge couplings and three Yukawa coupling matrices to those for
the three gauge couplings, nineteen scalar-fermion-fermion
coupling matrices and four Higgs-higgsino-gaugino couplings.
We should mention that, in addition, there are numerous quartic scalar 
couplings arising from the $D$-terms or the $F$-terms in the scalar potential.
Although these are given by the ``squares'' of gauge or superpotential 
Yukawa couplings in the SUSY limit, they will also renormalize 
differently below the SUSY thresholds. We do not discuss the evolution 
of these couplings which are less important phenomenologically. 
Fortunately, these couplings do not enter the evolution of the 
dimensionless couplings that we do consider in this paper.
We derive the RGEs for the dimensionful
parameters, the scalar masses (matrices in the case of squarks), the
trilinear coupling matrices and the SSB $B\mu$ parameter, which have
their own complications, in a companion paper \cite{dimensionful}.

Since, as we mentioned above, the threshold effects necessarily involve
SUSY breaking, we derive our RGEs from the RGEs in a general ({\it i.e.}
not necessarily supersymmetric) field theory that have been worked out
in the literature. We use the seminal papers by Machacek and Vaughn
\cite{mv} to derive the RGEs for the dimensionless couplings, while we
use the work by Luo, Wang and Xiao \cite{luo} to derive those for the
dimensionful couplings of the MSSM. These studies both use the two
component spinor formalism. In contrast, most phenomenologists are much
more familiar with the four-component formalism. To facilitate the use
of the general RGEs, we re-cast the formulae for the RGEs for
dimensionless couplings into four-component notation in
Sec.~\ref{formalism}. In Sec.~\ref{threshold} we discuss the transition to
the ``mass basis'' necessary for the implementation of the MSSM
thresholds. The actual derivation of the RGEs is carried out in
Sec.~\ref{derive}. In Sec.~\ref{anal} we show numerical examples of the
solutions to some of the RGEs and discuss the impact of the new effects we
have included in our analysis. We summarize in Sec.~\ref{summary}. The
set of RGEs for the dimensionless couplings is listed in the
Appendix. 

\section{Formalism} \label{formalism}
The one-loop RGEs for gauge couplings are well known and take the 
form \cite{oneloopg},
\begin{equation} \label{eq:gaugerge}
\left(4\pi\right)^{2}\left.\beta_{g}\right|_{1-\mathrm{loop}}=
-g^{3}\left[\frac{11}{3}C(G)-\frac{2}{3}\sum_{fermions}S(R_{F})
-\frac{1}{3}\sum_{scalars}S(R_{S})\right]\;,
\end{equation}
where $C(G)$ is the quadratic Casimir for the adjoint representation of
the associated Lie algebra, and $S(R_F)$ and $S(R_S)$, respectively, are
the Dynkin indices for the representations $R_F$, $R_S$ under which the
fermions and (complex) scalars transform. For the Lie algebra of
$SU(N)$, $C(G)=N$, while $S(R) =1/2$ for the fundamental $N$-dimensional
representation, and $S(R)=N$ for the adjoint representation. For the
$U(1)_Y$ gauge coupling $g'$, $C(G)=0$ while $S(R)=(Y/2)^{2}$. Heavy
scalars and fermions are decoupled simply by excluding them from the
sums on the right-hand-side of (\ref{eq:gaugerge}).

The derivation of the RGEs for ``Yukawa-type'' couplings of scalars with
fermions is more involved. 
As discussed in Sec.~\ref{intro} our strategy for deriving the RGEs is
to first cast the general two-component results in Ref.~\cite{mv} and
Ref.~\cite{luo} into four-component notation, match the parameters with
those of the MSSM Lagrangian density, and write down the RGEs. Toward
this end, we begin with the Lagrangian density written in terms of
\textit{two-component} spinor fields $\psi_p$ and {\it real} scalar fields
$\phi_a$, as
\begin{equation}\begin{split}\label{eq:twolag}
\mathcal{L}_{(2)}= &\ i\psi^\dagger_p\sigma^\mu D_\mu\psi_p + \frac{1}{2}D_\mu\phi_a D^\mu\phi_a -\frac{1}{4}F_{\mu\nu A}F^{\mu\nu}_A\\
&-\frac{1}{2}\left[\left(\mathbf{m}_{f}\right)_{pq}\psi^{T}_{p}\zeta\psi_{q}+\mathrm{h.c.}\right]-\frac{1}{2!}\mathbf{m}^{2}_{ab}\phi_{a}\phi_{b}\\
&-\left(\frac{1}{2}\by^a_{pq}\psi^T_p\zeta\psi_q\phi_a+\mathrm{h.c.}\right)- \frac{1}{3!}\mathbf{h}_{abc}\phi_{a}\phi_{b}\phi_{c}-\frac{1}{4!}\bm{\lambda}_{abcd}\phi_{a}\phi_{b}\phi_{c}\phi_{d}\;.
\end{split}\end{equation}
The anti-symmetric matrix $\zeta\equiv i\sigma_2$, which is included to make
the spinor bilinears Lorentz invariant, also ensures that the Yukawa
coupling matrices $\mathbf{Y}^a$ (labelled by the scalar field that couples to
the spinors) are symmetric in the fermion field type indices $p, q$. The spinor
mass matrices $\mathbf{m}_f$ are likewise symmetric, but not necessarily
Hermitian. The coefficients $\mathbf{m}_{ab}$, $\mathbf{h}_{abc}$ and 
$\bm{\lambda}_{abcd}$ are real and completely symmetric under permutations of
all their indices.\footnote{The factor $\frac{1}{2}$ in front of the Yukawa
  couplings, which is a reflection of the symmetry of the Yukawa coupling
  matrices, was included in Ref.~\cite{luo} but not in Ref.~\cite{mv}.}

We now turn to the corresponding Lagrangian density in the
\textit{four-component} notation with Dirac and Majorana spinor fields and {\it
complex} scalar fields, $\Phi_{a}$, usually used by phenomenologists. The familiar
four-component Dirac spinor $\Psi_D$ has unrelated left- and right-chiral
components and is, therefore, made up by two independent two-component
spinors $\psi_L$ and $\psi_R$ according to $\Psi_D \equiv (\psi_L,
-\zeta\psi_R^*)$ (where we really mean to write this equation as one for
the column matrix $\Psi_D$). In contrast, the left- and right- chiral
components of a Majorana spinor which, by definition, satisfies $\Psi_M
= C\bar{\Psi}_M^T$ are related. Thus, for example, a Majorana spinor may
be written in terms of a single two-component spinor as 
$\Psi_M \equiv (\psi_L,-\zeta\psi_L^*)$ or alternatively, as
$(\zeta\psi_R^{*},\psi_R)$.  The general form of the Lagrangian 
density that we work with in the context of the MSSM with a conserved 
$R$-parity then takes the form,
\begin{equation}\begin{split}\label{eq:fourlag}
\mathcal{L}_{(4)} = &\ \frac{i}{2}\bar{\Psi}_{j} \gamma^\mu D_\mu \Psi_j + \left(D_{\mu}\Phi_{a}\right)^{\dagger}\left(D^{\mu}\Phi_{a}\right) - \frac{1}{4}F_{\mu\nu A}F^{\mu\nu}_A\\
&-\frac{1}{2}\left[\left(\mathbf{m}_{X}\right)_{jk}\bar{\Psi}_{Mj}\Psi_{Mk}+i\left(\mathbf{m}'_{X}\right)_{jk}\bar{\Psi}_{Mj}\gamma_{5}\Psi_{Mk}\right]+\left[\frac{1}{2!}\bm{\mathcal{B}}_{ab}\Phi_a\Phi_b+\mathrm{h.c.}\right]-\mathbf{m}^{2}_{ab}\Phi^\dagger_a\Phi_b\\
&-\left[\left(\bu^{1}_{a}\right)_{jk}\bar{\Psi}_{Dj}P_L\Psi_{Dk}\Phi_a+\left(\bu^{2}_{a}\right)_{jk}\bar{\Psi}_{Dj}P_L\Psi_{Dk}\Phi^\dagger_a\right.\\
&\quad\ +\left(\bv_{a}\right)_{jk}\bar{\Psi}_{Dj}P_L\Psi_{Mk}\Phi_a+\left(\bw_{a}\right)_{jk}\bar{\Psi}_{Mj}P_L\Psi_{Dk}\Phi^\dagger_a\\
&\left.\quad\ +\frac{1}{2}\left(\bx^{1}_{a}\right)_{jk}\bar{\Psi}_{Mj}P_L\Psi_{Mk}\Phi_a+\frac{1}{2}\left(\bx^{2}_{a}\right)_{jk}\bar{\Psi}_{Mj}P_L\Psi_{Mk}\Phi^\dagger_a+\mathrm{h.c.}\right]\\
&+\left[\frac{1}{2!}\Phi^\dagger_a\mathbf{H}_{abc}\Phi_b\Phi_c+\mathrm{h.c.}\right]-\frac{1}{2!}\frac{1}{2!}\bm{\Lambda}_{abcd}\Phi^\dagger_a\Phi^\dagger_b\Phi_c\Phi_d-\left[\frac{1}{3!}\bm{\Lambda}'_{abcd}\Phi^{\dagger}_{a}\Phi_{b}\Phi_{c}\Phi_{d}+\mathrm{h.c.}\right]\;.
\\
\end{split}\end{equation}
In (\ref{eq:fourlag}), Yukawa couplings of Dirac fields to scalars
$\Phi_a$ are denoted by the matrices $\bu^{1}_{a}$ and $\bu^{2}_{a}$,
while couplings of scalars to one Majorana and one Dirac field are
denoted by the matrices $\bv_{a}$ and $\bw_{a}$. These matrices have no
particular symmetry (or hermiticity) properties under interchange of the
fermion field indices $j$ and $k$. These indices label the fermion field
type (quark, lepton, gaugino, higgsino) and also carry information of
flavour and other quantum numbers (\textit{e.g.} weak isospin and
colour). On the other hand, the matrices $\bx_a^1$ and $\bx_a^2$ 
that couple scalars to two Majorana fields are
symmetric under $j \leftrightarrow k$ because of the symmetry properties
\cite{wss} of the Majorana spinor bilinears that appear in
(\ref{eq:fourlag}).  The scalar mass squared matrix $\mathbf{m}^2$ is
Hermitian in the scalar field indices, $a,b$, while the matrix
$\bm{\mathcal{B}}$ is symmetric.  The Majorana fermion mass matrices
$\mathbf{m}_X$ and $\mathbf{m}'_X$ are both Hermitian and symmetric. The
matrix $\mathbf{m}_X$ is the familiar mass matrix, while $\mathbf{m}'_X$
is the coefficient of the $CP$-violating Majorana fermion bilinear \cite{wss},
whose effects in the two-component spinor language show up as phases in
the entries of the fermion mass matrix $\mathbf{m}_f$ in
(\ref{eq:twolag}).  The trilinear and quartic scalar couplings
$\bm{\Lambda}$, $\bm{\Lambda}'$ and $\mathbf{H}$ are symmetric under
interchanges $a \leftrightarrow b$ and/or $c \leftrightarrow d$ for
$\mathbf{\Lambda}$, under interchanges of $b,c,d$ for
$\mathbf{\Lambda}'$, and finally under $b\leftrightarrow c$ for
$\mathbf{H}$.

Before proceeding further, we should point out that the Lagrangian
density in (\ref{eq:fourlag}) is not the most general one that we can
write. For instance, we have not included mass terms for Dirac
fermions. Also, in writing the Yukawa interactions of fermions, we have
not included terms with the operators $$\overline{\Psi}_M P_{L}\Psi_D\Phi, \
\ {\rm and}\ \ \overline{\Psi}_D P_L \Psi_M \Phi^\dagger\;.$$ Dirac fermion
masses appear in the MSSM only upon the spontaneous breakdown of
electroweak symmetry, while the omitted Yukawa terms do not appear in
the MSSM with $R$-parity conservation because Dirac fermions then carry
a baryon or lepton number which is assumed to be conserved by these
renormalizable operators. Likewise, we have also written only the subset of 
all scalar interactions that we need for our analysis of the MSSM.

Upon substitution of the four-component spinors in terms of their
two-component cousins, and decomposing the complex field $$\Phi_a =
\frac{\Phi_{a_{R}}+ {\rm i}\Phi_{a_{I}}}{\sqrt{2}}$$ into its real and
imaginary pieces, both of which are real fields, we can re-cast
(\ref{eq:fourlag}) into the general form of (\ref{eq:twolag}), involving
only two-component spinors and real scalar fields. This then allows us
to translate the parameters in (\ref{eq:fourlag}) into the corresponding
parameters in (\ref{eq:twolag}). We can now use the general results for
the RGEs for the dimensionless \cite{mv} and dimensionful \cite{luo}
parameters of a general quantum field theory to obtain the RGEs for the
dimensionless Yukawa coupling matrices as well as for the mass and
trilinear coupling parameters that appear in (\ref{eq:fourlag}). 
We then use these to obtain the RGEs for the parameters of the 
MSSM, including the effects of decouplings of various sparticles. 

Since, our focus in this paper is on the RGEs for the dimensionless
parameters contained in the matrices $\bu^{1}_{a}$, $\bu^{2}_{a}$, $\bv_{a}$,
$\bw_{a}$, $\bx^{1}_{a}$ and $\bx^{2}_{a}$, we begin by re-casting the
corresponding terms into two spinor component notation to find,
\begin{equation}\label{eq:fourlagtwo}\begin{split}
\mathcal{L}_{(4)}\ni\ &-\left[\frac{1}{2}(\vec{\psi})^{T}\zeta\frac{1}{\sqrt{2}}\left(\begin{array}{ccc}\mathbf{0}&\left({\bu^{1}_{a}}^T+{\bu^{2}_{a}}^T\right)&{\bw_{a}}^T\\\left(\bu^{1}_{a}+\bu^{2}_{a}\right)&\mathbf{0}&\bv_{a}\\\bw_{a}&{\bv_{a}}^T&\left(\bx^{1}_{a}+\bx^{2}_{a}\right)\end{array}\right)(\vec{\psi})\Phi_{a_{R}}\right.\\[10pt]
&\left.+\frac{1}{2}(\vec{\psi})^{T}\zeta\frac{i}{\sqrt{2}}\left(\begin{array}{ccc}\mathbf{0}&\left({\bu^{1}_{a}}^T-{\bu^{2}_{a}}^T\right)&-{\bw_{a}}^T\\\left(\bu^{1}_{a}-\bu^{2}_{a}\right)&\mathbf{0}&\bv_{a}\\-\bw_{a}&{\bv_{a}}^T&\left(\bx^{1}_{a}-\bx^{2}_{a}\right)\end{array}\right)(\vec{\psi})\Phi_{a_{I}}\right]+\mathrm{h.c.}
\end{split}\end{equation}
where $$\vec{\psi}\equiv\left(\begin{array}{c}\psi_L\\\psi_R\\\psi_M\end{array}\right),$$ 
with the Dirac spinor $\Psi_D \equiv (\psi_L,-\zeta\psi_R^*)$ and the
Majorana spinor $\Psi_M \equiv (\psi_M,-\zeta\psi_M^*)$.
Notice that the two-component spinor fields are coupled to {\it real}
scalar fields, $\Phi_{a_R}$ and $\Phi_{a_I}$ via symmetric (but not
necessarily Hermitian) matrices as expected. We can now use the
one-loop RGE for the Yukawa matrices $\by$ that appear in (\ref{eq:twolag}) 
\cite{mv,luo}\footnote{Eq.~(\ref{eq:mv}) is slightly modified from that in
    Ref.~\cite{mv}. We have written the $Y_2^T(F)$ instead of
    $Y_2^\dagger(F)$ in the first term and symmetrized the trace with
    respect to $a$ and $b$. The second modification also appears in
    Ref.~\cite{luo}, while the first one preserves the symmetry of the
    Yukawa coupling matrix.},
\begin{equation}\begin{split}\label{eq:mv}
\left(4\pi\right)^2\left.\bbet_Y^a\right|_{1-\mathrm{loop}}=&\tfrac{1}{2}\left[\by^T_2(F)\by^a+\by^a\by_2(F)\right]+2\by^b\by^{a\dagger}\by^b\\[2pt]
&+\by^b\mathrm{Tr}\left\{\tfrac{1}{2}\left(\by^{b\dagger}\by^a+\by^{a\dagger}\by^b\right)\right\}-3g^2\left\{\mathbf{C}_2(F),\by^a\right\}\;, 
\end{split}\end{equation}
to obtain the one-loop $\beta$ functions for the various dimensionless
couplings in (\ref{eq:fourlag}). Here, ${\mathbf Y}_2(F)={\mathbf
Y}^{b\dagger}{\mathbf Y}^b$ and ${\mathbf C}_2(F)={\bf t}^A{\mathbf
t}^A$, where ${\bf t}^A$ are the group generators
in the reducible representation that includes all the fermion
fields. Clearly, the matrices $\mathbf{Y}^a$, $\mathbf{Y}_2(F)$ and
$\mathbf{C}_2(F)$ all have the same dimensionality, determined by the
total number of two-component fermion fields in the system.  We also
draw the reader's attention to the fact that in the 
four-component notation the sum over $a,b,...$ runs over the different
complex scalar fields, whereas in the two-component notation, we must
not only sum over the various field types, but also separately over the
real and imaginary parts of the same complex field. In other words, we
have two distinct Yukawa matrices $\by^a$, one when $a$ refers to
$\Phi_{a_{R}}$ and the other when $a$ refers to $\Phi_{a_{I}}$. The trace that
appears above is a sum over the {\it fermion types}.

The $\beta$-functions for the Yukawa coupling matrices in
(\ref{eq:fourlag}) can then be readily found to be,
\begin{equation}\label{eq:U1}\begin{split}
\left(4\pi\right)^2\left.\bbet_{U^1_a}\right|_{1-\mathrm{loop}}=&\frac{1}{2}\left[\left(\bu^1_b\bu^{1\dagger}_b+\bu^2_b\bu^{2\dagger}_b+\bv_b\bv^\dagger_b\right)\bu^1_a+\bu^1_a\left(\bu^{1\dagger}_b\bu^1_b+\bu^{2\dagger}_b\bu^2_b+\bw^\dagger_b\bw_b\right)\right]\\
&+2\left[\bu^1_b\bu^{2\dagger}_a\bu^2_b+\bu^2_b\bu^{2\dagger}_a\bu^1_b+\bv_b\bx^{2\dagger}_a\bw_b\right]\\
&+\bu^1_b\mathrm{Tr}\left\{\left(\bu^{1\dagger}_b\bu^1_a+\bu^{2\dagger}_a\bu^2_b\right)+\frac{1}{2}\left(\bx^{1\dagger}_b\bx^1_a+\bx^{2\dagger}_a\bx^2_b\right)
\right\}\\
&+\bu^2_b\mathrm{Tr}\left\{\left(\bu^{2\dagger}_b\bu^1_a+\bu^{2\dagger}_a\bu^1_b\right)+\frac{1}{2}\left(\bx^{2\dagger}_b\bx^1_a+\bx^{2\dagger}_a\bx^1_b\right)\right\}\\
&-3g^2\left[\bu^1_a\mathbf{C}^L_2(F)+\mathbf{C}^R_2(F)\bu^1_a\right]\\[10pt]
&+\frac{1}{2}\bv_a\left(\bv^\dagger_b\bu^1_b+\bx^{2\dagger}_b\bw_b\right)+2\bu^1_b\bw^\dagger_a\bw_b+\bu^{1}_{b}\mathrm{Tr}\left\{\bw^\dagger_a\bw_b+\bv^\dagger_b\bv_a\right\}\;, 
\end{split}\end{equation}
\begin{equation}\label{eq:U2}\begin{split}
\left(4\pi\right)^2\left.\bbet_{U^2_a}\right|_{1-\mathrm{loop}}=&\frac{1}{2}\left[\left(\bu^1_b\bu^{1\dagger}_b+\bu^2_b\bu^{2\dagger}_b+\bv_b\bv^\dagger_b\right)\bu^2_a+\bu^2_a\left(\bu^{1\dagger}_b\bu^1_b+\bu^{2\dagger}_b\bu^2_b+\bw^\dagger_b\bw_b\right)\right]\\
&+2\left[\bu^1_b\bu^{1\dagger}_a\bu^2_b+\bu^2_b\bu^{1\dagger}_a\bu^1_b+\bv_b\bx^{1\dagger}_a\bw_b\right]\\
&+\bu^2_b\mathrm{Tr}\left\{\left(\bu^{2\dagger}_b\bu^2_a+\bu^{1\dagger}_a\bu^1_b\right)+\frac{1}{2}\left(\bx^{2\dagger}_b\bx^2_a+\bx^{1\dagger}_a\bx^1_b\right)\right\}\\
&+\bu^1_b\mathrm{Tr}\left\{\left(\bu^{1\dagger}_b\bu^2_a+\bu^{1\dagger}_a\bu^2_b\right)+\frac{1}{2}\left(\bx^{1\dagger}_b\bx^2_a+\bx^{1\dagger}_a\bx^2_b\right)\right\}\\
&-3g^2\left[\bu^2_a\mathbf{C}^L_2(F)+\mathbf{C}^R_2(F)\bu^2_a\right]\\[10pt]
&+\frac{1}{2}\left(\bu^{2}_{b}\bw^\dagger_b+\bv_b\bx^{1\dagger}_b\right)\bw_a+2\bv_{b}\bv^\dagger_a\bu^2_b+\bu^{2}_{b}\mathrm{Tr}\left\{\bw^\dagger_b\bw_a+\bv^\dagger_a\bv_b\right\}
\end{split}\end{equation}
Within the MSSM with $R$-parity conservation, the last line of each
of these equations vanishes because $\mathbf{W}_a$ and $\mathbf{V}_a$
vanish for the $a$ values for which $\mathbf{U}^1_a$ and
$\mathbf{U}^2_a$ are non-zero.
Continuing,
\begin{equation}\label{eq:X1}\begin{split}
\left(4\pi\right)^2\left.\bbet_{X^1_a}\right|_{1-\mathrm{loop}}=&\frac{1}{2}\left[\left(\bw_{b}\bw^{\dagger}_b+\bv^{T}_{b}\bv^{*}_{b}+\bx^{1}_{b}\bx^{1\dagger}_{b}+\bx^{2}_{b}\bx^{2\dagger}_{b}\right)\bx^{1}_{a}\right.\\
&\left.\quad+\bx^{1}_{a}\left(\bw^{*}_{b}\bw^{T}_{b}+\bv^{\dagger}_{b}\bv_{b}+\bx^{1\dagger}_{b}\bx^{1}_{b}+\bx^{2\dagger}_{b}\bx^{2}_{b}\right)\right]\\
&+2\left[\bw_{b}\bu^{2\dagger}_{b}\bv_{b}+\bv^{T}_{b}\bu^{2*}_{a}\bw^{T}_{b}+\bx^{1}_{b}\bx^{2\dagger}_{a}\bx^{2}_{b}+\bx^{2}_{b}\bx^{2\dagger}_{a}\bx^{1}_{b}\right]\\
&+\bx^1_b\mathrm{Tr}\left\{\left(\bu^{1\dagger}_b\bu^1_a+\bu^{2\dagger}_a\bu^2_b\right)+\frac{1}{2}\left(\bx^{1\dagger}_b\bx^1_a+\bx^{2\dagger}_a\bx^2_b\right)
\right\}\\
&+\bx^2_b\mathrm{Tr}\left\{\left(\bu^{2\dagger}_b\bu^1_a+\bu^{2\dagger}_a\bu^1_b\right)+\frac{1}{2}\left(\bx^{2\dagger}_b\bx^1_a+\bx^{2\dagger}_a\bx^1_b\right)\right\}\\
&-3g^2\left[\bx^1_a\mathbf{C}^L_2(F)+\mathbf{C}^R_2(F)\bx^1_a\right]\\[10pt]
&+\frac{1}{2}\left[\left(\bw_{b}\bu^{2\dagger}_{b}+\bx^{1}_{b}\bv^{\dagger}_{b}\right)\bv_{a}+\bv^{T}_a\left(\bu^{2*}_{b}\bw^{T}_{b}+\bv^{*}_{b}\bx^{1}_{b}\right)\right]\\
&+2\left[\bw_{b}\bw^{\dagger}_{a}\bx^{1}_{b}+\bx^{1}_{b}\bw^{*}_{a}\bw^{T}_{b}
\right]+\bx^{1}_{b}\mathrm{Tr}\left\{\bw^\dagger_a\bw_b+\bv^\dagger_b\bv_a\right\}
\end{split}\end{equation}
\begin{equation}\label{eq:X2}\begin{split}
\left(4\pi\right)^2\left.\bbet_{X^2_a}\right|_{1-\mathrm{loop}}=&\frac{1}{2}\left[\left(\bw_{b}\bw^{\dagger}_{b}+\bv^{T}_{b}\bv^{*}_{b}+\bx^{1}_{b}\bx^{1\dagger}_{b}+\bx^{2}_{b}\bx^{2\dagger}_{b}\right)\bx^{2}_{a}\right.\\
&\left.\quad+\bx^{2}_{a}\left(\bw^{*}_{b}\bw^{T}_{b}+\bv^{\dagger}_{b}\bv_{b}+\bx^{1\dagger}_{b}\bx^{1}_{b}+\bx^{2\dagger}_{b}\bx^{2}_{b}\right)\right]\\
&+2\left[\bw_{b}\bu^{1\dagger}_{a}\bv_{b}+\bv^{T}_{b}\bu^{1*}_{a}\bw^{T}_{b}+\bx^{1}_{b}\bx^{1\dagger}_{a}\bx^{2}_{b}+\bx^{2}_{b}\bx^{1\dagger}_{a}\bx^{1}_{b}\right]\\
&+\bx^2_b\mathrm{Tr}\left\{\left(\bu^{2\dagger}_b\bu^2_a+\bu^{1\dagger}_a\bu^1_b\right)+\frac{1}{2}\left(\bx^{2\dagger}_b\bx^2_a+\bx^{1\dagger}_a\bx^1_b\right)\right\}\\
&+\bx^1_b\mathrm{Tr}\left\{\left(\bu^{1\dagger}_b\bu^2_a+\bu^{1\dagger}_a\bu^2_b\right)+\frac{1}{2}\left(\bx^{1\dagger}_b\bx^2_a+\bx^{1\dagger}_a\bx^2_b\right)\right\}\\
&-3g^2\left[\bx^2_a\mathbf{C}^L_2(F)+\mathbf{C}^R_2(F)\bx^2_a\right]\\[10pt]
&+\frac{1}{2}\left[\left(\bv^{T}_{b}\bu^{1*}_{b}+\bx^{2}_{b}\bw^{*}_{b}\right)\bw^{T}_{a}+\bw_{a}\left(\bu^{1\dagger}_{b}\bv_{b}+\bw^{\dagger}_{b}\bx^{2}_{b}\right)\right]\\
&+2\left[\bv^{T}_{b}\bv^{*}_{a}\bx^{2}_{b}+\bx^{2}_{b}\bv^{\dagger}_{a}\bv_{b}
\right]+\bx^{2}_{b}\mathrm{Tr}\left\{\bw^\dagger_b\bw_a+\bv^\dagger_a\bv_b\right\}\;.
\end{split}\end{equation}
The last two lines of each of (\ref{eq:X1}) and (\ref{eq:X2}) vanish in
the $R$-parity conserving MSSM, again because $\mathbf{W}^a$ and
$\mathbf{V}^a$ are zero. Finally,
\begin{equation}\label{eq:Va}\begin{split}
\left(4\pi\right)^2\left.\bbet_{V_a}\right|_{1-\mathrm{loop}}=&\frac{1}{2}\left[\left(\bu^{1}_{b}\bu^{1\dagger}_{b}+\bu^{2}_{b}\bu^{2\dagger}_{b}+\bv_{b}\bv^{\dagger}_{b}\right)\bv_{a}\right.\\
&\left.\quad+\bv_{a}\left(\bw^{*}_{b}\bw^{T}_{b}+\bv^{\dagger}_{b}\bv_{b}+\bx^{1\dagger}_{b}\bx^{1}_{b}+\bx^{2\dagger}_{b}\bx^{2}_{b}\right)\right]\\
&+2\left[\bu^{1}_{b}\bw^{\dagger}_{a}\bx^{2}_{b}+\bu^{2}_{b}\bw^{\dagger}_{a}\bx^{1}_{b}+\bv_{b}\bw^{*}_{a}\bw^{T}_{b}\right]\\
&+\bv_b\mathrm{Tr}\left\{\bw^{\dagger}_{a}\bw_{b}+\bv^{\dagger}_{b}\bv_{a}\right\}-3g^2\left[\bv_a\mathbf{C}^L_2(F)+\mathbf{C}^R_2(F)\bv_a\right]\\[10pt]
&+\frac{1}{2}\left[\left(\bu^{2}_{b}\bw^{\dagger}_{b}+\bv_{b}\bx^{1\dagger}_{b}\right)\left(\bx^{1}_{a}+\bx^{2}_{a}\right)+\left(\bu^{1}_{a}+\bu^{2}_{a}\right)\left(\bu^{1\dagger}_{b}\bv_{b}+\bw^{\dagger}_{b}\bx^{2}_{b}\right)\right]\\
&+2\left[\bu^{2}_{b}\left(\bu^{1\dagger}_{a}+\bu^{2\dagger}_{a}\right)\bv_{b}+\bv_{b}\left(\bx^{1\dagger}_{a}+\bx^{2\dagger}_{a}\right)\bx^{2}_{b}\right]\\
&+\bv_{b}\mathrm{Tr}\left\{\left[\bu^{1\dagger}_{b}\left(\bu^{1}_{a}+\bu^{2}_{a}\right)+\left(\bu^{1\dagger}_{a}+\bu^{2\dagger}_{a}\right)\bu^{2}_{b}\right]\right.\\
&\left.\qquad\qquad+\frac{1}{2}\left[\bx^{1\dagger}_{b}\left(\bx^{1}_{a}+\bx^{2}_{a}\right)+\left(\bx^{1\dagger}_{a}+\bx^{2\dagger}_{a}\right)\bx^{2}_{b}\right]\right\}
\end{split}\end{equation}
\begin{equation}\label{eq:Wa}\begin{split}
\left(4\pi\right)^2\left.\bbet_{W_a}\right|_{1-\mathrm{loop}}=&\frac{1}{2}\left[\left(\bw_{b}\bw^{\dagger}_{b}+\bv^{T}_{b}\bv^{*}_{b}+\bx^{1}_{b}\bx^{1\dagger}_{b}+\bx^{2}_{b}\bx^{2\dagger}_{b}\right)\bw_{a}\right.\\
&\left.\quad+\bw_{a}\left(\bu^{1\dagger}_{b}\bu^{1}_{b}+\bu^{2\dagger}_{b}\bu^{2}_{b}+\bw^{\dagger}_{b}\bw_{b}\right)\right]\\
&+2\left[\bv^{T}_{b}\bv^{*}_{a}\bw_{b}+\bx^{1}_{b}\bv^{\dagger}_{a}\bu^{2}_{b}+\bx^{2}_{b}\bv^{\dagger}_{a}\bu^{1}_{b}\right]\\
&+\bw_b\mathrm{Tr}\left\{\bw^{\dagger}_{b}\bw_{a}+\bv^{\dagger}_{a}\bv_{b}\right\}-3g^2\left[\bw_a\mathbf{C}^L_2(F)+\mathbf{C}^R_2(F)\bw_a\right]\\[10pt]
&+\frac{1}{2}\left[\left(\bw_{b}\bu^{2\dagger}_{b}+\bx^{1}_{b}\bv^{\dagger}_{b}\right)\left(\bu^{1}_{a}+\bu^{2}_{a}\right)+\left(\bx^{1}_{a}+\bx^{2}_{a}\right)\left(\bv^{\dagger}_{b}\bu^{1}_{b}+\bx^{1\dagger}_{b}\bw_{b}\right)\right]\\
&+2\left[\bw_{b}\left(\bu^{1\dagger}_{a}+\bu^{2\dagger}_{a}\right)\bu^{1}_{b}+\bx^{1}_{b}\left(\bx^{1\dagger}_{a}+\bx^{2\dagger}_{a}\right)\bw_{b}\right]\\
&+\bw_{b}\mathrm{Tr}\left\{\left[\bu^{2\dagger}_{b}\left(\bu^{1}_{a}+\bu^{2}_{a}\right)+\left(\bu^{1\dagger}_{a}+\bu^{2\dagger}_{a}\right)\bu^{1}_{b}\right]\right.\\
&\left.\qquad\qquad+\frac{1}{2}\left[\bx^{2\dagger}_{b}\left(\bx^{1}_{a}+\bx^{2}_{a}\right)+\left(\bx^{1\dagger}_{a}+\bx^{2\dagger}_{a}\right)\bx^{1}_{b}\right]\right\}
\end{split}\end{equation}
The last four lines of each of (\ref{eq:Va}) and (\ref{eq:Wa}) do not
contribute in the $R$-parity conserving MSSM because each of
$\mathbf{U}^1_a$, $\mathbf{U}^2_a$, $\mathbf{X}^1_a$ and
$\mathbf{X}^2_a$ vanish for the values of $a$ for which $\mathbf{V}_a$
and $\mathbf{W}_a$ do not. 

The quantities $\mathbf{C}^L_2(F)$ and
$\mathbf{C}^R_2(F)$ that appear in Eq.~(\ref{eq:U1})-(\ref{eq:Wa}) are
the quadratic Casimirs $\mathbf{C}_2(F)$ separated into contributions
from left-handed, right-handed and Majorana fermions, so that
$$\mathbf{C}_2(F)=diag\left(\mathbf{C}_2^L(F), \mathbf{C}_2^R(F),
\mathbf{C}_2^M(F)\right)\;.$$ The traces that appear in
(\ref{eq:U1})--(\ref{eq:Wa}) denote a sum over fermion types, which 
sometimes, but not necessarily, are the trace over fermion flavours when
we refer to the MSSM. Note also that in the context of the MSSM, the index $a$
may itself include a flavour (or even colour) index, for instance 
when $\Phi_a=\tilde{c}_L$.

The reader may legitimately wonder what we have gained by writing the
RGE which took the much simpler-looking form (\ref{eq:mv}) in the much more
cumbersome-looking and definitely much longer forms given by
Eq.~(\ref{eq:U1})-(\ref{eq:Wa}).  The reasons for doing so are
two-fold. First, many authors are more familiar with the MSSM couplings
written in four-component notation, so that the longer four-component
equations can be more directly used. Second, and perhaps the more
important reason is that {\it much of the work that would be needed to
obtain the MSSM RGEs, even starting with the couplings in two-component
notation, has already been done when the RGEs are written in as in
(\ref{eq:U1})-(\ref{eq:Wa}).} The many terms that appear here are
indeed present in the MSSM RGEs (as we will see) and starting with the
compact notation of (\ref{eq:mv}) only means that part of the work has
to be repeated each time we use it to get the RGE that we need.

Before closing this section, we remark that the RGEs for the
dimensionless gauge couplings and those for the dimensionless
``Yukawa-like'' couplings form a closed set, uncoupled to the RGEs for
the dimensionful parameters. Dimensionless quartic couplings of scalars
also do not enter these RGEs. Since we have made no assumption about
supersymmetry, this is true even with threshold effects
included.\footnote{Of course, in this case, information about the SUSY
spectrum enters via the location of the thresholds.}  In the next
section, we turn to a discussion of how threshold effects are to be
included in the MSSM RGEs. Of course, since we want to work to two-loop
accuracy, it suffices to include the threshold effects only in the
one-loop terms, which is the reason for our focus on the one-loop RGEs
of a general quantum field theory. We mention here that the authors of
Ref.~\cite{mv} and~\cite{luo}, which is our starting point, use the
$\overline{\mathrm{MS}}$ renormalization scheme rather than the
$\overline{\mathrm{DR}}$ scheme better suited for the analysis of
radiative corrections in supersymmetric theories. We can, however, use
their general expressions (only for the one-loop part) to obtain our
RGEs since there is no scheme-dependence at this level. We can then
augment these with the two-loop terms (without threshold corrections) of
the MSSM RGEs \cite{martv} to obtain two-loop accuracy in the
$\overline{\mathrm{DR}}$ scheme.

\section{Particle Decoupling}\label{threshold}

We are now equipped to derive the RGEs for the dimensionless couplings 
in any quantum field theory with an arbitrary set of Dirac and Majorana
spinor fields together with spin zero fields (with non-trivial gauge
quantum numbers) whose Lagrangian can be written in the form
(\ref{eq:fourlag}). This is, as we have already noted, not the most
general form for the Lagrangian density since some field operators, such
as mass terms for Dirac fermions or dimension four operators of the form 
$\overline{\Psi}_M P_L\Psi_D\Phi$ and $\overline{\Psi}_D P_L
\Psi_M\Phi^\dagger$, that do not occur in the MSSM with $R$-parity
conservation imposed have been omitted. Since (\ref{eq:fourlag}) makes
no reference to supersymmetry, we can continue to use it even when SUSY
is broken via the inclusion of SSB terms. The procedure for obtaining
the one-loop RGEs is now straightforward. We simply write down the 
Lagrangian density for the MSSM, including the SSB terms consistent with
Lorentz and MSSM gauge symmetries and the assumed conservation of
$R$-parity, compare the dimensionless couplings with those in
(\ref{eq:fourlag}) to construct the matrices $\mathbf{U}^1_a$,
$\mathbf{U}^2_a$,$\cdots$,$\mathbf{W}_a$, and then use
(\ref{eq:U1})-(\ref{eq:Wa}) to get the required RGEs. This program has
two complications that we must take care of in its
implementation. Although we have referred to these in Sec.~\ref{intro},
it does not hurt to repeat them here.
\begin{itemize}
\item For $Q$ values above all SUSY thresholds, the effective theory is
  supersymmetric, and couplings of SM particles are related by SUSY to
  corresponding sparticle couplings, so that the particle and sparticle
  couplings evolve the same way. This is, however, not necessarily the
  case below the SUSY thresholds. Then, while the $\beta$-function for
  say the hypercharge gauge coupling still depends only on the
  hypercharge gauge coupling (because of Ward identities, its one-loop
  evolution is governed by just the ``vacuum polarization
  corrections''), the $\beta$-functions for the corresponding
  ``Yukawa-like'' couplings $\tilde{\bf{g}}'^{\tilde{f}}$ of the bino to
  the fermion-sfermion pair will, in general, depend on gauge as well as
  {\it all} ``Yukawa'' couplings, including those from super-potential
  interactions. There is no principle that precludes this once
  supersymmetry is broken. However, above all SUSY thresholds, this
  $\beta$ function must reduce to that for the gauge coupling, so that
  all dependence on super-potential Yukawa couplings must cancel. This
  provides a non-trivial check on our formulae. Thus our system must be
  extended to include the couplings $\tilde{\bf{g}}^{\Phi}_{i}$ as well
  as the couplings $\tilde{\bf{f}}^{\Phi}_{i}$ the additional
  ``Yukawa-like'' couplings of gauginos and higgsinos, respectively,
  which as we have noted become matrices in the flavour-space. 

\item To implement SUSY and Higgs particle thresholds, we need to have
  an idea of the spectrum as well as the couplings in the corresponding
  mass basis. Of course, the exact spectrum is only
  obtained after extracting the weak scale values of all SSB parameters,
  and then diagonalizing the various mass matrices. Since the positions
  of the thresholds only enter logarithmically in the solutions to the
  RGEs for the various gauge and ``Yukawa-like'' couplings, it suffices
  to include these in a reasonable approximation. Obviously, this 
  entails rewriting the interactions in terms of the approximate mass
  basis for the fields. A discussion of how we
  do so forms the subject of the remainder of this section.
\end{itemize}

\subsection{The Mass basis and MSSM thresholds}\label{subsec:massbasis}

In addition to the sparticles of the SM, the MSSM includes squarks and
sleptons, gluinos, charginos and neutralinos, and the additional spin
zero particles $A$, $H$ and $H^\pm$ in its Higgs sector. At a scale $Q$
larger than the masses of all these particles, the RGEs are given by
those of the MSSM, and as we reduce $Q$, these morph into those of
effective theories that interpolate between the MSSM and the SM, and
ultimately at the scale $Q \sim m_t$, reduce to those of the SM,
assuming that all sparticles and additional Higgs bosons are
significantly heavier than $m_t$. This will serve as another check of
our formulae. 

Assuming for the moment that $\tf_L$ and $\tf_R$ are also approximate
sfermion mass eigenstates, {\it i.e.} that mixing among the sfermions is
modest, decoupling the sfermions is straightforward. Simply exclude
any sfermion field (and all its couplings) in the effective theory at a
scale $Q$ below its mass, and evaluate the RGEs using this truncated
theory. The same is true for gluinos which, being colour octets, cannot
mix with any other fermion since $SU(3)_C$ is unbroken. The decoupling
of charginos, neutralinos and the additional Higgs bosons is more
complicated because these are frequently strongly mixed. 

Our procedure for decoupling these fields is guided by the observation
that threshold effects that we are trying to evaluate are significant
only if there are several well separated mass scales $M_i$ in the spectrum. If
this is not the case, then we can decouple all the additional particles
at a common scale $M_{\rm SUSY}$, since any error would only be of
${\cal O}\left({\frac{1}{16\pi^2}}\ln{\frac{M_{\rm SUSY}}{M_i}}\right)$
and would $\rightarrow 0$ as $M_i \rightarrow M_{\rm SUSY}$. 

In the Higgs boson sector, threshold effects are, therefore, important
only if the mass scale of the additional bosons $\gg m_h$, in which case
$m_A \simeq m_H \simeq m_{H^\pm}\gg m_h < m_t$. We can then decouple
all the additional spin-zero bosons at a common scale that we take to be
$m_H$, below which $h$ plays the role of the SM Higgs boson. To
implement the decoupling, we must rewrite the Lagrangian couplings in
terms of the (approximate) mass eigenstates contained in the
$SU(2)$ doublets\footnote{We use the notation of Ref.~\cite{wss} where
the left-chiral up-type Higgs superfield ${\hat{H}}_u$ transforms as a
{\bf 2} of $SU(2)$, while the left-chiral down-type Higgs superfield
transforms as a ${\bf 2}^*$. The doublets that make up the linear combinations in
(\ref{eq:hrot}) and (\ref{eq:Hrot}) both transform as a ${\bf 2}$ of
$SU(2)$ and have a positive weak hypercharge.}
\begin{eqnarray}
\label{eq:hrot}\left(\begin{array}{c}G^{+}\\\mathsf{h}\end{array}\right)&=\sn\left(\begin{array}{c}h^{+}_{u}\\[5pt]h^{0}_{u}\end{array}\right)+\cs\left(\begin{array}{c}h^{-*}_{d}\\[5pt]h^{0*}_{d}\end{array}\right)\\[5pt] 
\label{eq:Hrot}\left(\begin{array}{c}H^{+}\\\mathcal{H}\end{array}\right)&=\cs\left(\begin{array}{c}h^{+}_{u}\\[5pt]h^{0}_{u}\end{array}\right)-\sn\left(\begin{array}{c}h^{-*}_{d}\\[5pt]h^{0*}_{d}\end{array}\right),
\end{eqnarray}
where the electrically neutral, {\it complex} fields $\mathsf{h}$ and
$\mathcal{H}$ are given by,
\begin{eqnarray*}
\mathsf{h}&=\frac{h+iG^{0}}{\sqrt{2}}\\
\mathcal{H}&=\frac{-H+iA}{\sqrt{2}}\;,
\end{eqnarray*}
and $\sn=\sin{\beta}$ and $\cs=\cos{\beta}$. Above $Q=m_H$, the theory
includes two Higgs doublets which we may take to be $h_u$ and $h_d$, or
{\it any two orthogonal} combinations of these. However, at $Q=m_H$, it
is specifically the doublet (\ref{eq:Hrot}) that decouples, and the
effective theory includes just one spin-zero doublet which couples to
quarks and leptons via Yukawa coupling matrices
$\bm{\lambda}_u=\bdf_u\sin\beta$,
$\bm{\lambda}_d=\bdf_d\cos\beta$ and $\bm{\lambda}_e =
\bdf_{e}\cos\beta$, with these relations imposed at $Q=m_H$. For $Q<m_{H}$,
we will be interested in the evolution of the $\bm{\lambda}_{i}$'s which are the 
couplings of the doublet (\ref{eq:hrot}) that remains in the theory. We
stress again that supersymmetry is broken, and the decoupling of the
second doublet makes no statement about the decoupling of the
corresponding higgsinos. The effective theory below $Q=m_H$ may or may
not contain higgsinos, depending on, as we will see next, the value of
$\mu$. 

In the chargino-neutralino sector, threshold corrections are 
small if $|\mu|$ and the SSB gaugino masses all have a magnitude $\sim
M_Z$, since their effects will then persist down to the weak
scale. Threshold effects are significant only when at least one of these
is hierarchically larger than the weak scale, in which case
gaugino-higgsino mixing effects are small, and the gaugino-higgsino mass
eigenstates can be approximated by the bino, the wino-triplet and the
two Majorana higgsino doublet fields,\footnote{ $\psi_{h_u}$ and
$\psi_{h_d}$ are the fermionic components of the $SU(2)$ doublet
left-chiral superfields ${\hat{H}}_u$ and ${\hat{H}}_d$. Following
Ref.~\cite{wss}, we denote the charged and neutral components of $\psi_{h_u}$
by $(\psi_{h_u^+}, \psi_{h_u^0})$ and of $\psi_{h_d}$ by 
 $(\psi_{h_d^-}, \psi_{h_d^0})$. $\psi_{h_u^+}$ is a Majorana spinor
whose left-chiral component
destroys positively charged higgsinos, while its right-chiral
component destroys negatively charged higgsinos, and similarly for
$\psi_{h_d^-}$.}
\begin{equation}\label{eq:inorot}
\tilde{h}_{1,2}=\frac{\psi_{h_{d}}\mp\psi_{h_{u}}}{\sqrt{2}}\;.
\end{equation}
To implement decoupling, we must rewrite interactions of gauginos and
higgsinos in the basis consisting of binos, winos and these two higgsino
doublets.  Above the scale of the higgsinos, we could have worked with
any two orthogonal linear combinations of the higgsino doublets, whereas
below $Q=|\mu|$, both higgsinos decouple from the effective theory. We
note here that we have to perform different rotations in the spin-zero and
spin-$\frac{1}{2}$ Higgs sectors to go to the mass basis. This appears
to be different\footnote{We mention though that because we choose to decouple
both higgsinos at the common scale $|\mu|$, this rotation of the
higgsino fields is irrelevant for practical purposes.}
from what has been done in the literature
\cite{ramond,sakis}. There is one
additional complication that enters when we rewrite the Lagrangian
density in the (approximate) ``mass basis''. Some of the fermion fields
have negative eigenvalues, and we are led to make
field re-definitions \cite{wss} $\Psi_k \to (i\gamma_5)^{\theta_k}
\Psi_k$, where $\theta_k = 0 (1)$ if the corresponding mass eigenvalue
is positive (negative), on Majorana spinors to get their masses in
canonical form. While this introduces additional $\gamma_5$'s and
phases in the interactions, we have checked the RGEs are independent of
these, as may be expected since these RGEs do not depend on the masses.

Finally, let us return to mixing effects in the sfermion sector. We have
ignored these in our analysis. For the squarks of the first two
generations, flavour-physics constraints typically restrict the size of
inter-generation mixing effects (in the basis where the corresponding
quark Yukawa couplings are diagonal). There is, however, no principle
that dictates intra-generation mixing to be small, even though this is
usually assumed to be the case for first and second generation fermions
in many models, while third generation sfermions are expected to mix via
their Yukawa couplings\footnote{Substantial intra-generational mixing
may occur via the trilinear sfermion-fermion-Higgs boson SSB parameter
which is, after all, independent of the corresponding super-potential
Yukawa coupling.}. This said, barring large, accidental cancellations we
would expect that the physical sfermion masses are approximately of the
same order, in which case threshold effects from intra-generation mixing
are not important, or the diagonal SSB terms are themselves hierarchical
so that the unmixed sfermions $\tf_L$ and $\tf_R$ are approximate mass
eigenstates, and can be decoupled at their physical masses.

\section{Application to the MSSM}\label{derive}

\subsection{Interactions}
We now use the RGEs for the general quantum field theory that we
obtained in Sec.~\ref{formalism} to derive the RGEs for the dimensionless
couplings of the $R$-parity conserving MSSM. In the notation of 
Ref.~\cite{wss}, the superpotential for the MSSM is given by,
\begin{equation}\label{eq:super}
\hat{f}=\tilde{\mu}\hat{H}^a_u\hat{H}_{da}+(\bdf_u)_{ij}\epsilon_{ab}\hat{Q}^a_i\hat{H}^b_u\hat{U}^c_j+(\bdf_d)_{ij}\hat{Q}^a_i\hat{H}_{da}\hat{D}^c_j+(\bdf_e)_{ij}\hat{L}^a_i\hat{H}_{da}\hat{E}^c_j\;,
\end{equation}
while the SSB terms may be written as,
\begin{equation} \label{eq:soft}\begin{split}
\mathcal{L}_{\rm SSB}=\ &-\left\{\tilde{u}^\dagger_{Lk}(\mathbf{m}^2_{Q})_{kl}\tilde{u}_{Ll}
+\tilde{d}^\dagger_{Lk}(\mathbf{m}^2_{Q})_{kl}\tilde{d}_{Ll}
+\tilde{u}^\dagger_{Rk}(\mathbf{m}^2_{U})_{kl}\tilde{u}_{Rl}
+\tilde{d}^\dagger_{Rk}(\mathbf{m}^2_{D})_{kl}\tilde{d}_{Rl}\right.\\
&\quad\ +\tilde{\nu}^\dagger_{Lk}(\mathbf{m}^2_{L})_{kl}\tilde{\nu}_{Ll}
+\tilde{e}^\dagger_{Lk}(\mathbf{m}^2_{L})_{kl}\tilde{e}_{Ll}
+\tilde{e}^\dagger_{Rk}(\mathbf{m}^2_{E})_{kl}\tilde{e}_{Rl}\\[5pt]
&\left.\quad\ +m^2_{H_u}h^{0\dagger}_uh^0_u+m^2_{H_u}h^{+\dagger}_uh^+_u
+m^2_{H_d}h^{0\dagger}_dh^0_d+m^2_{H_d}h^{-\dagger}_dh^-_d\right\}\\[5pt]
&+\left\{\tilde{u}^\dagger_{Rk}(\mathbf{a}^T_u)_{kl}\tilde{u}_{Ll}h^0_u
-\tilde{u}^\dagger_{Rk}(\mathbf{a}^T_u)_{kl}\tilde{d}_{Ll}h^+_u
+\tilde{d}^\dagger_{Rk}(\mathbf{a}^T_d)_{kl}\tilde{d}_{Ll}h^0_d\right.\\
&\left.\quad\ +\tilde{d}^\dagger_{Rk}(\mathbf{a}^T_d)_{kl}\tilde{u}_{Ll}h^-_d
+\tilde{e}^\dagger_{Rk}(\mathbf{a}^T_e)_{kl}\tilde{\nu}_{Ll}h^-_d
+\tilde{e}^\dagger_{Rk}(\mathbf{a}^T_e)_{kl}\tilde{e}_{Ll}h^0_d+\mathrm{h.c.}\right\}\\[5pt]
&-\frac{1}{2}\left[M_1\bar{\lambda}_0\lambda_0+M_2\bar{\lambda}_p\lambda_p+M_3\bar{\tilde{g}}_A\tilde{g}_A\right]\\[5pt]
&-\frac{i}{2}\left[M'_1\bar{\lambda}_0\gamma_{5}\lambda_0+M'_2\bar{\lambda}_p\gamma_{5}\lambda_p+M'_3\bar{\tilde{g}}_A\gamma_{5}\tilde{g}_A\right]\\[5pt]
&+\left\{b(h^+_uh^-_d+h^0_uh^0_d)+\mathrm{h.c.}\right\}.
\end{split}\end{equation}
where the matrix indices $i,j,\cdots$ label the (s)particle flavours. 
The sfermion mass squared matrices $\mathbf{m}^{2}_{Q}$, 
$\mathbf{m}^{2}_{U}$, $\dots$ are Hermitian, whereas the Yukawa
coupling matrices $\mathbf{f}_{u,d,e}$ that appear in the
superpotential, as well as the trilinear SSB scalar coupling matrices
$\mathbf{a}_{u,d,e}$ have no hermiticity or symmetry property. Finally, the
gaugino mass parameters $M_a$ and $M'_a$ are real, while the SSB Higgs
mass parameter may, in general, be complex. Since the RGEs for the
dimensionless couplings do not depend on any dimensionful parameters, we do
not really need $\mathcal{L}_{\rm SSB}$ for their derivation. The SSB
parameters only enter the analysis in that they determine the positions
of the various thresholds.  

In order to be able to compare our results (above all thresholds) with
the standard results already in the literature \cite{martv,bbo,yamada,jackjones} 
we
need the following conversion between the notation used here and that in
Ref.~\cite{martv}: $\bdf\equiv\mathbf{Y}^{T}$,
$\mathbf{a}\equiv-\mathbf{h}^{T}$, $b\equiv-B$. Moreover the {\it
complex} gaugino mass parameters ($M_a$) in Ref.~\cite{martv} are
equivalent to $M_a-iM'_a$ in (\ref{eq:soft}).

It is clear that in order to obtain the RGEs, we need to extract the
various matrices $\mathbf{U}^{1,2}_\Phi$, $\mathbf{X}^{1,2}_\Phi$,
$\mathbf{V}_\Phi$ and $\mathbf{W}_\Phi$, whose elements are the
couplings between the various scalars, $\Phi$, and fermions of the MSSM. Thus
fermion-fermion-scalar interactions play an important role in our
derivation.  The interactions of quarks and squarks with Higgs bosons
and higgsinos may be obtained from the superpotential, and are given by,
\begin{equation}\label{eq:higgsinoint}\begin{split}
\mathcal{L}\ni\ &-\left[\bar{u}_j(\bdf_u)^T_{ji}h^0_uP_Lu_i-\bar{u}_j(\bdf_u)^T_{ji}h^+_uP_Ld_i+\bar{d}_j(\bdf_d)^T_{ji}h^-_dP_Lu_i+\bar{d}_j(\bdf_d)^T_{ji}h^0_dP_Ld_i\right.\\
&\left.\quad\ +\bar{e}_j(\bdf_e)^T_{ji}h^-_dP_L\nu_i+\bar{e}_j(\bdf_e)^T_{ji}h^0_dP_Le_i+\mathrm{h.c.}\right]\\
&-\left[\bar{\Psi}_{h^0_u}\tilde{u}^\dagger_{Rj}(\tilde{\bdf}^{u_R}_u)^T_{ji}P_Lu_i
+\bar{u}_{j}(\tilde{\bdf}^{Q}_u)^T_{ji}\tilde{u}_{Li}P_L\Psi_{h^0_u}
-\bar{\Psi}_{h^+_u}\tilde{u}^\dagger_{Rj}(\tilde{\bdf}^{u_R}_u)^T_{ji}P_Ld_i
-\bar{u}_{j}(\tilde{\bdf}^{Q}_u)^T_{ji}\tilde{d}_{Li}P_L\Psi_{h^+_u}\right.\\
&\quad\ +\bar{\Psi}_{h^-_d}\tilde{d}^\dagger_{Rj}(\tilde{\bdf}^{d_R}_d)^T_{ji}P_Lu_i
+\bar{d}_{j}(\tilde{\bdf}^{Q}_d)^T_{ji}\tilde{u}_{Li}P_L\Psi_{h^-_d}
+\bar{\Psi}_{h^0_d}\tilde{d}^\dagger_{Rj}(\tilde{\bdf}^{d_R}_d)^T_{ji}P_Ld_i
+\bar{d}_{j}(\tilde{\bdf}^{Q}_d)^T_{ji}\tilde{d}_{Li}P_L\Psi_{h^0_d}\\
&\quad\ +\bar{\Psi}_{h^-_d}\tilde{e}^\dagger_{Rj}(\tilde{\bdf}^{e_R}_e)^T_{ji}P_L\nu_i
+\bar{e}_{j}(\tilde{\bdf}^{L}_e)^T_{ji}\tilde{\nu}_{Li}P_L\Psi_{h^-_d}
+\bar{\Psi}_{h^0_d}\tilde{e}^\dagger_{Rj}(\tilde{\bdf}^{e_R}_e)^T_{ji}P_Le_i
+\bar{e}_{j}(\tilde{\bdf}^{L}_e)^T_{ji}\tilde{e}_{Li}P_L\Psi_{h^0_d}\\
&\quad\ \left.+\mathrm{h.c.}\right]
\end{split}\end{equation}
We have written these interactions with a left chiral projector
sandwiched between the four-component spinors to facilitate comparison
with Eq.~(\ref{eq:fourlag}). Notice that the couplings of matter
fermions to Higgs bosons are denoted the same way as the superpotential Yukawa
matrices. In contrast, we have denoted the couplings of the
fermion-sfermion system by a matrix with a twiddle on it, labelled by the
scalar that enters the corresponding interaction. Above the scale of all
particles and Higgs boson thresholds, the effective theory is the MSSM
and these $\tilde{\mathbf{f}}^\Phi$ couplings are the same as the corresponding
superpotential couplings $\mathbf{f}$, but may be different below this
scale. For instance, if $|\mu| < m_H$, the low energy theory just below
$Q=m_H$ contains two higgsinos, but just the one scalar doublet in
(\ref{eq:hrot}). The effective theory below $Q=m_{H}$ then contains only the 
couplings $\bm{\lambda}_{u,d,e}$ along with the couplings 
$\tilde{\mathbf{f}}^\Phi_{u,d,e}$, but no $\mathbf{f}_{u,d,e}$. As we will see
the evolution of $\tilde{\mathbf{f}}^\Phi_{u,d,e}$ differs from the 
``would-have-been'' evolution of $\mathbf{f}_{u,d,e}$ for this range 
of $Q$.

The remaining fermion-fermion-scalar interactions are those with the
gauginos, and are given by,
\begin{equation}\label{eq:gauginoint}\begin{split}
\mathcal{L}\ni\ &-\frac{1}{\sqrt{2}}\left\{\left(\tilde{u}^\dagger_{Lj},\tilde{d}^\dagger_{Lj}\right)\mathbf{G}_{Q}P_L\left(\begin{array}{c}u_i\\d_i\end{array}\right)+\left(\tilde{\nu}^\dagger_{Lj},\tilde{e}^\dagger_{Lj}\right)\mathbf{G}_{L}P_L\left(\begin{array}{c}\nu_i\\e_i\end{array}\right)\right.\\[5pt]
&\left.\qquad\ +\bar{u}_j\bgtpur_{ji}(-\tfrac{4}{3})\tilde{u}_{Ri}P_L\lambda_0+\bar{d}_j\bgtpdr_{ji}(\tfrac{2}{3})\tilde{d}_{Ri}P_L\lambda_0+\bar{e}_j\bgtper_{ji}(2)\tilde{e}_{Ri}P_L\lambda_0+\mathrm{h.c.}\right\}\\[5pt]
&-\sqrt{2}\left\{(\tilde{\mathbf{g}}_{s}^{\tilde{q}_{L}})_{ji}(-i)^{\theta_{\tilde{g}}}\tilde{q}^\dagger_{Lj}\bar{\tilde{g}}_{A}\tfrac{\lambda_A}{2}P_Lq_i
-(\tilde{\mathbf{g}}_{s}^{\tilde{q}_{R}})_{ji}(-i)^{\theta_{\tilde{g}}}
\bar{q}_{j}\tfrac{\lambda_A}{2}P_L \tg_A \tilde{q}_{Ri} +\mathrm{h.c.}\right\}\\
&-\frac{1}{\sqrt{2}}\left\{\left(h^{+\dagger}_u,h^{0\dagger}_u\right)\mathbf{G}_{h_{u}}P_L\left(\begin{array}{c}\Psi_{h^+_u}\\\Psi_{h^0_u}\end{array}\right)+\left(h^{-\dagger}_d,h^{0\dagger}_d\right)\mathbf{G}_{h_{d}}P_L\left(\begin{array}{c}\Psi_{h^-_d}\\\Psi_{h^0_d}\end{array}\right)+\mathrm{h.c.}\right\}\;,\\
\end{split}\end{equation}
where we have defined the following matrices:
\begin{equation}
\mathbf{G}_{Q}=\left(\begin{array}{cc}\bgtq_{ji}\bar{\lambda}_3+\frac{1}{3}\bgtpq_{ji}\bar{\lambda}_0&\bgtq_{ji}\left(\bar{\lambda}_1-i\bar{\lambda}_2\right)\\[2pt]\bgtq_{ji}\left(\bar{\lambda}_1+i\bar{\lambda}_2\right)&-\bgtq_{ji}\bar{\lambda}_3+\frac{1}{3}\bgtpq_{ji}\bar{\lambda}_0\end{array}\right)
\end{equation}
\begin{equation}
\mathbf{G}_{L}=\left(\begin{array}{cc}\bgtl_{ji}\bar{\lambda}_3-\bgtpl_{ji}\bar{\lambda}_0&\bgtl_{ji}\left(\bar{\lambda}_1-i\bar{\lambda}_2\right)\\[2pt]\bgtl_{ji}\left(\bar{\lambda}_1+i\bar{\lambda}_2\right)&-\bgtl_{ji}\bar{\lambda}_3-\bgtpl_{ji}\bar{\lambda}_0\end{array}\right)
\end{equation}
\begin{equation}\label{eq:Ghu}
\mathbf{G}_{h_{u}}=\left(\begin{array}{cc}\gthu\bar{\lambda}_3+\gtphu\bar{\lambda}_0&\gthu\left(\bar{\lambda}_1-i\bar{\lambda}_2\right)\\[2pt]\gthu\left(\bar{\lambda}_1+i\bar{\lambda}_2\right)&-\gthu\bar{\lambda}_3+\gtphu\bar{\lambda}_0\end{array}\right)
\end{equation}
\begin{equation}\label{eq:Ghd}
\mathbf{G}_{h_{d}}=\left(\begin{array}{cc}-\gthd\bar{\lambda}_3-\gtphd\bar{\lambda}_0&\gthd\left(-\bar{\lambda}_1-i\bar{\lambda}_2\right)\\[2pt]\gthd\left(-\bar{\lambda}_1+i\bar{\lambda}_2\right)&\gthd\bar{\lambda}_3-\gtphd\bar{\lambda}_0\end{array}\right)\;.
\end{equation}
Here, $\lambda_0$ and $\lambda_p$ denote the (Majorana) hypercharge and
$SU(2)$ gauginos. Just as the higgsino couplings in
(\ref{eq:higgsinoint}) may differ from the corresponding Higgs boson
couplings, below the various thresholds the effective theory is no
longer supersymmetric, and the couplings of gauginos may deviate from
the corresponding gauge boson couplings. Moreover, unlike gauge
couplings, the renormalization of these fermion-fermion-scalar
interactions depends also on the couplings $\mathbf{f}$ and
$\tilde{\mathbf{f}}^{\Phi}$, so that the $\tilde{\mathbf{g}}^{\Phi}$'s
can acquire a non-trivial flavour structure, as allowed for in
(\ref{eq:gauginoint}) above. Above the scale of the highest thresholds,
however, the RGEs for these $\tilde{\bf{g}}^{\Phi}$-type couplings must
coincide with those for the gauge couplings (in the matrix sense).

\subsection{Gauge Coupling RGEs}
The one-loop RGEs for gauge couplings (including threshold effects) are
well known. We will, however, write these for completeness. As explained
in Sec.~\ref{intro}, we will decouple particles using the ``step
approximation'' \cite{ramond,sakis}. Toward this end, for any particle
$\mathcal{P}$, we define 
\begin{eqnarray*}
\theta_\mathcal{P} =&1 \ \ {\rm if} \ Q > M_{\mathcal{P}}, \\
  & 0 \ \ {\rm if} \ Q < M_{\mathcal{P}}\;.
\end{eqnarray*}
Throughout this paper, we will assume that $Q>m_t$ so that the $SU(2)$
gauge symmetry is unbroken in the effective theory, and that there are
three matter generations.  Applying (\ref{eq:gaugerge}) for the MSSM
particle content, and using
$$N_{\tilde{f}}=\sum^{3}_{i=1}\theta_{\tilde{f}_{i}},$$ leads to the
familiar gauge coupling RGEs,
\begin{equation}\begin{split}\label{eq:g1bet}
\left(4\pi\right)^{2}\left.\beta_{g_{1}}\right|_{1-\mathrm{loop}}=&g^{3}_{1}\left[4+\frac{1}{30}N_{\tilde{Q}}+\frac{4}{15}N_{\tilde{u}_{R}}+\frac{1}{15}N_{\tilde{d}_{R}}+\frac{1}{10}N_{\tilde{L}}+\frac{1}{5}N_{\tilde{e}_{R}}\right.\\
&\left.\quad\;+\frac{1}{10}\left(\h+\Hh\right)+\frac{1}{5}
\left(\theta_{{\tilde{h}}_1}+\theta_{{\tilde{h}}_2}\right)\right]
\end{split}\end{equation}
\begin{equation}\begin{split}\label{eq:g2bet}
\left(4\pi\right)^{2}\left.\beta_{g}\right|_{1-\mathrm{loop}}=&g^{3}\left[-\frac{22}{3}+4+\frac{1}{2}N_{\tilde{Q}}+\frac{1}{6}N_{\tilde{L}}
+\frac{1}{6}\left(\h+\Hh\right)+\frac{1}{3}\left(\theta_{{\tilde{h}}_1}+\theta_{{\tilde{h}}_2}\right)+\frac{4}{3}\theta_{\tilde{W}}\right]
\end{split}\end{equation}
\begin{equation}\begin{split}\label{eq:g3bet}
\left(4\pi\right)^{2}\left.\beta_{g_{s}}\right|_{1-\mathrm{loop}}=&g^{3}_{s}\left[4+\frac{1}{3}N_{\tilde{Q}}+\frac{1}{6}N_{\tilde{u}_{R}}+\frac{1}{6}N_{\tilde{d}_{R}}+2\theta_{\tilde{g}}-11\right]\;.
\end{split}\end{equation}
Here, $g_{1}$ is the scaled hypercharge gauge coupling that unifies
with the $SU(2)$ and $SU(3)$ couplings when the MSSM is embedded in a
GUT: $g'^{2}=\frac{3}{5}g^{2}_{1}$. We mention that although we have
shown two distinct higgsino thresholds above, since we are working in
the approximation that both higgsinos have the mass $|\mu|$, we will
from now on write $$\theta_{\tilde{h}}=\theta_{\tilde{{h}}_1}=
\theta_{\tilde{{h}}_2}\;.$$

\subsection{RGEs for Yukawa and Yukawa-type Couplings}

Turning now to the RGEs for fermion-fermion-scalar couplings, we see
that there is a large number of such ``Yukawa-type'' couplings in the
Lagrangian.  These are: the usual Yukawa couplings to Higgs bosons,
$\bdf_{u}$, $\bdf_{d}$, and $\bdf_{e}$; the couplings of higgsinos to
the various fermion-sfermion pairs,
$\bft^{Q}_{u}$, $\bft^{Q}_{d}$, $\bft^{L}_{e}$, $\bft^{u_{R}}_{u}$,
$\bft^{d_{R}}_{d}$, $\bft^{e_{R}}_{e}$; hypercharge  gaugino couplings,
$\bgt'^{Q}$, $\bgt'^{L}$, $\bgt'^{u_{R}}$, $\bgt'^{d_{R}}$,
$\bgt'^{e_{R}}$, $\tilde{g}'^{h_{u}}$, $\tilde{g}'^{h_{d}}$; the $SU(2)$ gaugino
couplings, $\bgt^{Q}$, $\bgt^{L}$, $\tilde{g}^{h_{u}}$, $\tilde{g}^{h_{d}}$; and
finally, the
gluino couplings,  $\bgt_{s}^{Q}$, $\bgt_{s}^{u_{R}}$,
$\bgt_{s}^{d_{R}}$. While the RGEs for the gauge couplings still form a
closed set, those for the usual matter fermion Yukawa couplings do not. 
It is only the RGEs for \textit{all} these fermion-fermion-scalar 
couplings that together with those for the gauge couplings form
a closed system. These new couplings  
must coincide with the corresponding gauge or usual Yukawa coupling in
the SUSY limit, {\it i.e.} above the scale of the highest threshold. This
serves as the boundary condition for the RGEs for these new couplings.

To implement decoupling of particles as described in Sec.~\ref{threshold},
 we must construct the matrices $\bu^{1}_{\Phi}$, $\bu^{2}_{\Phi}$, 
 $\bv_{\Phi}$, $\bw_{\Phi}$, $\bx^1_{\Phi}$ and $\bx^2_{\Phi}$, 
 for all $\Phi$, in the (approximate) mass basis for the various
 fermion fields.  We choose this basis to comprise of the Dirac fermions
 $\{u_{i},d_{i},\nu_{i},e_{i}\}$ together with the Majorana fermions
 $\{{\tilde{h}}_1^0,{\tilde{h}}_2^0,{\tilde{h}}_1^\pm,{\tilde{h}}_2^\pm,
 \lambda_{0},\lambda_{1},\lambda_{2},\lambda_{3},\tilde{g}_{A}\}$.  Here,
 ${\tilde{h}}_{1,2}^0$ and ${\tilde{h}}_{1,2}^\pm$ are the neutral and
 charged components of the Majorana spinor ${\tilde{h}}_{1,2}$
 introduced in (\ref{eq:inorot}).\footnote{As already mentioned, since
 we are working with a common threshold for higgsinos, we could have
 stayed in the original basis. We will see below that performing this
 rotation (by an arbitrary angle) gives us another check on our
 procedure. We also draw attention to the fact that we have combined the
 higgsinos of the ${\bf 2}$ and ${\bf 2^*}$ representations, so that
 although ${\tilde{h}}_{1,2}^\pm$ are Majorana spinors, their chiral
 components do not annihilate a definite sign of the charge.} The
 subscript $i$ (which runs from 1 to 3) for the Dirac fermions is a
 flavour index which can be suppressed if the Yukawa terms are written
 as matrices in flavour space. This means that in the MSSM, $\bu^{1}_{\Phi}$
 and $\bu^{2}_{\Phi}$ will be $(4\times4)$ blocks of $(3\times 3)$ matrices when
 $\Phi$ is one of the Higgs bosons in (\ref{eq:hrot}) or (\ref{eq:Hrot}).
 Similarly, since flavour is carried by the sfermion scalar
 index, $\Phi=\tf_i$, $\bv_{\tilde{f}_{i}}$ will be a $(4\times9)$ matrix where 
 the number of rows  can be further expanded to 
 show each of the three matter fermion flavours. Likewise, $\bw_{\tilde{f}_{i}}$ a
 $(9\times4)$ matrix where now the number of columns can be similarly expanded to
 exhibit these flavours. Thus, when fully written out, $\bv_{\tilde{f}_{i}}$ 
 is a $(12\times9)$ matrix, while $\bw_{\tilde{f}_{i}}$ is a $(9\times12)$ matrix; 
 see (\ref{eq:vuli})--(\ref{eq:swuli}).
 Finally, $\bx^{1}_{\Phi}$ and $\bx^{2}_{\Phi}$ will both be $(9\times9)$ matrices.

Within the MSSM with $R$-parity conservation, the matrices
$\bu_{\Phi}^{1,2}$ and $\bx_{\Phi}^{1,2}$ are non-zero only for
$\Phi=\{\mathsf{h},\mathcal{H},G^{+},H^{+}\}$. They can be readily worked
out by comparing (\ref{eq:fourlag}) with the MSSM Yukawa interactions in 
(\ref{eq:higgsinoint}). It would take up too much space to display each
of these matrices explicitly, so we will only present one example for each
type of matrix.  If
$\Phi=\mathsf{h}$ we have, 
\begin{equation}
\bu^{2}_{\mathsf{h}}=\left(\begin{array}{cccc}
\mathbf{0}&\mathbf{0}&\mathbf{0}&\mathbf{0}\\
\mathbf{0}&\cs\bdf^{T}_{d}&\mathbf{0}&\mathbf{0}\\
\mathbf{0}&\mathbf{0}&\mathbf{0}&\mathbf{0}\\
\mathbf{0}&\mathbf{0}&\mathbf{0}&\cs\bdf^{T}_{e}\\
\end{array}\right)\;, 
\end{equation}
where we use a bold-type $\mathbf{0}$ to indicate a $(3\times3)$ matrix
in flavour space whose entries are all zero.
The corresponding $\bu^{1}_{\mathsf{h}}$ matrix has
its only non-vanishing entry in the $(1,1)$ block. The matrix
$\bx^{2}_{\mathsf{h}}$ takes the form,
\begin{equation}
\bx^{2}_{\mathsf{h}}=\frac{1}{2}\left(\begin{array}{cc}
\mathbf{0}_{(4\times4)}& 
\bm{\mathcal{X}}^{2}_{\mathsf{h}}\\[5pt]
\bm{\mathcal{X}}^{2T}_{\mathsf{h}}
&\mathbf{0}_{(5\times5)} 
\end{array}\right)\;,
\end{equation}
where
\begin{equation}
\bm{\mathcal{X}}^{2}_{\mathsf{h}}=\left(\begin{array}{ccccc}
-\sn\gtphu&0&0&\sn\gthu&0\\ 
\sn\gtphu&0&0&-\sn\gthu&0\\ 
0&-\sn\gthu&-i\sn\gthu&0&0\\ 
0&\sn\gthu&i\sn\gthu&0&0 
\end{array}\right)\;.
\end{equation}
The eight $\bx^{1}_{\Phi}$ and $\bx^{2}_{\Phi}$ matrices can be obtained from 
(\ref{eq:gauginoint}) in conjunction with (\ref{eq:Ghu}) and (\ref{eq:Ghd}).
The reader will need to write these as symmetric matrices using 
the symmetry properties of Majorana spinor bilinears. Note that all the 
entries in the 
last row and last column of these matrices are zero because 
the Higgs boson has no coupling to gluinos.\footnote{We have 
suppressed the colour index which would otherwise expand the last
row/column.} It is also clear from the MSSM Lagrangian that 
$\bx^{1}_{\Phi}$ contains only the $\gtphd$ and $\gthd$
couplings, while $\bx^{2}_{\Phi}$ contains only $\gtphu$ and $\gthu$.

The $\bv_{\Phi}$ and $\bw_{\Phi}$ matrices, on the other hand, 
are non-zero only for
$\Phi=\{\tilde{u}_{L},\tilde{d}_{L},\tilde{e}_{L},\tilde{\nu}_{L},\tilde{u}_{R},\tilde{d}_{R},\tilde{e}_{R}\}$. 
From the interactions in (\ref{eq:higgsinoint}) and (\ref{eq:gauginoint}) 
we can see that for left-handed sfermions,
$\bv_{\tilde{f}}$ contains just $\tilde{{\bf f}}^{\Phi}$-type couplings while $\bw_{\tilde{f}}$ contains 
only $\tilde{\bf{g}}^{\Phi}$-type couplings. The opposite is the case for the right-handed 
sfermions. In the case where $\tilde{f}=\tilde{u}_{Li}$, we have
\begin{equation}\label{eq:vuli}
\bv_{\tilde{u}_{Li}}=\frac{1}{\sqrt{2}}\left(\begin{array}{cc}
\bm{\mathcal{V}}_{\tilde{u}_{Li}}&\mathbf{0}_{(6\times5)}\\
\mathbf{0}_{(6\times4)}&\mathbf{0}_{(6\times5)}\\
\end{array}\right)\;,
\end{equation}
with
\begin{equation}\label{eq:svuli}
\bm{\mathcal{V}}_{\tilde{u}_{Li}}=\left(\begin{array}{ccccc}
-(\bft^{Q}_{u})^{T}_{\bullet i}&(\bft^{Q}_{u})^{T}_{\bullet i}
&\mathbf{0}_{(3\times1)}&\mathbf{0}_{(3\times1)}\\[5pt]
\mathbf{0}_{(3\times1)}&\mathbf{0}_{(3\times1)}
&(\bft^{Q}_{d})^{T}_{\bullet i}&(\bft^{Q}_{d})^{T}_{\bullet i}\end{array}\right)\;.
\end{equation}
As alluded to above, since the scalar labels on 
$\bv_{\tilde{f}}$ and $\bw_{\tilde{f}}$ carry \textit{sfermion} flavour, 
we need to be careful when writing these if we choose not to 
expand out the \textit{matter fermion} flavour.
Here we denote this suppressed quark flavour index by the $\bullet$, 
so that $(\bft^{Q}_{u})^{T}_{\bullet i}$ is really a $(3\times1)$
column matrix (since the index $i$ is fixed by the \textrm{squark}
flavour). Similarly
\begin{equation}\label{eq:wuli}
\bw_{\tilde{u}_{Li}}=\frac{1}{\sqrt{2}}\left(\begin{array}{cc}
\mathbf{0}_{(4\times6)}&\mathbf{0}_{(4\times6)}\\
\bm{\mathcal{W}}_{\tilde{u}_{Li}}&\mathbf{0}_{(5\times6)}\\
\end{array}\right)\;,
\end{equation}
with
\begin{equation}\label{eq:swuli}
\bm{\mathcal{W}}_{\tilde{u}_{Li}}=\left(\begin{array}{cc}
\frac{1}{3}\bgtpq_{i\bullet}&\mathbf{0}_{(1\times3)}\\[5pt]
\mathbf{0}_{(1\times3)}&\bgtq_{i\bullet}\\[5pt]
\mathbf{0}_{(1\times3)}&-i\bgtq_{i\bullet}\\[5pt]
\bgtq_{i\bullet}&\mathbf{0}_{(1\times3)}\\[5pt]
2(\frac{\bm{\lambda}_{A}}{2})\bgtsq_{i\bullet}&\mathbf{0}_{(1\times3)}\end{array}\right)\;.
\end{equation}
Notice that the positions of the suppressed fermion index and the (fixed) 
sfermion flavour index, $i$, are swapped between the matrices 
(\ref{eq:svuli}) and (\ref{eq:swuli}).
Once again, we have suppressed the colour index on the quark (and
squark) field. Also note that the entry from couplings to the gluino enter
only via the $\bw_{\tilde{f}}$ matrix for $\tq_L$, and only via 
the $\bv_{\tilde{f}}$ matrix for $\tq_R$, for all flavours of squarks.


We are now ready to proceed with the derivation of the RGEs. Before we
begin our real derivation, we perform a pedagogical exercise to
facilitate comparison with earlier literature and also provide checks on
our procedure. Since earlier studies did not make the distinction
between couplings of quarks to gauge and Higgs bosons (or $g$'s and
$\bdf$'s) and the corresponding couplings ($\bgt$'s and $\bft$'s) of the
quark-squark system to the gauginos and higgsinos, we first derive the
RGE for $(s\bdf_{u})$ without separating out $\bft$'s and $\bgt$'s from
$\bdf$'s and $g$'s, respectively. It is also instructive to illustrate 
the additional
complexity that results from keeping distinct thresholds for $A$, $H$
and $H^\pm$, and an arbitrary rotation angle for the higgsinos instead of
the $45^\circ$ rotation used in
(\ref{eq:inorot}); {\it i.e.}
\begin{equation}
\Psi_{{\tilde{h}}_1^0}=(i\gamma_5)^{\theta_1}(\csp\Psi_{h_d^0}-\snp\Psi_{h_u^0})
\;,\qquad
\Psi_{{\tilde{h}}_2^0}=\left(i\gamma_5\right)^{\theta_2}\left(\snp\Psi_{h_d^0}+\csp\Psi_{h_u^0}\right)\;,
\end{equation}
and
\begin{equation}
\Psi_{{\tilde{h}}_1^\pm}=(i\gamma_5)^{\theta_1}(\csp\Psi_{h_d^-}-\snp\Psi_{h_u^+})\;,\qquad \Psi_{{\tilde{h}}_2^\pm}=\left(i\gamma_5\right)^{\theta_2}\left(\snp\Psi_{h_d^-}+\csp\Psi_{h_u^+}\right)\;,
\end{equation}
where $\snp=\sin{\beta'}$ (set equal to $\beta$ in Ref.~\cite{ramond})
so that the rotation is by an independent angle to the scalars. The
factors $(i\gamma_5)^{\theta_{1,2}}$ allow for the possibility that
either higgsino could have had a negative mass eigenvalue, as noted
below (\ref{eq:inorot}).
As we saw earlier, the RGE is not dependent on where or
whether we include the $(i\gamma_{5})^\theta$ factors. We stress again
that we do not mean the equation that immediately follows to be the
correct RGE, but write it only to facilitate comparison, and to
illustrate some issues. Ignoring the separation of twiddle terms 
on the couplings in the Lagrangian we find,
\begin{equation} \label{eq:fake}\begin{split}
{\left(4\pi\right)}^2\frac{d{\left(\sn \bdf_u\right)}_{ij}}{dt}=&\frac{3s}{2}\left[\frac{\sn^2}{3}\left(\h+\Go+\Gp\right)+\frac{\cs^2}{3}\left(\Hh+\A+\Hp\right)\right](\bdf_u\bdf^\dagger_u\bdf_u)_{ij}\\
&+\frac{\sn}{2}\left[\left\{\snp^2\left(\shram+\shc\right)+\csp^2\left(\sH+\sHc\right)\right\}\sqk{(\bdf_u\bdf^\dagger_u)}_{ik}{(\bdf_u)}_{kj}\right.\\
&\left.\qquad+\left\{\snp^2\shram+\csp^2\sH\right\}\suk{(\bdf_u)}_{ik}{(\bdf^\dagger_u\bdf_u)}_{kj}\right]\\
&+\sn\left[\sn^2\left(\h-\Go\right)+\cs^2\left(\Hh-\A\right)\right](\bdf_u\bdf^\dagger_u\bdf_u)_{ij}\\
&+\frac{\sn}{2}\left[\cs^2\Gp+\sn^2\Hp-4\cs^2\left(\Gp-\Hp\right)\right](\bdf_d\bdf^\dagger_d\bdf_u)_{ij}\\
&+\frac{\sn}{2}\left[\left\{\csp^2\shc+\snp^2\sHc\right\}\sdk{(\bdf_d)}_{ik}(\bdf^\dagger_d\bdf_u)_{kj}\right]\\
&+\sn{(\bdf_u)}_{ij}\left[3\left(\sn^2\h+\cs^2\Hh\right)\mathrm{Tr}\{\bdf^\dagger_u\bdf_u\}+\cs^2\left(\h-\Hh\right)\mathrm{Tr}\{3\bdf^\dagger_d\bdf_d+\bdf^\dagger_e\bdf_e\}\right]\\
&-{(\bdf_u)}_{ij}\left[\frac{3}{5}g^2_1\left\{\sn\frac{17}{12}-\sn\left(\frac{1}{36}\sq+\frac{4}{9}\su\right)\sbi\right.\right.\\
&\qquad\qquad\qquad\ -\left(\frac{\sn}{2}\left[{\left(\snp\sn+\csp\cs\right)}^2\shram+{\left(\snp\cs-\csp\sn\right)}^2\sH\right]\h\right.\\
&\qquad\qquad\qquad\ +\frac{\cs}{2}\left(\snp\cs-\csp\sn\right)\left(\snp\sn+\csp\cs\right)\left(\shram-\sH\right)\Hh\\
&\left.\left.\qquad\qquad\qquad\ +\left[\snp\left(\snp\sn+\csp\cs\right)\shram-\csp\left(\snp\cs-\csp\sn\right)\sH\right]\left\{\frac{1}{3}\sq-\frac{4}{3}\su\right\}\right)\sbi\right\}\\
&\qquad\quad\quad\;\;+g^2_2\left\{\sn\frac{9}{4}-\sn\frac{3}{4}\sq\swi-\left(\frac{\sn}{2}\left[{\left(\snp\sn+\csp\cs\right)}^2\shram+{\left(\snp\cs-\csp\sn\right)}^2\sH\right.\right.\right.\\
&\left.\qquad\qquad\qquad\quad+2\left(\snp^2\sn^2+\csp^2\cs^2\right)\shc+2\left(\snp^2\cs^2+\csp^2\sn^2\right)\sHc\right]\h\\
&\qquad\qquad\qquad\quad+\frac{\cs}{2}\left[\left(\snp\cs-\csp\sn\right)\left(\snp\sn+\csp\cs\right)\left(\shram-\sH\right)\right.\\
&\left.\qquad\qquad\qquad\qquad\quad+2\sn\cs\left(\snp^2-\csp^2\right)\left(\shc-\sHc\right)\right]\Hh\\
&\qquad\qquad\qquad\quad-\left[\snp\left(\snp\sn+\csp\cs\right)\shram-\csp\left(\snp\cs-\csp\sn\right)\sH\right.\\
&\left.\left.\left.\qquad\qquad\qquad\qquad+2\sn\left(\snp^2\shc+\csp^2\sHc\right)\right]\sq\right)\swi\right\}\\
&\left.\qquad\quad\quad\;\;+g^2_3\left\{8\sn-\sn\frac{4}{3}\left(\sq+\su\right)\sgl\right\}\right]
\end{split}\end{equation}
A sum over repeated flavour indices $k,l,\dots$ is implied including
over those with three repeated indices, one of which is a 
$\theta_{\tilde{q}_{k}}$, \textit{e.g.} 
$\sqk{(\bdf_u\bdf^\dagger_u)}_{ik}{(\bdf_u)}_{kj}$ on line two.
Several comments are worth noting. 
\begin{itemize}

\item In writing the RGE as in (\ref{eq:fake}) where we have ignored the
  differences between the usual gauge/Yukawa couplings and their twiddle
  counterparts, we have retained distinct thresholds for each of the
  Higgs bosons as well as for the higgsinos. This is partly to
  facilitate comparison with Ref.~\cite{ramond}, and partly to indicate
  the origins of the various terms.

\item We see that if we take the MSSM limit, where all the $\theta$'s are
  set equal to unity, we recover the usual, well known MSSM RGE
  \cite{bbo,yamada,martv,jackjones}. This is evident if the rotation angle
  that defines the doublets in (\ref{eq:hrot}) and (\ref{eq:Hrot}) is
  treated as a scale independent parameter. In this case, the
  right-hand-side of (\ref{eq:fake}) is proportional to $\sin\beta$ which
  cancels out leaving the MSSM RGE for $\mathbf{f}_u$.\footnote{If
  instead we consider scale-dependent rotations, then the rotated fields
  themselves have an explicit scale dependence. The RGEs in
  Ref.~\cite{mv,luo} that were our starting point assume, of course, no
  such scale dependence of fields. However, we know that above all thresholds, the
  rotation (which is merely a field redefinition) cannot change the
  RGEs. In other words, there must be additional compensating terms in
  the RGEs which effectively allow us to take the $\sin\beta$ out of the
  derivative on the left-hand-side, so the RGE once again reduces to the
  MSSM RGE.} Notice also that in the MSSM limit, the dependence on the
  higgsino rotation angle, $\beta'$, also disappears as it must.

\item If, on the other hand, we set all the $\theta$'s other than
  $\theta_h$, $\theta_{G^0}$ and $\theta_{G^+}$ to zero, the
  theory includes just the SM fermions, gauge bosons and the single
  Higgs doublet (\ref{eq:hrot}), that couples to SM quarks and leptons
  via the coupling matrices $\bm{\lambda}_u=\mathbf{f}_u\sin\beta$,
  $\bm{\lambda}_d=\mathbf{f}_d\cos\beta$ and
  $\bm{\lambda}_e=\mathbf{f}_e\cos\beta$. It is then easy to see that we
  recover the relevant RGE for the Yukawa couplings in the SM \cite{ramond1}. Indeed,
  all dependence on the $\mathbf{f}$'s in (\ref{eq:fake}) disappears in favour of
  dependence on the SM $\bm{\lambda}$'s.

\item Since $SU(2)_L$ remains a symmetry of the theory at low energy, we
  should expect that the couplings related to one another by $SU(2)_L$
  should have the same RGE. We have checked that this is indeed the
  case, {\it but only if we set a common value for all the heavy Higgs boson
  thresholds, and an independent common value for the higgsino
  thresholds.} This should not be surprising since any splitting between
  the extra Higgs bosons, or between the higgsinos, is an $SU(2)_L$
  breaking effect (and would entail introducing even more couplings into
  the low energy theory). We do not regard as reliable those terms in
  (\ref{eq:fake}) that come from $SU(2)$ breaking effects because we are 
  then using an inconsistent approximation.

\item Below the mass scale of the heavy Higgs bosons we may expect that
  since there is just one Higgs boson doublet in the theory,
  only the $\bm{\lambda}_{u,d,e}$ couplings enter. We can easily see
  that this is not the case if higgsinos are lighter than $m_H$. There
  will still be two higgsino doublets in the low energy theory, and
  these  couple via the matrices $\tilde{\mathbf{f}}_{u,d,e}^\Phi$
  (notice the twiddle on the $\mathbf{f}$). 

\item We have compared (\ref{eq:fake}) using a common threshold for the
  higgsinos with the corresponding RGE in Ref.~\cite{ramond}. We find
  agreement for all but the terms involving $SU(2)_L$ and $U(1)_Y$ gauge
  couplings. We find, 
\begin{equation}\nonumber\begin{split}
{\left(4\pi\right)}^2\frac{d{\left(s \bdf_u\right)}_{ij}}{dt}&\ni\ -s{(\bdf_u)}_{ij}\left[\frac{3}{5}g^2_1\left\{\frac{17}{12}-\left(\frac{1}{36}\sq+\frac{4}{9}\su\right)\sbi\right.\right.\\
&\left.\qquad\qquad\quad\qquad\quad-\left(\frac{1}{2}\h+\frac{1}{3}\sq-\frac{4}{3}\su\right)\shram\sbi\right\}\\
&\left.\qquad\qquad\quad\ \ +g^2_2\left\{\frac{9}{4}-\frac{3}{4}\sq\swi-\left(\frac{3}{2}\h-3\sq\right)\shram\swi\right\}\right]\;, 
\end{split}\end{equation}
to be contrasted with~\cite{ramond},
\begin{equation}\nonumber\begin{split}
{\left(4\pi\right)}^2\frac{d{\left(s \bdf_u\right)}_{ij}}{dt}&\ni\ -s{(\bdf_u)}_{ij}\left[\frac{3}{5}g^2_1\left\{\frac{17}{12}+\frac{3}{4}\shram-\left(\frac{1}{36}\sq+\frac{4}{9}\su+\frac{1}{4}\shram\right)\sbi\right\}\right.\\
&\left.\qquad\qquad\quad\ \ +g^2_2\left\{\frac{9}{4}+\frac{9}{4}\shram-\frac{3}{4}\left(\sq+\shram\right)\swi\right\}\right]\;.
\end{split}\end{equation}
There is a similar point of disagreement in the RGEs for the down-type
Yukawa RGEs.\footnote{The SM as well as the MSSM limits come out
right. Since an even number of SUSY particles couple at any vertex, we do
not, however, understand how $\theta_{\tilde{h}_{i}}$ could enter without being
multiplied by a second $\theta$ for another SUSY particle.}

\item We found it difficult to perform a corresponding comparison with
Ref.~\cite{sakis} where there the RGEs appear to be written without
doing any rotation of the Higgs fields, so that we could not abstract
the relationship between the Higgs bosons $h$, $A$, $H$ and $H^\pm$, and
the thresholds that appear in their RGE. For the same
reason, we could not see how to reduce this RGE to 
the SM RGE below the
scale of all sparticles and heavy Higgs bosons. The MSSM limit is,
however, correctly obtained. 
\end{itemize}

We reiterate that (\ref{eq:fake}) is not the correct RGE to use. 
If we derive this RGE keeping the distinction between $g_i$ and 
$\tilde{\mathbf{g}}^{\Phi}_{i}$ and between $\mathbf{f}_{u,d,e}$ and
$\tilde{\mathbf{f}}_{u,d,e}^\Phi$, and independently set common thresholds for
the heavy Higgs scalars ($\Hh$) and the higgsinos ($\sh$), along with a common threshold
at $m_h$ for the ``light higgs doublet'' ($\h$), we find that the RGE for the
coupling of the up-type quarks to Higgs bosons becomes,
\begin{equation}\label{eq:real}\begin{split}
{\left(4\pi\right)}^2\frac{d(\sn\bdf_u)_{ij}}{dt}=&\frac{\sn}{2}\left\{3\left[\sn^2\h+\cs^2\Hh\right](\bdf_u\bdf_u^\dagger)_{ik}+\left[\cs^2\h+\sn^2\Hh\right](\bdf_d\bdf_d^\dagger)_{ik}\right.\\
&\left.\quad+4\cs^2\left[-\h+\Hh\right](\bdf_d\bdf_d^\dagger)_{ik}\right\}(\bdf_u)_{kj}\\
&+\sn(\bdf_u)_{ik}\left[\sh\sql\bftuq^\dagger_{kl}\bftuq_{lj}+\frac{4}{9}\sbi\sul\bgtpur^*_{kl}\bgtpur^T_{lj}+\frac{4}{3}\sgl\sul\bgtsur^*_{kl}\bgtsur^T_{lj}\right]\\
&+\frac{\sn}{4}\left[2\sh\suk\bftur_{ik}\bftur^\dagger_{kl}+2\sh\sdk\bftdr_{ik}\bftdr^\dagger_{kl}+3\swi\sqk\bgtq^T_{ik}\bgtq^*_{kl}\right.\\
&\left.\qquad+\frac{1}{9}\sbi\sqk\bgtpq^T_{ik}\bgtpq^*_{kl}+\frac{16}{3}\sgl\sqk\bgtsq^T_{ik}\bgtsq^*_{kl}\right](\bdf_u)_{lj}\\
&+\sn\sh\sqk\left[-3\swi\gthus\bgtq^T_{ik}+\frac{1}{3}\sbi\gtphus\bgtpq^T_{ik}\right]\bftuq_{kj}\\
&-\frac{4}{3}\sn\sbi\sh\suk\gtphus\bftur_{ik}\bgtpur^T_{kj}\\
&+\sn(\bdf_u)_{ij}\left[\left(\sn^2\h+\cs^2\Hh\right)\mathrm{Tr}\{3\bdf_u^\dagger\bdf_u\}+\cs^2\left(\h-\Hh\right)\mathrm{Tr}\{3\bdf_d^\dagger\bdf_d+\bdf_e^\dagger\bdf_e\}\right]\\
&+\frac{\sn}{2}\sh(\bdf_u)_{ij}\left\{3\swi\left[\mgthusq\left(\sn^2\h+\cs^2\Hh\right)+\mgthdsq\left(\cs^2\h-\cs^2\Hh\right)\right]\right.\\
&\left.\qquad\qquad\qquad+\sbi\left[\mgtphusq\left(\sn^2\h+\cs^2\Hh\right)+\mgtphdsq\left(\cs^2\h-\cs^2\Hh\right)\right]\right\}\\
&-\sn(\bdf_u)_{ij}\left[\frac{17}{12}g'^2+\frac{9}{4}g^2_2+8g^2_3\right]
\end{split}\end{equation}
Here, as with (\ref{eq:fake}) and in all the RGEs that follow, including
those in the Appendix, repeated flavour indices $k,l,\dots$  are summed
over, including indices repeated thrice, once in a $\theta$ and twice
in couplings.

It is easy to see the reduction to the MSSM, and as before, to the SM
RGE for the Yukawa coupling matrix $\bm{\lambda}_u$. Indeed, for $Q <
m_H$, this RGE becomes the RGE for the coupling, $\bm{\lambda}_u=
\sin{\beta}\ {\bf{f}}_u$, of up-type quarks to the light Higgs doublet in
(\ref{eq:hrot}), even if higgsinos, gauginos or matter sfermions remain
in the low energy theory.  The factors of $\sin\beta$ and $\cos\beta$ in
the first pair of curly parentheses can clearly be absorbed to turn all the
$\bf{f}$'s into the corresponding $\bm{\lambda}$'s. This is not, however,
true in the next set of terms in square brackets where there is just 
one factor of $\sin\beta$ that
combines with the $\left({\bf{f}}_u\right)_{ik}$ outside the square
bracket to yield $\left(\bm{\lambda}_u\right)_{ik}$. Notice, however, that
the remaining couplings on this line have twiddles on them, and do not
correspond to the quark-quark-gauge/Higgs boson couplings. Indeed, it is
interesting to see that we get just the right powers of $\sin\beta$ and
$\cos\beta$ on the right-hand-side for all the MSSM couplings of matter
fermions to Higgs bosons to reduce to those of the SM when all $\theta$'s
other than $\theta_{h}$ vanish. Notice also that
the $\tilde{\bf{g}}$ couplings have a non-trivial matrix
structure. Since it is these couplings (and not the corresponding gauge
boson couplings) that directly enter the decays of squarks and sleptons,
it behooves us to keep careful track of these in a study of
flavour-physics of sparticles.

While it is straightforward to derive the MSSM RGEs starting from the 
general forms in (\ref{eq:U1})--(\ref{eq:Wa}), we reiterate that the reader 
should keep clear the distinction between the use of $i,j,\dots$ as 
fermion field type indices in the general form from their use as flavour 
indices in the MSSM RGEs such as (\ref{eq:real}). Thus, in contrast to the 
trace in (\ref{eq:real}), the trace in (\ref{eq:U1}) refers to a sum over the 
fermion field types, not just flavour. This means that not all trace 
terms in (\ref{eq:U1})--(\ref{eq:Wa}) lead to a trace in the RGEs of the MSSM, 
and further, sometimes the trace in the MSSM RGEs may originate in terms 
that are not traces in the general RGEs. We illustrate this with some 
examples. For instance, in the derivation of (\ref{eq:real}), starting from 
(\ref{eq:U1}):
\begin{itemize}
\item Since the matrix indices $j,k$ on $(\bu^{1}_{a})_{jk}$ and 
  $(\bu^{2}_{a})_{jk}$ denote Dirac fermion type and flavour, there will 
  be a different matrix index for each quark and lepton flavour. As 
  a result the term 
  $\bu^{1}_{b}\mathrm{Tr}\{\bu^{1\dagger}_{b}\bu^{1}_{a}+\bu^{2\dagger}_{a}\bu^{2}_{b}\}$ 
  in (\ref{eq:U1}) leads to the trace over flavours found in 
  (\ref{eq:real}).
  
\item The matrix indices in $(\bx^{1}_{a})_{jk}$ and 
  $(\bx^{2}_{a})_{jk}$ refer to Majorana fermions, and so do not carry 
  any flavour. Tracing over these in (\ref{eq:U1}) can, therefore, never 
  lead to a trace in the MSSM RGEs. This trace term results in the 
  term that immediately follows the trace term in (\ref{eq:real}).

\item The situation is also different for the $(\bv_{a})_{jk}$ and 
  $(\bw_{a})_{jk}$ in (\ref{eq:U1}). For $\bv_{a}$, flavour is located 
  in the $a$ and $j$ 
  indices and for $\bw_{a}$ in the $a$ and $k$ indices. When 
  there is a sum over squarks or sleptons (\textit{i.e.} a sum over the scalar 
  index), we include a $\theta_{\tilde{q}_{i}}$, where $i$ is a flavour 
  index, to keep only the active sfermions at that scale. 
  This means that we can keep usual matrix multiplication with 
  the proviso that the contribution from each squark flavour is 
  associated with a different $\theta$. For example, the sum over 
  left-handed squarks in $2\bv_{b}\bx^{2\dagger}_{a}\bw_{b}$ leads to 
  $-3\sn\sh\sqk\swi\gthus\bgtq^T_{ik}\bftuq_{kj}$.
\end{itemize}

As a different example, consider the RGE for $\bgtpur$ derived using 
(\ref{eq:Va}):
\begin{equation}\begin{split}\label{eq:gurtilde}
{\left(4\pi\right)}^2\frac{d\bgtpur_{ij}}{dt}=&\left[\sn^2\h+\cs^2\Hh\right](\bdf_u^T\bdf_u^*)_{ik}\bgtpur_{kj}\\
&+\left[\frac{4}{9}\sbi\suk\bgtpur_{ik}\bgtpur^\dagger_{kl}+\frac{4}{3}\sgl\suk\bgtsur_{ik}\bgtsur^\dagger_{kl}\right.\\
&\left.\qquad+\sh\sqk\bftuq^T_{ik}\bftuq^*_{kl}\right]\bgtpur_{lj}\\
&+\frac{1}{2}\sbi\bgtpur_{ij}\left[\frac{1}{3}\sql\bgtpq^\dagger_{kl}\bgtpq_{lk}+\sll\bgtpl^\dagger_{kl}\bgtpl_{lk}+\frac{8}{3}\suk\bgtpur^\dagger_{kl}\bgtpur_{lk}\right.\\
&\left.\qquad\qquad\qquad+\frac{2}{3}\sdk\bgtpdr^\dagger_{kl}\bgtpdr_{lk}+2\sek\bgtper^\dagger_{kl}\bgtper_{lk}\right]\\
&+\frac{1}{2}\sbi\sh\bgtpur_{ij}\left\{\left[\sn^2\h+\cs^2\Hh\right]\mgtphusq+\left[\cs^2\h+\sn^2\Hh\right]\mgtphdsq\right\}\\
&-3\sbi\sh\left[\sn^2\h+\cs^2\Hh\right]\gtphu(\bdf_u)^T_{ik}\bftur^*_{kj}\\
&-\sh\sqk\bftuq^T_{ik}\bgtpq_{kl}\bftur^*_{lj}+2\sh\suk\bgtpur_{ik}\bftur^T_{kl}\bftur^*_{lj}\\
&+\suk\bgtpur_{ik}\left[\frac{8}{9}\sbi\bgtpur^\dagger_{kl}\bgtpur_{lj}+\frac{8}{3}\sgl\bgtsur^\dagger_{kl}\bgtsur_{lj}\right]\\
&-\bgtpur_{ij}\left[\frac{4}{3}g'^2+4g^2_3\right]
\end{split}\end{equation}
Here, the terms involving a trace over flavours may not be immediately  
evident. One such trace term, found on line four, is
$$\frac{1}{2}\sbi\bgtpur_{ij}\left[\frac{1}{3}\sql\bgtpq^\dagger_{kl}\bgtpq_{lk}\right]\;,$$ 
in which the trace is not explicitly written since we need to keep 
information about the position of the squark thresholds.
We point out the following:
\begin{itemize}
\item The terms involving the trace over flavours in (\ref{eq:gurtilde})
  do not originate from the trace term in (\ref{eq:Va}). The trace in
  (\ref{eq:gurtilde}) can only originate when the external quark and
  squark flavour indices on the left-hand side occur in the
  \textit{same} $\bv_{a}$ or $\bw_{a}$ matrix on the right-hand side. We
  see upon inspection this is the case only for the term in the second
  line of (\ref{eq:Va}), which necessarily leads to the trace over
  flavours in (\ref{eq:gurtilde}). An analogous discussion applies to
  the derivation of the RGE for, say $\bgtpqnb$, starting from
  Eq.~(\ref{eq:Wa}). Note also that when we are checking reduction to
  the MSSM, gaugino coupling matrices such as $\bgtpur$ reduce to
  $(g'\times\bm{\dblone})$, so that we
  obtain an extra factor of three when we take a trace over the flavours.

\item When inserting thetas, we must, of course, only put thetas which 
  correspond to summed internal indices. To obtain, for example, the term 
  $-\sh\sqk\bftuq^T_{ik}\bgtpq_{kl}\bftur^*_{lj}$, which comes from
  $2\bv_{b}\bw^{*}_{a}\bw^{T}_{b}$ in (\ref{eq:Va}), where we have picked out 
  the left-handed squarks in the
  sum over $b$, we insert a $\sqk$, with $k$ summed over squark flavours.
  Similarly, we insert a $\sh$ which comes from picking out the 
  higgsino terms in the sum over Majorana fermions.\footnote{We also sum over 
  quark flavours, $l$, but there is no corresponding theta 
  since we are always working at scales above $m_{t}$. Therefore, all quarks are active
  over the whole range of validity of our RGEs.} Finally, even though we have $\bfturnb$
  in our example term, there is no $\theta_{{\tilde{u}}}$ because the $\tilde{u}_{R}$ 
  squark is the same as the one which appears on the left-hand side,
  and hence is necessarily an active squark.

\item In the 1-loop RGEs for $\bgtsurnb$, $\bgtsdrnb$ and $\bgtsqnb$ there are no 
  contributions from $\bx^{1}$ or $\bx^{2}$, which is to be expected since, 
  as mentioned before, the Higgs boson does not couple to the gluino.
  
\item Special care must be taken when considering interactions with gluinos 
  to correctly evaluate the sum over colours. For example, the RGE for 
  $\bgtsurnb$ has some terms which may initially seem to lead to a trace 
  which includes $\bgtpurnb$. However, the term in question is 
  $$\sim\delta_{dc}(\frac{\lambda_{A}}{2})_{cd}\bgtpur^{\dagger}_{kl}\bgtsur_{lk}\;,$$
  where $l,k$ are flavour indices and $c,d$ are colour indices. This term 
  is clearly zero, since $\mathrm{Tr}\{\frac{\lambda_{A}}{2}\}=0$.
  
\end{itemize}

The RGEs for all the dimensionless couplings 
can be similarly obtained. We list these in the Appendix. 

\section{Solutions to the RGEs and flavour violation in the MSSM}\label{anal}

Since our ultimate goal is to examine flavour-violation in sparticle
interactions, and in sparticle decays in particular, we are naturally
led to examine the flavour structure of various couplings, renormalized
at the scale of the sparticle mass, {\it i.e} at scales $Q$ typically
between $\sim100$~GeV and a few TeV.

The couplings of neutral gauge bosons remain flavour-diagonal under
renormalization group evolution even at higher loops since any
inter-generation couplings would violate current conservation. This does
not mean that their evolution is independent of Yukawa couplings: Yukawa
couplings can, and do, enter the evolution of gauge couplings via
two-loop contributions to the RGEs, but only as a ``flavour-blind
trace''. As a result of this, and the fact that the charged current weak
interaction only couples to the left-handed fermions and their
superpartners, inter-generation couplings in the charged current
weak-interaction are completely determined by the Kobayashi-Maskawa
\cite{KM} (KM) matrix.  

We  thus turn to an  examination of the flavour structure of
fermion-fermion-scalar interactions. These include the familiar Yukawa
couplings of quarks with Higgs bosons, as well as the couplings of the
quark-squark system with gluinos, charginos and neutralinos. Although we
will confine ourselves to the quark/squark sector, our considerations
readily extend to the lepton sector. In this case, however, the MSSM
would need to be extended to include effects from the flavour structure
of the singlet neutrino/sneutrino mass matrix and interactions.

It may appear simplest to work in a basis where the Yukawa coupling matrices
in (\ref{eq:higgsinoint}) are diagonal at one chosen scale, so that 
quarks are mass eigenstates,
rather than in the ``current quark basis'' which we have used to
write the superpotential (\ref{eq:super}) and the interactions
(\ref{eq:soft}) -- (\ref{eq:gauginoint}) as well as the RGEs. The two
bases are related by the transformations, 
\begin{eqnarray}
{\bf{V}}_L(u)u^M_L = u_L,  & {\bf{V}}_L(d)d^M_L = d_L;\label{eq:Yukrot}\\
{\bf{V}}_R(u)u^M_R = u_R,   & {\bf{V}}_R(d)d^M_R = d_R\; \nonumber
\end{eqnarray}
where the unitary matrices ${\bf{V}}_L(u)$ and ${\bf{V}}_R(u)$ [${\bf{V}}_L(d)$ and ${\bf{V}}_R(d)$]
diagonalize the Yukawa coupling matrix ${\bf{f}}_u$ [${\bf{f}}_d$] via 
\begin{equation}
{\bf{V}}^T_L(u) {\bf{f}}_u {\bf{V}}^{*}_R(u) = {\bf{f}}_u^{\rm diag}, \ \
{\bf{V}}^T_L(d) {\bf{f}}_d {\bf{V}}^{*}_R(d) = {\bf{f}}_d^{\rm diag}\;,
\label{eqn:rotate} 
\end{equation} 
Of course, the matrices ${\bf{f}}_u^{\rm diag}$
and ${\bf{f}}_d^{\rm diag}$ are diagonal only at one scale, which we take 
to be $Q=m_{t}$.
The KM matrix that enters the interactions of $W^\pm$ bosons with quarks
is then given by 
$${\bf{K}}= {\bf{V}}_L^\dagger(u){\bf{V}}_L(d)\;.$$ 
The problem with choosing to work in the quark mass basis
is that using different transformations for the left-handed up and
down quarks breaks the $SU(2)_L$ symmetry. As a result, interactions
of say the charged and neutral Higgs bosons with quarks are no longer simply
related by an $SU(2)_L$ transformation, but involve an additional
matrix (which in the case of the MSSM is the KM matrix).\footnote{This
matrix is the KM matrix for the MSSM where one of the
Higgs doublets couples to up-type quarks, and the other to down-type
quarks. In the general two Higgs doublet model, this is no longer the
case.} 

It is convenient to work instead in the basis where just one of up- or
down-type (but not both) Yukawa couplings are diagonal at $m_{t}$. The
interactions and the RGEs that we have written in the last section
continue to hold in this special ``current quark'' basis, provided we
define the squark fields as the super-partners of these quark fields. In
the following, we will choose the basis so that the up quark Yukawa
couplings are diagonal at $m_{t}$, {\it i.e.} the up-type quarks are in
their mass basis at this scale.  {\it This does not imply that the up-type
squarks are
also in the mass eigenstate basis. } This would be the case in the exact
SUSY limit, but the SSB squark mass matrices and the trilinear SSB
scalar couplings may independently violate flavour in the squark sector.
In writing our RGEs with independent thresholds for the squarks we have,
however, assumed that mixing between up-type squarks is small in this
basis so that $\tu_{Li}$ and $\tu_{Ri}$ are approximately also mass
eigenstates, and likewise for down-type squarks.

With this preamble, let us turn to an examination of the flavour
violating dimensionless couplings of the MSSM. Let us begin with the
Yukawa couplings of quarks to neutral Higgs bosons. We will assume that
we have diagonalized the SM Yukawa coupling matrices at the scale
$Q=m_t$, and examine the solutions for the various elements of the up
quark Yukawa coupling matrix for larger values of $Q$. We thus need to
solve the system of RGEs listed in the Appendix. Notice that though the
RGEs for the the dimensionful SSB parameters are decoupled from the RGEs
for dimensionless couplings, the weak scale values of the dimensionful
parameters nonetheless enter our analysis in that they determine the
various thresholds. This dependence is, fortunately, only logarithmic
and it suffices to approximate the location of the thresholds as
discussed in Sec.~\ref{threshold}. To solve the RGEs we rotate the
Yukawa coupling matrices to the current basis using (\ref{eqn:rotate}),
with matrices ${\bf{V}}_{L,R}(q)$ chosen to reproduce the KM matrix.
Next, we evolve these along with the values of gauge couplings 
to the scale $Q=M_{\rm GUT}$ using the RGEs with all the $\theta$'s set
to unity, {\it i.e.} the MSSM RGEs. We then evolve back down to $Q=m_t$,
but this time including thresholds and the couplings with
twiddles.\footnote{This procedure would be somewhat modified if instead
of choosing the location of heavy Higgs boson and sparticle thresholds
``by hand'' as we do here, these were to be obtained using GUT scale
boundary conditions for SSB parameters. Universal boundary conditions,
such as the ones used in mSUGRA, can of course be used in any sfermion
basis, but in all other cases we have to be specific about the choice of
basis in which we specify these SSB boundary conditions. This will be
discussed in more detail in Ref. \cite{dimensionful}. As an aside, we
also note that to get the solutions of the RGEs in this paper, it is
only necessary to evolve beyond the highest threshold during the
iterations to obtain convergence. We have evolved all the way to $M_{\rm
GUT}$ only because we anticipate that this will be the $Q$ value where
SSB boundary conditions will be specified in many realistic scenarios.}
We reset the gauge couplings and current-basis Yukawa coupling matrices
back to their input values, run back up to $M_{\rm GUT}$, and iterate
the procedure until it converges with the required precision. We can
then read off the couplings in any basis at the desired values of $Q$.

Rather than show numerical results for
many cases in many SUSY models that have been considered, we have chosen
to illustrate our analysis for a simplified scenario where the sfermions
and the electroweak gauginos are all at a mass scale $\sim 600$~GeV
while the heavy Higgs bosons and gluinos have a mass $\sim 2$~TeV. Thus,
our effective theory is supersymmetric for $Q> 2$~TeV, includes only
sfermions, charginos and neutralinos together with SM particles for
600~GeV $<Q <$ 2~TeV, and is the SM for $Q < 600$~GeV. In the
intermediate range between 600~GeV and 2~TeV, we will have to separately
examine the couplings ${\tilde{\bf{g}}}^\Phi$ and ${\tilde{\bf{g}}}'^\Phi$
(but not ${\tilde{\bf{g}}}_s^\Phi$) and ${\tilde{\bf{f}}}^\Phi_{u,d,e}$ as
these split off from the usual gauge and Yukawa couplings.

\subsection{Quark Yukawa couplings}\label{subsec:quarkyukawacouplings}

The values of the running quark mass parameters determine the Yukawa
coupling matrices in the mass basis. These can be rotated to an arbitrary
current basis using (\ref{eq:Yukrot}), or to the particular basis with
diagonal up-type Yukawa couplings by choosing the four matrices
suitably. In models where the SSB interactions are flavour-blind (at any
one scale) the four matrices are separately unphysical, and physics is determined by
just the KM matrix. In more general scenarios though, physical
quantities will depend on more than just the KM matrix. In our numerical
illustration, we have used the program ISAJET \cite{isajet} to extract
quark Yukawa couplings at the scale $Q=m_t$ in their mass
basis.\footnote{ISAJET currently does not include flavour mixing among the
quarks. It does, however, include important radiative corrections to the
relationship between quark masses and the Yukawa couplings. We expect
that because this flavour mixing is small, its effect on the quark
masses will also not be large. In any case, we expect to incorporate
these RGEs into ISAJET, at which stage flavour mixing effects can be
included in the iterative procedure used to extract the Yukawa
couplings. Since our discussion here is meant only to be illustrative, it is
not crucial that we use absolutely precise values for the Yukawa couplings.}
Starting from this boundary condition, we have iteratively solved
the system of RGEs for a toy scenario where the SUSY thresholds are
taken to be
clustered either at 600~GeV ($M_1$, $M_2$, $|\mu|$ and $m_{\tf}$), or at
2~TeV ($m_H$, $m_{\tg}$), as mentioned in the previous paragraph. 

\begin{figure}\begin{centering}
\includegraphics[width=.7\textwidth]{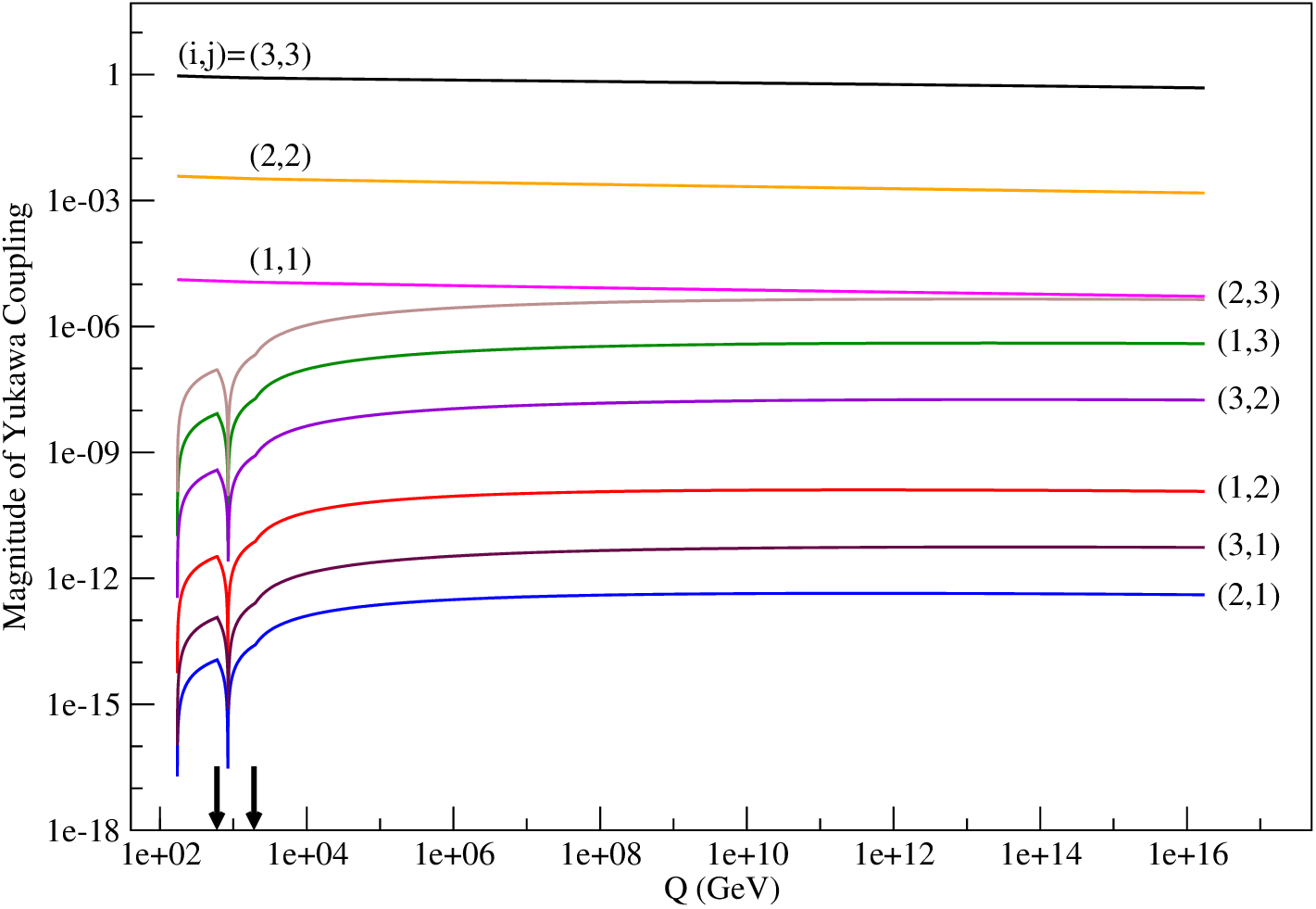}
\caption{Evolution of the magnitudes of the elements of the up-quark
Yukawa coupling matrix for the MSSM with thresholds, shown by arrows,
clustered at $600$~GeV and $2$~TeV as discussed in the text. Above
$m_{H}$ $(=2\ \mathrm{TeV})$ we plot $\left|(\bdf_{u})_{ij}\right|$
whereas below $m_{H}$, where the effective theory includes just one
scalar Higgs doublet, we plot
$\left|(\bm{\lambda}_{u})_{ij}\right|/\sin{\beta}$ which is equal to
$\left|(\bdf_{u})_{ij}\right|$ at $Q=m_{H}=2$~TeV.}
\label{fig:absf}
\end{centering}
\end{figure}

The results of this calculation for the elements of ${\bf{f}}_u$ are shown
in Fig.~\ref{fig:absf}.
Specifically, we show $\left|\left({\bdf}_u\right)_{ij}\right|$ if $Q>
m_H$, and $\left|\left(\bm{\lambda}_u\right)_{ij}\right|/\sin\beta $
(which joins continuously to the the corresponding element of $\bdf_u$ at
$Q=m_H$) for $Q<m_H$.  Since the KM matrix includes a complex phase
which we take to be $60^\circ$ (and since the matrices in
(\ref{eq:Yukrot}) can themselves be complex), these elements are, in
general, complex numbers.\footnote{We have solved these equations in a
randomly chosen current basis, {\it i.e.} one with a random choice of
the matrices in (\ref{eq:Yukrot}), and then rotated back to the basis
where up-type quark Yukawa couplings are diagonal at $Q=m_t$.  Our
results are independent of the initial choice of current basis as they
should be, providing a non-trivial check on the code.} In
Fig.~\ref{fig:absf}, we have plotted the absolute values of these
elements, in the basis where the up-quark Yukawa coupling matrix is
diagonal at $Q=m_{t}$, versus the energy scale $Q$.  In our calculation, we include
two-loop terms albeit without threshold corrections which are
numerically completely negligible as even the two-loop corrections have
only a small effect on our results.  The three roughly horizontal lines
show the scale dependence of the diagonal Yukawa couplings. This is
largely governed by the strong interaction gauge coupling terms in the
RGE, and for large $Q$ values scales as a calculable power of this gauge
coupling.\footnote{For the (3,3) element, contributions from the third
generation Yukawa coupling are also significant.} There are kinks in
these curves at 600~GeV and 2~TeV, but these are not visible on the
scale of this figure: see, however, Fig.~\ref{fig:yuk}. The off-diagonal
elements of (the scaled) $|\bm{\lambda}_u|$ all start off at zero at
$Q=m_{t}$, but rapidly increase to values that though small, may as in
the case of $\left({\bf{f}}_u\right)_{23}$ become as large as
$10^{-6}$. The most striking thing about the figure is the dip in all
the off-diagonal elements at a value of $Q$ in the few hundred GeV
range, close to the expected value of squark masses in our example. An
understanding of this feature is of more than mere academic interest,
since we shall see that it also appears in the corresponding
$\tilde{\bf{f}}$-couplings, which directly enter the amplitudes for flavour
violating squark decays.

To understand this curious feature, let us first consider a further
simplification where the two SUSY thresholds coalesce into a single one
at $Q=M$. We will, of course, return to the case in Fig.~\ref{fig:absf}
shortly. For $Q>M$, the RGE for ${\bf{f}}_u$ becomes the corresponding
MSSM RGE
\begin{equation}\begin{split}\label{eq:mssmrge}
{\left(4\pi\right)}^2\frac{d(\bdfm_{u})}{dt}=&3(\bdfm_{u})(\bdfm_{u})^{\dagger}(\bdfm_{u})+\mathbf{K}^{*}(\bdfm_{d})(\bdfm_{d})^{\dagger}\mathbf{K}^{T}(\bdfm_{u})\\
&+(\bdfm_{u})\mathrm{Tr}\{3(\bdfm_{u})^{\dagger}(\bdfm_{u})\}\\
&-(\bdfm_{u})\left[\frac{13}{15}g^{2}_{1}+3g^{2}_{2}+\frac{16}{3}g^{2}_{3}\right]
\end{split}\end{equation}
while for $Q<M$, we have the SM RGE for the Yukawa
coupling matrix,
\begin{equation}\begin{split}\label{eq:smrge}
{\left(4\pi\right)}^2\frac{d(\bdlm_{u})}{dt}=&\frac{3}{2}(\bdlm_{u})(\bdlm_{u})^{\dagger}(\bdlm_{u})-\frac{3}{2}\mathbf{K}^{*}(\bdlm_{d})(\bdlm_{d})^{\dagger}\mathbf{K}^{T}(\bdlm_{u})\\
&\hspace{-.2cm}+(\bdlm_{u})\mathrm{Tr}\{3(\bdlm_{u})^{\dagger}(\bdlm_{u})+3(\bdlm_{d})^{\dagger}(\bdlm_{d})+(\bdlm_{e})^{\dagger}(\bdlm_{e})\}\\
&\hspace{-.2cm}-(\bdlm_{u})\left[\frac{17}{20}g^{2}_{1}+\frac{9}{4}g^{2}_{2}+8g^{2}_{3}\right]
\end{split}\end{equation}
The KM matrix $\mathbf{K}$ appears in these equations because we have
written them in the basis with {\it all} Yukawa couplings diagonal and
real at the scale $Q=m_t$.

That the real or imaginary parts of the off-diagonal elements all have a
zero is simple to understand. First, we observe that since
$\bm{\lambda}_u$ and $\bm{\lambda}_d$ start off as real and diagonal at
$Q=m_t$, off-diagonal elements develop only because the KM matrix has
off-diagonal components. Also, the imaginary parts of
$\bm{\lambda}_{u,d}$ develop only via the phase in the KM matrix.  Next,
notice that the ``seed of flavour-violation'', 
{\it i.e.} the term involving the KM matrix in the
RGEs,  which is also the seed for
the imaginary part of the off-diagonal elements, 
enters with opposite signs in the SM and in the MSSM RGEs. This means
that if this seed term causes the real or imaginary part of any off-diagonal
element of say the up-type Yukawa coupling to evolve with a negative 
slope in the SM, it would cause the
same element to evolve with a positive slope in the MSSM, and {\it
vice-versa}.  The evolution for $m_t < Q <M$ is governed by the SM RGE,
which causes the real part of an off-diagonal element to evolve 
to say negative values. However, for $Q>M$, the evolution is governed
by the MSSM RGE,
and the real piece of this Yukawa coupling now begins to evolve in the
opposite direction, so that it first becomes less negative, then passes
through zero and continues to positive values. Of course, at $Q=M$, we
must remember to switch from the SM couplings $\bm{\lambda}_{u,d,e}$ to
the MSSM couplings ${\bf{f}}_{u,d,e}$.  The imaginary piece of the
off-diagonal Yukawa couplings similarly passes through zero. However, we
see from the sharp dips in Fig.~\ref{fig:absf} that the real and
imaginary parts have a zero at (almost) {\it the same value} of $Q$, and
furthermore, this location appears independent of the choice of the
flavour indices $i$ and $j$. To understand this, we need to analyse the
equations further.

In the following we will collectively denote the solutions of these
equations by $\bf{U}$, $\bf{D}$ and $\bE$, {\it i.e} ${\bf{U}}=\bdf_u$
for $Q>M$, while ${\bf{U}}=\bm{\lambda}_u$ for $Q<M$, and analogously
for $\bf{D}$ and $\bE$. For example, the RGE for $\mathbf{U}$ in the
basis where both up- and down- type quark matrices are diagonal at
$Q=m_t$ can be written in the form,
\begin{equation}\begin{split}\label{eq:URGE}
{\left(4\pi\right)}^2\frac{d\bu_{ij}}{dt}=&A_{1}\bu_{ik}\bu^{\dagger}_{kl}\bu_{lj}+A_{2}(\mathbf{K}^{*}\bd\bd^{\dagger}\mathbf{K}^{T}\bu)_{ij}\\
&+\bu_{ij}\mathrm{Tr}\{A_{3}\bu^{\dagger}\bu+A_{4}\bd^{\dagger}\bd+A_{5}\bE^{\dagger}\bE\}+\bu_{ij}\mathcal{G}(\vec{\alpha}(t))\;.
\end{split}\end{equation}
Here, $\vec{\alpha}(t)$ denotes the collection of the three gauge
couplings. Of course, the coefficients $A_{i}$ and the function
$\mathcal{G}$ differ for the SM and the MSSM.  As already mentioned, the
evolution of the diagonal elements of $\bf{U}$ and $\bf{D}$ is
dominantly governed by the last term in these equations which contains
the gauge couplings. If we drop all other terms on the right-hand-side,
the RGE for the diagonal Yukawa terms can be easily integrated, and we
find,
\begin{equation} \label{eq:diag}
\bu_{ii}(t) = \bu_{ii}(t_0)F(\vec{\alpha}(t)), \ \ {\rm no \ sum \ on} \ i. 
\end{equation}
It is important to note that {\it the function $F$ is
independent of the quark flavour}.\footnote{In the approximation where
  we retain just the last gauge coupling terms in (\ref{eq:mssmrge}) and
  (\ref{eq:smrge}), these equations take the form,
  $$(4\pi)^2\frac{d\bu_{ii}}{dt}=\bu_{ii}(a_1g_1^2+a_2g^2+a_3g_s^2). $$ They
  can be readily integrated to give, 
$$\bu_{ii}(t)=\bu_{ii}(t_0)\Pi_{r=1}^{3}\left(g_r\right)^{a_r/b_r},$$
  where $b_r$ is the coefficient of the one-loop gauge coupling
  $\beta$-function, $\beta_r(g_r)=\frac{1}{16\pi^2}b_rg_r^3$.}

 Note now that the trace as well as the gauge
coupling terms are flavour-independent. Assuming that the diagonal
Yukawas saturate the trace (remember that we just argued that these
diagonal terms are
a function only of the gauge couplings $\vec{\alpha}(t)$), we can combine
these terms so that the second line on the RHS of (\ref{eq:URGE})
can be written as $$\bu_{ij}(t)f(t),$$
{\it again with $f$ being independent of flavour.} Next, let us look at the 
term involving the KM matrix. Noting that the diagonal elements of
$\bf{U}$ and $\bf{D}$ are much larger than the off-diagonal ones, 
we can write,
$$\left(\mathbf{K}^* \bf{D}\bf{D}^\dagger \mathbf{K}^T\bf{U}\right)_{ij}\simeq \sum_k
\textbf{K}^*_{ik}{\bf{D}}_{kk}(t){\bf{D}}_{kk}(t)\textbf{K}_{jk}{\bf{U}}_{jj}(t),$$ where
$i$ and $j$ are unsummed, fixed indices. In writing this, we have used
the reality of the diagonal elements of the Yukawa matrices under
renormalization group evolution (discussed below).  From  the analogue
of (\ref{eq:diag}) for ${\bf{D}}(t)$, we see that this term has the
structure,
$$\mathbf{M}_{ij}(t_0)G(t), $$ with 
$\mathbf{M}_{ij}(t_0) = 
\textbf{K}^*_{ik}{\bf{D}}_{kk}(t_0){\bf{D}}_{kk}(t_0)\textbf{K}_{jk}
{\bf{U}}_{jj}(t_0)$, and $G$ a {\it flavour-independent} function of
just the gauge couplings, which themselves depend just on $t$. 
Ignoring, for the moment, the first term in the RGE for $\bu$
(the one cubic in $\bf{U}$) we see that we 
have managed to decouple the RGEs for 
 ${\bf{U}}_{ij}$ which (if the first term is dropped) take the form,
$$D\bu_{ij}\equiv
\frac{d\bu_{ij}}{dt}=\mathbf{M}_{ij}(t_0)G(t)+f(t)\bu_{ij}(t), $$
where $f(t)$ and $G(t)$ are known functions of $t$. The solution 
to this may be written as,
$${\bf{U}}_{ij}(t)= \mathbf{M}_{ij}(t_0)\frac{1}{D-f}G(t)+{\bf{U}}_{ij}^0(t),$$
where $\frac{1}{D-f}$ is the resolvent operator, and 
$$\bu_{ij}^0(t)=\bu_{ij}^0(t_0) \exp\left(\int_{t_0}^t
f(t)dt\right)$$ is the solution to the corresponding homogenous
differential equation. For the SM evolution starting at $Q=m_t$, 
$\bu_{ij}^0(t_0)$ vanishes (if $i\ne j$), and we are left with the solution of interest,
\begin{equation}\label{eq:yukfin}
{\bf{U}}_{ij}(t)=\mathbf{M}_{ij}(t_0){\cal{H}}(t),
\end{equation}
where the function ${\cal{H}}(t)\equiv \frac{1}{D-f}G(t)$ is {\it independent
of flavour.} All the flavour information is contained in the ``boundary
value'', $\mathbf{M}_{ij}(t_0)$. To clarify, we use the boundary
conditions on the SM Yukawa couplings to obtain $\bu$ up to the (common)
SUSY threshold. The values of $\bu$ and $\bd$ at $Q=2$~TeV now 
serve as the boundary condition
for MSSM evolution. The central point of this analysis is that the
flavour-dependence is completely captured in the first factor
$\mathbf{M}_{ij}(t_0)$, where $t_0$ corresponds to $Q=m_t$ for SM
evolution, and to the SUSY threshold for MSSM evolution.

To the extent that we can ignore  the first ``cubic term'' in the
$\bf{U}$ RGEs, we
see that the flavour structure of the evolution of $\bm{\lambda}_{ij}$ is
given by (\ref{eq:yukfin}). We make the following observations.
\begin{itemize}
\item This structure is independent of the coefficients in the various
  terms in the RGEs as well as independent of which matrices enter into
  the trace. The same structure thus obtains for both SM and MSSM evolution. 
  The important approximation was that the diagonal elements of
  the Yukawa matrix dominate in the basis where these are diagonal at
  the low scale.

\item Eq.~(\ref{eq:yukfin}) was obtained without any assumptions about
  the hierarchical structure of the KM matrix.
\end{itemize}

We can immediately obtain two corollaries from (\ref{eq:yukfin}).

\begin{enumerate}
\item $\frac{\bu_{ij}^R(t)}{\bu_{ij}^I(t)}$, where the
  superscripts $R$ and $I,$ respectively denote the real and imaginary
  parts, is independent of $t$.

\item $\frac{\bu_{ij}^{R/I}(t)}{\bu_{kl}^{R/I}(t)}$ is independent
  of $t$. 
\end{enumerate}
Our program is to start at $Q=m_t$ with the ``diagonal boundary conditions''
for the SM Yukawa couplings and evolve $\bm{\lambda}_{u,d,e}$ to the scale
$Q=m_H$. We then scale the up- and down-type Yukawa couplings by $1/\sin\beta$
and $1/\cos\beta$, respectively, and finally continue this evolution to
higher values of $t$ using the MSSM RGEs. By corollary 1 above we see that
when the real part of any off-diagonal Yukawa coupling vanishes (as it 
must for some value of $Q>M$), the imaginary part of this coupling
also vanishes. Corollary 2 then tells us that the other off-diagonal 
elements likewise vanish for this same value of $Q$. This is, of course, 
exactly the qualitative feature in
Fig.~\ref{fig:absf} that we started out to explain.

Before turning to other issues, let us briefly return to the first term
that we have ignored up to now. If we saturate the product
${\bf{U}}{\bf{U}}^\dagger{\bf{U}}$  
with
two diagonal elements in each of three terms and also remember that the
diagonal elements are real,  
{\it i.e.} write
$\left({\bf{U}}{\bf{U}}^\dagger{\bf{U}}\right)_{ij}\simeq
\bu_{ii}\bu_{ii}\bu_{ij}+\bu_{jj}\bu_{jj}\bu_{ij}+\bu_{ii}\bu_{jj}\bu_{ji}^*$, 
we find that, like terms
we have considered up to now, two of these terms do not couple
$\bu_{ij}$ with anything else, while the third term couples
$\bu_{ij}$ to $\bu_{ji}^*$. Eq.~(\ref{eq:yukfin}) is thus violated
only by this last term. We have indeed checked that the position of the
zeros are not exactly coincident but have a very tiny spread of a 
few GeV not visible
on the figure. We have also checked that increasing the coefficient of the
cubic term increases this spread, while altering the other coefficients
has no effect on it, in keeping  with expectations from our
analysis.\footnote{We have also numerically solved the RGEs using a
  fictitious KM matrix with large off-diagonal entries. We found the
  same behaviour as in the figure, in keeping with our observation that
  the analysis does not depend on the hierarchical structure of the KM matrix.}

Our analysis of Fig.~\ref{fig:absf}, up to now, has assumed a common
location for the threshold for all non-SM particles which is, of course,
not the case in the figure. It is, however, clear that the bulk of the
change in the slope of the curve occurs from the thresholds at 600~GeV
where the coefficient of the term involving the KM matrix changes its
sign, causing the magnitudes of the off-diagonal couplings to decrease
until they pass through zero, beyond which the magnitude starts
increasing once again. The kink in the various curves at $Q=2$~TeV marks
the position of the second threshold. It should, therefore, be clear
that the location of the zero is largely determined by the spectrum, and
is insensitive to other details.\footnote{We checked also that using a
fictitious KM matrix with large off-diagonal entries changes the
evolution of the off-diagonal couplings (for instance the relative size
of the imaginary parts) as expected but hardly affects the location of
the zero which is determined by the mass spectrum. As mentioned
previously, the location of the zeros is also largely insensitive to the
coefficients of the various terms in the RGEs, unless of course, a
coefficient happens to be so large that the solution blows up before the
zero can be reached.}

Finally, we remark that the imaginary parts of the off-diagonal
couplings (recall that these have their origin in the phase in the 
KM matrix) may,
depending on which matrix element we are considering, be of comparable
magnitude to the corresponding real part, or much smaller. The relative
size of the imaginary part is largely determined by the relative size of
the imaginary part in the corresponding element of the ``seed term'', 
$\sim \mathbf{K}^*{\bf{D}}{\bf{D}}^\dagger \mathbf{K}^T {\bf{U}}$ in the 
RGE. It is easy to see that this ``seed'' is real for the diagonal 
elements of ${\bf{U}}$ (or ${\bf{D}}$), which, therefore, continue to
remain real. For the case study in the figure, the (1,3) element has the
largest  imaginary part ${\cal O}(10^{-7})$, while the imaginary part of
other off-diagonal elements is smaller by 3--6 orders of magnitude over
most of the range of $Q$. 

We should mention that in a complete diagrammatic calculation there would
be other non-logarithmic corrections to the couplings not included in
our calculation. The corrections to the diagonal elements of the Yukawa
coupling matrices are already included in ISAJET using the formulae in
Ref.~\cite{pierce}, and so will be automatically included when these
RGEs are embedded into the ISAJET code. In contrast to a diagrammatic 
approach, our method (also the one used in ISAJET) automatically sums 
the potentially large logarithms that arise when the MSSM spectrum is 
significantly split. We also remark that in
Fig.~\ref{fig:absf} (as well as in the other figures below), we have
ignored any finite shifts in the coupling constants coming from the
matching of the two effective theories (with and without the heavy
particle) when we decouple particles at their mass scale
\cite{shifts}. This would potentially give jumps in the couplings as we
cross the various thresholds which must be taken into account to achieve
true two-loop accuracy. The RGEs that we have derived are, of course,
unaffected.

\subsection{Higgsino and gaugino couplings to quarks and squarks}

We now turn to a study of the evolution of the couplings of gauginos and
higgsinos to the quark-squark system. Above all thresholds, {\it i.e}
for $Q> m_{\tg}= m_H$ in our numerical study, these are equal to the
corresponding gauge and quark Yukawa couplings, but differ from them
for the range of $Q$ between the lower threshold at the sfermion,
higgsino and electroweak gaugino masses and the high threshold. Of
course, for $Q$ values below this lower threshold, the effective theory
is the SM and the couplings ${\tilde{\bf{g}}}$, ${\tilde{\bf{g}}}'$,
$\tilde{\bf{f}}_u$ and $\tilde{\bf{f}}_d$ cease to be meaningful
quantitates in our simple-minded approach with thresholds being
incorporated by step functions. Note also that because the gluino is the
heaviest SUSY particle in our illustration, there is no
${\tilde{\bf{g}}}_s$ coupling. Above all thresholds, supersymmetry 
relates gaugino and gauge couplings so that flavour violation via 
gaugino interactions is not allowed. For $Q<m_{H}$, the connection 
between the gauge couplings and gaugino couplings is broken, and 
flavour-violating interactions of gauginos are no longer forbidden by 
any symmetry. Although these $\tilde{\mathbf{g}}^{\Phi}$-type couplings 
start off proportional to unit matrices at $Q=m_{H}$, contributions 
to their renormalization group evolution from terms involving Higgs 
boson and higgsino couplings to quarks and squarks (see, \textit{e.g.} 
Eq.~(\ref{app:bgtpur}) of the Appendix), which cancel out if all 
thetas are set to unity, cause these to develop off-diagonal elements 
for $Q<m_{H}$.

These off-diagonal components of $\tilde{\bf{g}}$ and $\tilde{\bf{g}}'$
(and also of the corresponding Yukawa couplings) induce
flavour-violating decays, $\tq_i \to q_j\tz_k$, of squarks even if there
is no explicit flavour-mixing in the SSB sector ({\it i.e.} all flavour
violation is induced via superpotential Yukawa couplings as is the case
in many models), and hence are phenomenologically relevant. In this
case, the partial widths for flavour-violating decays will be small, so
that these decays will likely be most important for those squarks for
which flavour-conserving, tree-level two body decays are kinematically
forbidden. Thus, the neutralino in the flavour-violating decay is most
likely to be the LSP. Since the LSP is bino-like in many models, it
would seem that the couplings ${\tilde{\bf{g}}}'$ would be the most
relevant, but this would of course depend on the size of the
corresponding off-diagonal component, relative to the same component of
the other couplings. 

\begin{figure}\begin{centering}
$\begin{array}{c}
\includegraphics[width=.65\textwidth]{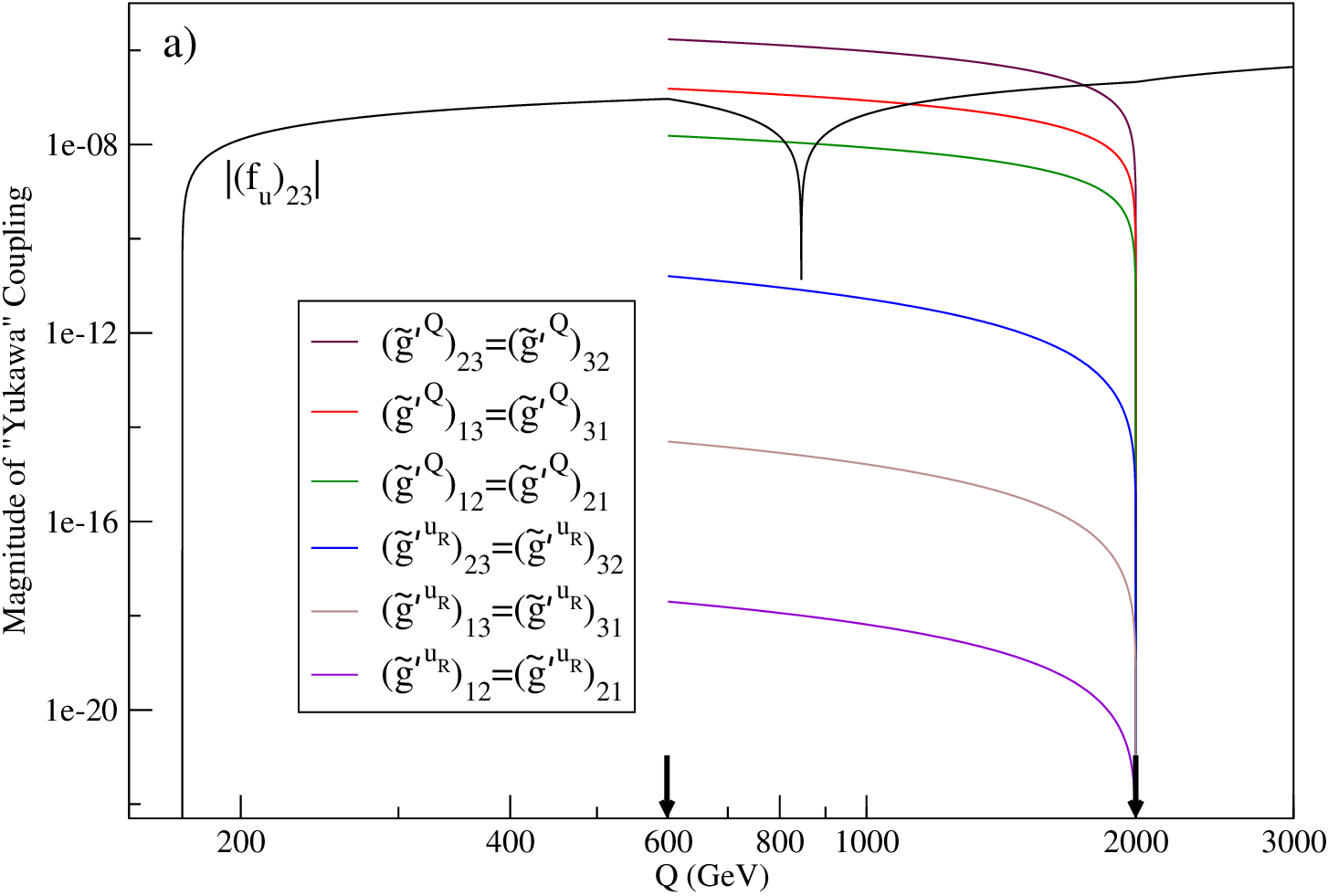}\\[35pt]
\includegraphics[width=.65\textwidth]{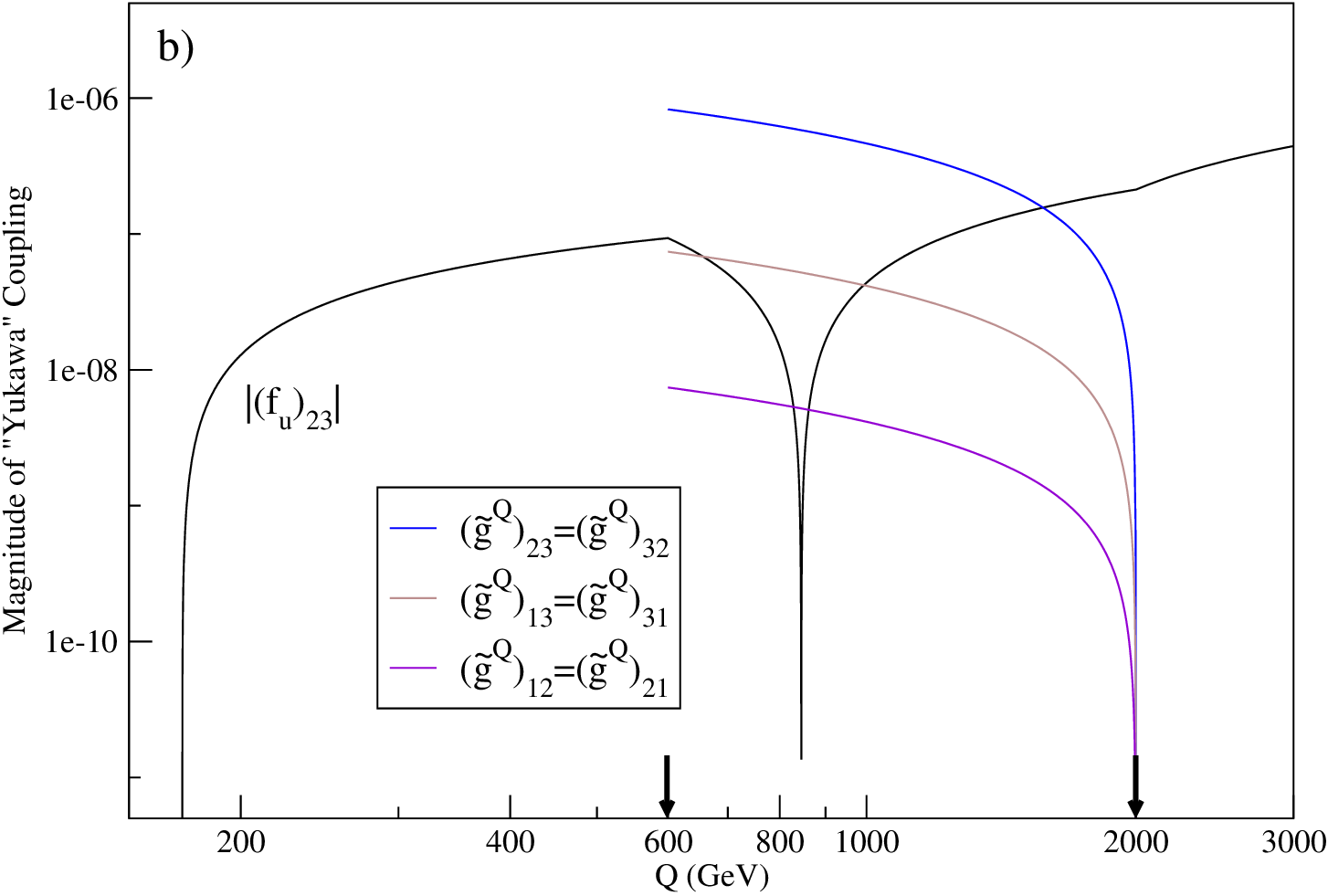}
\end{array}$
\caption{Evolution of the magnitudes of the complex off-diagonal elements
of the gaugino coupling matrices: a) $\bgtpqnb$ and $\bgtpurnb$, and b)
$\bgtqnb$ defined in the text for the MSSM, with the positions of the
sparticle and Higgs thresholds (again denoted by arrows) as in
Fig.~\ref{fig:absf}. Also shown for comparison is the magnitude of the
$(2,3)$ element, the one with the largest magnitude, of the up-quark
coupling matrix $\bdf_{u}$.  The legend is in the same order (from top
to bottom) as the curves. The magnitudes of the elements of the gaugino
coupling matrices are symmetric in the flavour indices to a good 
approximation, as explained in the text.}
\label{fig:gaugino}
\end{centering}
\end{figure}

We begin by showing the magnitudes of the off-diagonal components of the
electroweak gaugino-quark-squark couplings in Fig.~\ref{fig:gaugino} for
{\it a})~the matrices ${\tilde{\bf{g}}}'^{Q}$ and
${\tilde{\bf{g}}}'^{u_R}$, and {\it b})~the matrices
${\tilde{\bf{g}}}^{Q}$ (there is no ${\tilde{\bf{g}}}^{u_R}$). 
We obtain these from the corresponding evolution equations with the
boundary condition that these matrices are equal to the corresponding
gauge coupling times the unit matrix at $Q=m_H$. On the low $Q$ side, we
terminate these curves at the lower SUSY threshold below which these
couplings no longer exist. Also shown for comparison is the (2,3)
element, the one with the largest magnitude, 
of the up-quark Yukawa coupling matrix ${\bf{f}}_u$.  Remember
that we are plotting these matrix elements in the basis where the
up-quark (but not the down-quark) Yukawa couplings are diagonal at
$Q=m_t$. There are several features worth remarking about.
\begin{enumerate}
\item The off-diagonal elements of the gaugino coupling matrices vary
  over the same broad range as the corresponding elements of the Yukawa
  coupling matrices shown in Fig.~\ref{fig:absf}. Indeed the magnitude
  of some of these elements considerably exceed the largest element of
  the quark Yukawa coupling matrix. It would, therefore, be dangerous
  to simply disregard these when discussing sparticle flavour physics,
  particularly in models where Yukawa couplings are the sole source of
  flavour violation.

\item The off-diagonal couplings of the ${\tilde{\bf{g}}}'^Q$ matrix are
  several orders of magnitude larger than those of
  ${\tilde{\bf{g}}}'^{u_R}$. The reason for this can be traced to the
  RGEs for these listed in the Appendix. We see from (\ref{app:bgtpq})
  and (\ref{app:bgtpur}) that the evolution of the matrices
  ${\tilde{\bf{g}}}'^Q$ depends on the down-type Yukawa couplings, while
  that of  ${\tilde{\bf{g}}}'^{u_R}$ does not have any such
  contributions. In the basis that we are working in, these
  contributions are much larger than those from up-type Yukawa matrices
  (and larger than those from the gaugino coupling matrices which 
start off as unit matrices at $Q=m_H$), so that it is no surprise that the
  off-diagonal elements of  ${\tilde{\bf{g}}}'^{Q}$ attain larger values
  than those of ${\tilde{\bf{g}}}'^{u_R}$. A similar remark applies to
  the magnitudes of the elements of ${\tilde{\bf{g}}}^{Q}$ shown in
  frame {\it b}).

\item We also see from the figure that the magnitudes of the gaugino coupling
  matrices are symmetric under the interchange of the two indices. We
  have checked that this symmetry is not exact but (for our
  illustration) holds to a few parts per mille. The reason for the
  symmetry is that in the course of their evolution starting from a unit
  matrix at $Q=m_H$, the gaugino coupling matrices remain (approximately)
  Hermitian. This is simplest to see for the $\bgtpqnb$ matrices,
  retaining only the dominant terms involving the down-type Higgs and
  higgsino coupling matrices on the right-hand-side of
  (\ref{app:bgtpq}). If we now ignore the small difference between the
  Higgs and higgsino coupling matrices in the RGE, and keep in mind that
  the matrices $\bgtpqnb$ and $\bgtqnb$ on the right-hand-side are almost
  the unit matrices, it is not difficult to see that the gaugino
  coupling matrices $\bgtpqnb$ and $\bgtqnb$ remain Hermitian with, but only
  with, our approximations. A similar, but somewhat more involved,
  argument also holds for the off-diagonal elements of $\bgtpurnb$, where
  we have to look at all the terms in (\ref{app:bgtpur}) since, in the
  absence of down-type Yukawa matrices on the right-hand-side, no single
  term dominates. Nonetheless, we have checked that if the difference
  between Higgs boson and higgsino coupling matrices on the
  right-hand-side can be ignored, and the gaugino coupling matrices can
  be approximated by unit matrices, the renormalization group evolution
  preserves the (approximate) hermiticity of $\bgtpurnb$, explaining why
  the magnitudes of the corresponding off-diagonal elements are
  symmetric in the flavour indices. 
\end{enumerate}

\begin{figure}\begin{centering}
\includegraphics[width=.7\textwidth]{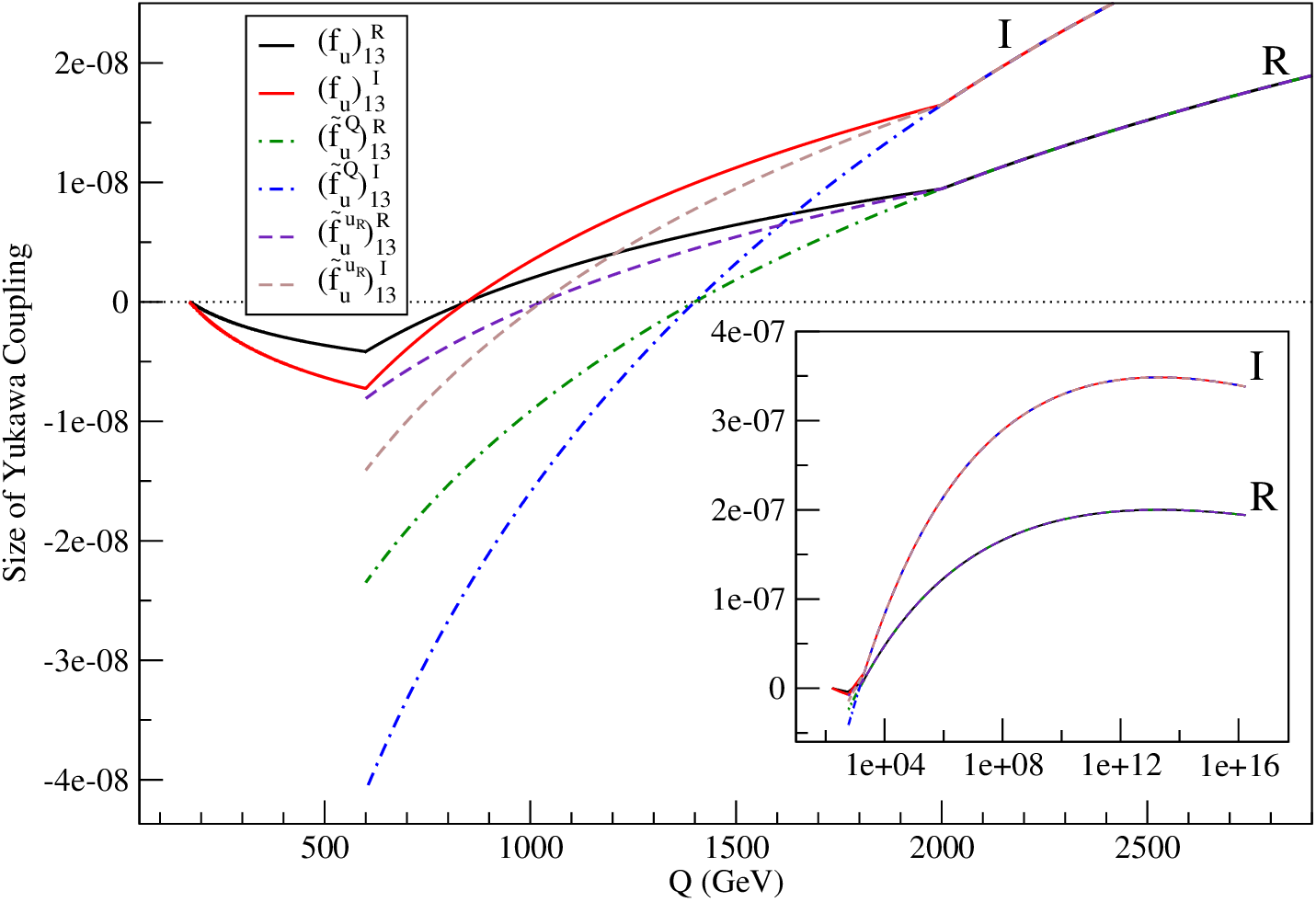}
\caption{The real (R) and imaginary (I) part of the $(1,3)$ element of
the up-quark Yukawa coupling matrix, $\bdf_{u}$, along with the
corresponding elements of the matrices $\bftuqnb$ and $\bfturnb$ for the
MSSM with the spectrum as in Fig.~\ref{fig:absf}. For the solid black
and red lines, we plot the elements of the matrix
$\bm{\lambda}_{u}/\sin{\beta}$ below $Q=m_{H}$, as in
Fig.~\ref{fig:absf}. The main figure zooms in on the low end of the
range of $Q$ where the Higgs boson and higgsino coupling differ from one
another, while the inset shows the evolution all the way to $M_{\rm GUT}$.}
\label{fig:higgsino}
\end{centering}
\end{figure}

Next, we turn to the off-diagonal elements of the $\tilde{\bf{f}}$-matrices. 
We know that these will deviate from the elements
of the corresponding Yukawa coupling matrix only below the threshold at
$Q=m_H$, but have magnitudes similar to the corresponding Yukawa
coupling matrix elements, whose absolute values are shown in
Fig.~\ref{fig:absf}. It is the {\it difference} between the couplings
of Higgs bosons and higgsinos that will be the main focus of our
attention. In Fig.~\ref{fig:higgsino}, we show the evolution of the real
and imaginary parts of the (1,3) element of (\textit{i.})~the Yukawa coupling
matrix, $\bdf_u$ for $Q>m_H$ and $\bm{\lambda}_u/\sin\beta$ (which
connects continuously to $\bdf_u$) for $Q<m_H$, (\textit{ii.})~the higgsino coupling
matrices $\bftuqnb$, and $\bfturnb$ whose evolution is shown in
(\ref{app:bftuq}) and (\ref{app:bftur}), respectively, of the Appendix.
There is no particular reason for our choice of the $(1,3)$ element (which
happens to have a comparable real and imaginary piece) for the
illustration in the figure. Here we have focussed on the lower end of $Q$ 
where the $\mathbf{f}_{u}$ and $\tilde{\textbf{f}}^{\Phi}_{u}$ couplings 
are different, while the inset shows the evolution all the way to
$M_{\rm GUT}$. 
Several points may be worthy of notice.
\begin{itemize}
\item For $Q>m_H$ where the effective theory is supersymmetric, we see
  that the real and imaginary parts of the Higgs boson and higgsino
  couplings separately come together as expected.

\item The reader can easily check that the ratio of the real and imaginary
  parts of the Higgs boson Yukawa couplings is independent of the scale
  as we had discussed in
  Sec.~\ref{subsec:quarkyukawacouplings}. 
\item For $Q<m_H$, the higgsino couplings are split from the
  corresponding Higgs boson couplings as well as from one another by a
  factor of several. For instance at the scale of squark masses, the
  real (imaginary) parts of the (1,3) element of both $\bftuqnb$ and
  $\bfturnb$ are quite different from the real (imaginary) parts of
  $({\bdf_u})_{13}$. It seems to us that the use of the evolved Higgs
  boson coupling in place of the corresponding higgsino coupling could
  be a poor approximation.
\item Notice that while the real and imaginary parts of $\bftuq_{13}$
  and $\bftur_{13}$ come to zero at the same point, the position of 
  the zero differs for the two couplings.
\end{itemize}

\begin{figure}\begin{centering}
\includegraphics[width=.7\textwidth]{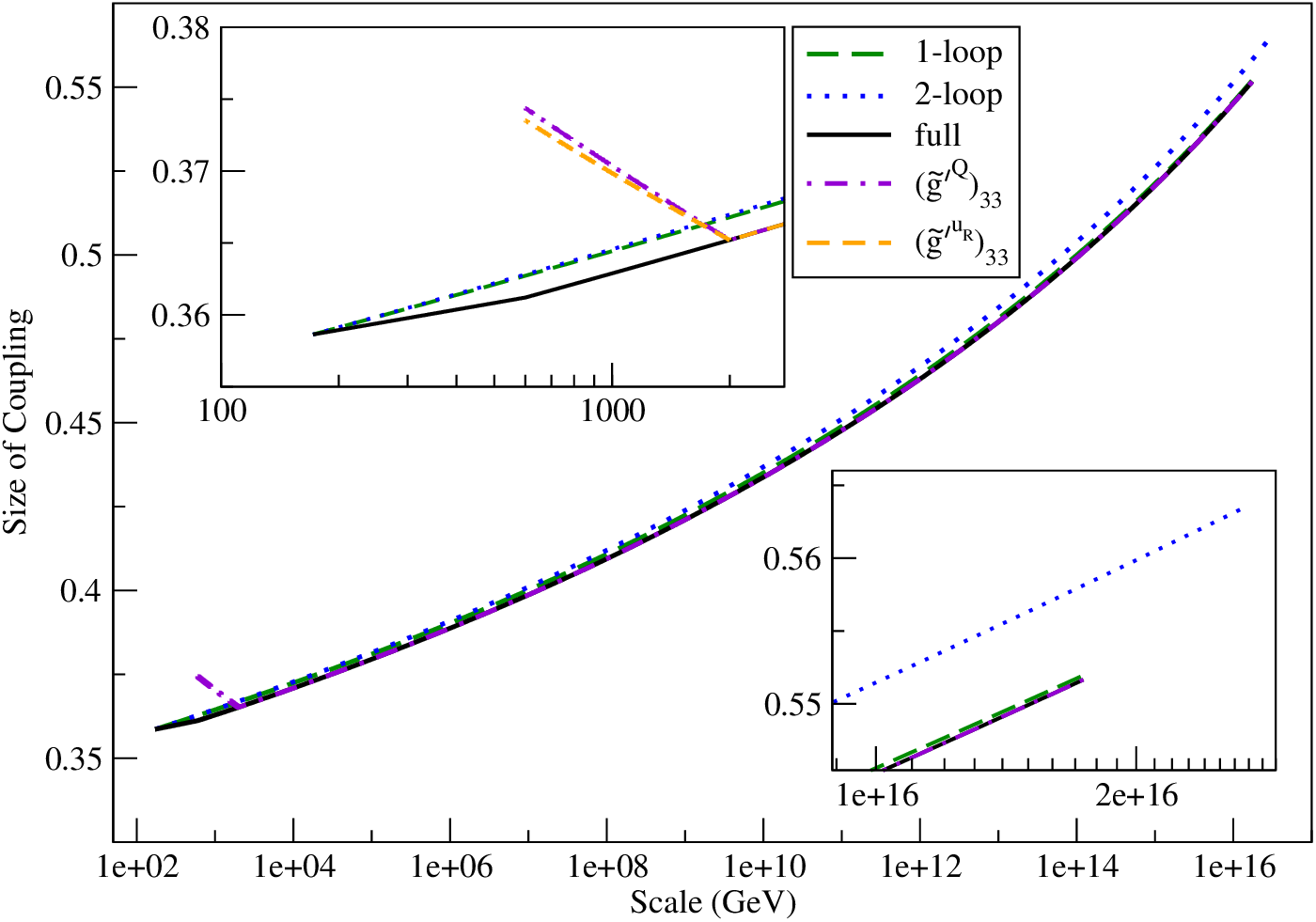}
\caption{Evolution of the hypercharge gauge coupling at the 1-loop 
(green dashed line) and 2-loop 
(blue dotted line) levels (excluding threshold effects), and with the
full calculation (solid black line). Also shown is the evolution of the
$(3,3)$ element of $\bgtpqnb$ (and $\bgtpurnb$ in the left inset). The 
main figure shows
the evolution of these couplings between $Q=m_{t}$ and $Q=M_{\rm GUT}$,
while the insets on the left and right zoom in on the range of $Q$
near the TeV scale and near $M_{\rm GUT}$ respectively.}
\label{fig:gauge}
\end{centering}
\end{figure}

Up to now we have focussed our attention on flavour off-diagonal
couplings. Before closing our discussion, we briefly consider the effect
of the thresholds on the evolution of flavour diagonal couplings. As an
illustration, we show the evolution of the hypercharge gauge coupling
$g'$ and the (3,3) elements hypercharge gaugino couplings ${\bf{g}}'^Q$
and ${\bf{g}}'^{u_R}$ in Fig.~\ref{fig:gauge}.

The black solid line denotes the result of our complete
calculation of the gauge coupling including both threshold and two loop
effects. Also shown by green dashed and blue dotted lines are the
corresponding results obtained at the one- and two-loop levels, but
ignoring threshold effects. The evolution
of the (3,3) element of $\bgtpqnb$ breaks away from the
evolution of $g'$ at $Q=m_H$, and is shown as the violet
dot-dashed curve that terminates at $Q=600$~GeV. The insets on the 
left and right show zooms
of these curves at the low and high ends, respectively. In the inset on
the left, we have also shown the evolution of the (3,3)  element of
 $\bgtpurnb$ as the orange short-dashed line. 
We have checked that
all the diagonal elements of both $\tilde{\mathbf{g}}'^{Q}$ and 
$\tilde{\mathbf{g}}'^{u_{R}}$ have a similar behaviour. Moreover,
the $(1,1)$ and $(2,2)$ elements all evolve essentially together 
with only the $(3,3)$ elements split from these due to top ``Yukawa''
couplings.\footnote{In our illustration where we have $\tan{\beta}=4$,
the curves for the $(1,1)$ and $(2,2)$ elements fall in-between the 
dot-dashed violet curve for $\bgtpq_{33}$ and the orange-dashed curve for
$\bgtpur_{33}$ in Fig.~\ref{fig:gauge}.}
The gauge coupling
curves use the measured value of the coupling at the low scale as the
boundary condition, while the gaugino coupling curves are obtained
assuming that the corresponding matrix equals the hypercharge coupling
times the unit matrix at $Q=m_H$. Several points are worth noting:
\begin{enumerate} 
\item Since the value of the gauge coupling is fixed at $Q=m_t$ in our
  illustration, the difference between one and two-loop evolution
  (without threshold effects) is seen at the high $Q$ end where the
  curves are terminated at $Q=M_{\rm GUT}$ defined as the point where the
  $SU(2)$ and the scaled hypercharge couplings meet. Notice that $M_{\rm
  GUT}$ is different in the two cases, as is the value of the gauge
  coupling at $Q=M_{\rm GUT}$. This relative difference is ${\cal
  O}\left({\frac{1}{16\pi^2}}\right)$, roughly the expected magnitude of a higher
  loop effect. The full calculation of the evolution depicted by the
  solid black line differs most from the other two curves at the low end,
  because we start off with the evolution using the $\beta$-function of
  the SM and, as we pass through the various thresholds, join up at
  $Q=m_H$ to the MSSM evolution. Beyond $Q=m_H$, the full and two-loop
  curves evolve with the same $\beta$-function, but the latter ends up
  lower because threshold effects caused it to start off lower at
  $Q=m_H$, as can be seen in the inset on the left. That it ends up so
  close to the one-loop curve is a coincidence, but its proximity
  reflects the conventional wisdom that two loop effects in the
  evolution are numerically comparable in magnitude to the threshold
  corrections in the evolution at one-loop.
 
\item Turning to the gaugino couplings at the low end, the most striking
  feature is that below $Q=m_H$, these evolve in the opposite direction
  to the gauge couplings. This behaviour can be understood if we
  recognize that for $Q< m_{\tg}$ {\em the evolution of the diagonal
  gaugino couplings now depends on the much larger gluon coupling even
  at the one-loop level}. As we can see from (\ref{app:bgtpq}), the
  terms involving ${\tilde{\bf{g}}}_s^Q$ cancel the terms depending on
  the QCD coupling constant $g_s$ if $Q>m_{\tg}$. However, for
  $Q<m_{\tg}$ the terms with $\theta_{\tg}$ are no longer operative, and
  this cancellation is incomplete, causing a large change in the
  $\beta$-function, and hence in the slope of the curve. It is striking
  to see that even though we have maintained both thresholds not far
  from the TeV scale, the corresponding gaugino and gauge couplings can
  develop a difference of~$\sim 4$\%. The existence of a difference
  between a gauge boson coupling and the corresponding gaugino coupling
  has been discussed in Ref. \cite{difference} (although without any
  flavour structure), where it was suggested that its determination at
  an $e^+e^-$ linear collider \cite{feng} would give an idea of
  the splitting between sparticle masses, even if the heavy sparticles are
  not kinematically accessible.
\end{enumerate}

\begin{figure}\begin{centering}
\includegraphics[width=.7\textwidth]{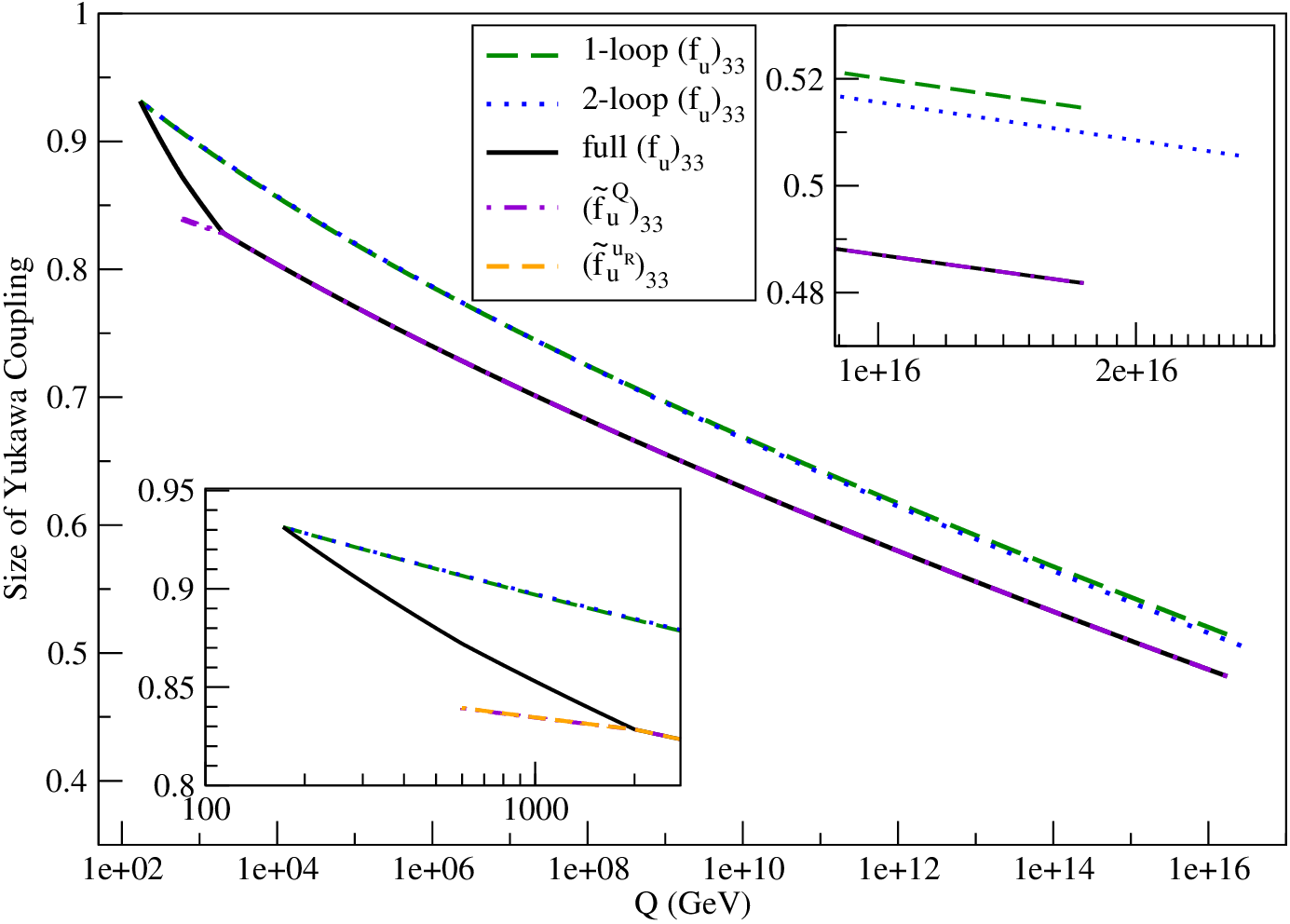}
\caption{Evolution of the $(3,3)$ element of the Yukawa coupling,  
$\bdf_{u}$, at the 1-loop (green dashed line) and 2-loop 
(blue dotted line) levels (excluding threshold effects), and with the
full calculation (solid black line). Also shown is the evolution of the
$(3,3)$ element of $\bftuqnb$ (and $\bfturnb$ in the left inset). The 
main figure shows
the evolution of these couplings between $Q=m_{t}$ and $Q=M_{\rm GUT}$,
while the insets on the left and right zoom in on the range of $Q$
near the TeV scale and near $M_{\rm GUT}$ respectively. As in 
Fig.~\ref{fig:absf}, we have plotted 
$\bm{\lambda}_{u}/\sin{\beta}$ below $Q=m_{H}$.}
\label{fig:yuk}
\end{centering}
\end{figure}

Finally, we turn to flavour-conserving Higgs and
higgsino interactions. As an illustration, we show the evolution of
the (3,3) elements of $\bdf_u$, of $\bftuqnb$, and of
$\bfturnb$ in Fig.~\ref{fig:yuk}.
Just as in the previous figure, we show results for the
complete calculation at the two-loop level, including threshold
effects as well as differences between the couplings of Higgs/gauge
bosons and of higgsinos/gauginos by the solid black line. We also show
the results that we obtain with just MSSM evolution all the way to
$Q=m_t$ using one-loop RGEs (the green dashed curve) and two-loop RGEs
(the blue dotted curve). In all three cases we start with the same
value for the Yukawa coupling at $Q=m_t$. Also shown by the violet
dot-dashed and orange short-dashed lines is the evolution of the (3,3)
element of the matrices $\bftuqnb$ and $\bfturnb$, respectively. The two
insets are similar to those in the previous figure. We note the
following:
\begin{enumerate}
\item We see that the complete calculation of the Yukawa coupling leads
to a large difference from the two-loop calculation without thresholds
over most of the range of $Q$. This is not new and is largely due to the
difference between the evolution of Yukawa couplings in the SM and in
the effective theories that interpolate between the SM and the
MSSM.\footnote{Notice in the left inset that the solid black line 
``curves'' significantly
between $Q=600$~GeV and $Q=m_t$. This highlights the advantage of this
approach (also used in the event generator ISAJET) which ``sums logs of
the ratio of the low and high threshold by solving the RGEs'', with that
sometimes used in the literature where MSSM evolution is used all the
way down to $Q=m_t$ and then corrected for ``via a single step''
evolution to take into account the difference between the running in the
MSSM and in the SM.} Below the kink at $Q=600$~GeV in the solid black
curve the evolution of the Yukawa coupling is as in the SM, while above
$Q= m_H$ it is as in the MSSM. Notice that in this case, threshold
effects are considerably larger than the difference between one- and
two-loop evolution. 
\item It can be clearly seen from the left-inset that the higgsino
  couplings evolve quite differently from the corresponding Higgs boson
  couplings once $Q$ is below $m_H$. Once again, this difference is
  largely due to incomplete cancellations in terms involving
  ``strong'' interaction couplings of gluons and gluinos, as may be seen
  from (\ref{app:bftuq}) and (\ref{app:bftur}) of the Appendix. 
\end{enumerate}

\section{Concluding Remarks}\label{summary}

RGEs provide the bridge that allows us to extract predictions of
theories with simple physical principles valid at very high energy
scales, many orders of magnitude larger than energies accessible in
experiments. Because of renormalization effects, these same simple principles
lead to complex predictions at accessible energies.  Since
supersymmetric theories allow sensible extrapolation to high energy,
RGEs have played a central role in the analysis of many supersymmetric
models, generally assumed to reduce to the MSSM (possibly augmented by
right-handed neutrino superfields) in the range between the weak 
and GUT or Planck scales.

In this, the first of a series of two papers, we have re-examined the
threshold corrections to the RGEs for the dimensionless couplings of the
MSSM, incorporating also the effects from flavour mixing of quarks and
squarks.  Above the scale of all new particle thresholds, the effective
theory is the MSSM, with just three gauge couplings and three different
Yukawa coupling matrices (that specify the interactions of matter
fermions with Higgs bosons) being the independent dimensionless couplings
of the theory. All other dimensionless couplings, for instance those of
gauginos or higgsinos to the fermion-sfermion system, or quartic scalar
couplings, are related to these by supersymmetry. These relations are,
however, no longer valid once supersymmetry is broken, so that then the
couplings to gauginos and higgsinos  will renormalize differently from
gauge and Yukawa couplings, respectively. In a consistent treatment of
threshold corrections, the RGEs for these couplings will therefore
differ from the RGEs for the gauge  and quark Yukawa couplings that are
available in the literature. 

We have adapted the RGEs for a general ({\it i.e.} non-supersymmetric)
field theory \cite{mv,luo} and rewritten them in four-component spinor
notation that we use when we obtain the RGEs of the MSSM. The details of our
procedure may be found in Sec.~\ref{formalism} and \ref{threshold}, and the 
application to the MSSM in Sec.~\ref{derive}. The complete set of RGEs for
the dimensionless couplings between SM particles and their superpartners
(but not for the quartic couplings of scalars) is given in the Appendix.
These quartic couplings are less important from a phenomenological
perspective, though some of these do enter the squark and slepton mass
matrices. The important thing for the present discussion is these do not
enter into the RGEs for the fermion-fermion-scalar couplings, which can
then be evolved independently from the quartic couplings.

We have presented some sample numerical results for the evolution of
various couplings in Sec.~\ref{anal}. The analysis here is meant only to
give a sampling of effects that, to our knowledge, have not been
previously pointed out in the literature. Since flavour physics in the
sparticle sector has been the main motivation for our analysis, we point
out that when threshold effects are included, gauginos are not different
from higgsinos in that neutral gauginos also develop flavour-changing
couplings to quarks and squarks. As seen from Fig.~\ref{fig:gaugino},
while these flavour-violating gaugino couplings vary over several orders
of magnitude, they can be comparable, or even larger, than the
corresponding couplings to Higgs bosons. An illustration of
flavour-violating higgsino couplings is shown in
Fig.~\ref{fig:higgsino}. The scale-dependence of these is quite
different from that of the usual Yukawa couplings of quarks to Higgs
bosons. We stress that the induced flavour violating couplings of
gauginos and higgsinos, evaluated at the scale of any particular squark
mass, will contribute to the amplitude for the flavour-violating decay
of this squark. Such a contribution is distinct from the usually
evaluated contribution from the induced flavour-violation in the squark
mass matrix. Both types of contributions need to be included. Finally,
we point out one last striking feature of our analysis. We see from
Fig.~\ref{fig:gauge} and Fig.~\ref{fig:yuk} that the inclusion of SUSY
threshold corrections leads to a difference of a few percent between the
flavour-diagonal electroweak couplings of gauginos and gauge bosons, and
of higgsinos and Higgs bosons, traced to SUSY QCD contributions in even
the 1-loop RGEs. This could have a noticeable effect on the evaluation
of sparticle masses as well as of thermal relic densities of neutralino
dark matter created in the Big Bang.

In summary, we have presented the RGEs of the dimensionless couplings of
the MSSM in this paper. In a follow-up paper under preparation
\cite{dimensionful}, we will present the RGEs for the dimensionful
parameters, again including threshold and flavour-mixing effects. This
complete set will facilitate the examination of flavour phenomenology in
SUSY models with arbitrary ans\"atze for flavour in the SSB sector.

\section*{Acknowledgments} We are grateful to H.~Baer, D.~Casta\~no,
A.~Dedes, S.~Martin, K.~Melnikov, A.~Mustafayev and M.~Vaughn for
clarifying comments and communications. We thank D.~Casta\~no for
sending us the unpublished erratum to Ref.~\cite{ramond}, and H.~Baer and
A.~Mustafayev for their comments on the manuscript. This research
was supported in part by a grant from the United States Department of Energy.

\section*{Appendix}
\begin{equation}\begin{split}
{\left(4\pi\right)}^2\frac{d(\sn\bdf_u)_{ij}}{dt}=&\frac{\sn}{2}\left\{3\left[\sn^2\h+\cs^2\Hh\right](\bdf_u\bdf_u^\dagger)_{ik}+\left[\cs^2\h+\sn^2\Hh\right](\bdf_d\bdf_d^\dagger)_{ik}\right.\\
&\left.\quad\ +4\cs^2\left[-\h+\Hh\right](\bdf_d\bdf_d^\dagger)_{ik}\right\}(\bdf_u)_{kj}\\
&+\sn(\bdf_u)_{ik}\left[\sh\sql\bftuq^\dagger_{kl}\bftuq_{lj}+\frac{4}{9}\sbi\sul\bgtpur^*_{kl}\bgtpur^T_{lj}+\frac{4}{3}\sgl\sul\bgtsur^*_{kl}\bgtsur^T_{lj}\right]\\
&+\frac{\sn}{4}\left[2\sh\suk\bftur_{ik}\bftur^\dagger_{kl}+2\sh\sdk\bftdr_{ik}\bftdr^\dagger_{kl}+3\swi\sqk\bgtq^T_{ik}\bgtq^*_{kl}\right.\\
&\left.\qquad+\frac{1}{9}\sbi\sqk\bgtpq^T_{ik}\bgtpq^*_{kl}+\frac{16}{3}\sgl\sqk\bgtsq^T_{ik}\bgtsq^*_{kl}\right](\bdf_u)_{lj}\\
&+\sn\sh\sqk\left[-3\swi\gthus\bgtq^T_{ik}+\frac{1}{3}\sbi\gtphus\bgtpq^T_{ik}\right]\bftuq_{kj}\\
&-\frac{4}{3}\sn\sbi\sh\suk\gtphus\bftur_{ik}\bgtpur^T_{kj}\\
&+\sn(\bdf_u)_{ij}\left[\left(\sn^2\h+\cs^2\Hh\right)\mathrm{Tr}{\left\{3\bdf_u^\dagger\bdf_u\right\}}+\cs^2\left(\h-\Hh\right)\mathrm{Tr}{\left\{3\bdf_d^\dagger\bdf_d+\bdf_e^\dagger\bdf_e\right\}}\right]\\
&+\frac{\sn}{2}\sh(\bdf_u)_{ij}\left\{3\swi\left[\mgthusq\left(\sn^2\h+\cs^2\Hh\right)+\mgthdsq\left(\cs^2\h-\cs^2\Hh\right)\right]\right.\\
&\left.\qquad\qquad\qquad+\sbi\left[\mgtphusq\left(\sn^2\h+\cs^2\Hh\right)+\mgtphdsq\left(\cs^2\h-\cs^2\Hh\right)\right]\right\}\\
&-\sn(\bdf_u)_{ij}\left[\frac{17}{12}g'^2+\frac{9}{4}g^2_2+8g^2_s\right]
\end{split}\end{equation}

\begin{equation}\begin{split}
{\left(4\pi\right)}^2\frac{d(\cs\bdf_d)_{ij}}{dt}=&\frac{\cs}{2}\left\{3\left[\cs^2\h+\sn^2\Hh\right](\bdf_d\bdf_d^\dagger)_{ik}+\left[\sn^2\h+\cs^2\Hh\right](\bdf_u\bdf_u^\dagger)_{ik}\right.\\
&\left.\quad\ +4\sn^2\left[-\h+\Hh\right](\bdf_u\bdf_u^\dagger)_{ik}\right\}(\bdf_d)_{kj}\\
&+\cs(\bdf_d)_{ik}\left[\sh\sql\bftdq^\dagger_{kl}\bftdq_{lj}+\frac{1}{9}\sbi\sdl\bgtpdr^*_{kl}\bgtpdr^T_{lj}+\frac{4}{3}\sgl\sdl\bgtsdr^*_{kl}\bgtsdr^T_{lj}\right]\\
&+\frac{\cs}{4}\left[2\sh\suk\bftur_{ik}\bftur^\dagger_{kl}+2\sh\sdk\bftdr_{ik}\bftdr^\dagger_{kl}+3\swi\sqk\bgtq^T_{ik}\bgtq^*_{kl}\right.\\
&\left.\qquad+\frac{1}{9}\sbi\sqk\bgtpq^T_{ik}\bgtpq^*_{kl}+\frac{16}{3}\sgl\sqk\bgtsq^T_{ik}\bgtsq^*_{kl}\right](\bdf_d)_{lj}\\
&+\cs\sh\sqk\left[-3\swi\gthds\bgtq^T_{ik}-\frac{1}{3}\sbi\gtphds\bgtpq^T_{ik}\right]\bftdq_{kj}\\
&-\frac{2}{3}\cs\sbi\sh\sdk\gtphds\bftdr_{ik}\bgtpdr^T_{kj}\\
&+\cs(\bdf_d)_{ij}\left[\sn^2\left(\h-\Hh\right)\mathrm{Tr}{\left\{3\bdf_u^\dagger\bdf_u\right\}}+\left(\cs^2\h+\sn^2\Hh\right)\mathrm{Tr}{\left\{3\bdf_d^\dagger\bdf_d+\bdf_e^\dagger\bdf_e\right\}}\right]\\
&+\frac{\cs}{2}\sh(\bdf_d)_{ij}\left\{3\swi\left[\mgthusq\left(\sn^2\h-\sn^2\Hh\right)+\mgthdsq\left(\cs^2\h+\sn^2\Hh\right)\right]\right.\\
&\left.\qquad\qquad\qquad+\sbi\left[\mgtphusq\left(\sn^2\h-\sn^2\Hh\right)+\mgtphdsq\left(\cs^2\h+\sn^2\Hh\right)\right]\right\}\\
&-\cs(\bdf_d)\left[\frac{5}{12}g'^2+\frac{9}{4}g^2_2+8g^2_s\right]
\end{split}\end{equation}

\begin{equation}\begin{split}
{\left(4\pi\right)}^2\frac{d(\cs\bdf_e)_{ij}}{dt}=&\frac{3}{2}\cs\left[\cs^2\h+\sn^2\Hh\right](\bdf_e\bdf_e^\dagger\bdf_e)_{ij}\\
&+\cs(\bdf_e)_{ik}\left[\sh\sll\bftel^\dagger_{kl}\bftel_{lj}+\sbi\sel\bgtper^*_{kl}\bgtper^T_{lj}\right]\\
&+\frac{\cs}{4}\left[2\sh\sek\bfter_{ik}\bfter^\dagger_{kl}+3\swi\slk\bgtl^T_{ik}\bgtl^*_{kl}+\sbi\slk\bgtpl^T_{ik}\bgtpl^*_{kl}\right](\bdf_e)_{lj}\\
&+\cs\sh\slk\left[-3\swi\gthds\bgtl^T_{ik}+\sbi\gtphds\bgtpl^T_{ik}\right]\bftel_{kj}\\
&-2\cs\sbi\sh\sek\gtphds\bfter_{ik}\bgtper^T_{kj}\\
&+\cs(\bdf_e)_{ij}\left[\sn^2\left(\h-\Hh\right)\mathrm{Tr}{\left\{3\bdf_u^\dagger\bdf_u\right\}}+\left(\cs^2\h+\sn^2\Hh\right)\mathrm{Tr}{\left\{3\bdf_d^\dagger\bdf_d+\bdf_e^\dagger\bdf_e\right\}}\right]\\
&+\frac{\cs}{2}\sh(\bdf_e)_{ij}\left\{3\swi\left[\mgthusq\left(\sn^2\h-\sn^2\Hh\right)+\mgthdsq\left(\cs^2\h+\sn^2\Hh\right)\right]\right.\\
&\left.\qquad\qquad\qquad+\sbi\left[\mgtphusq\left(\sn^2\h-\sn^2\Hh\right)+\mgtphdsq\left(\cs^2\h+\sn^2\Hh\right)\right]\right\}\\
&-\cs(\bdf_e)_{ij}\left[\frac{15}{4}g'^2+\frac{9}{4}g^2_2\right]
\end{split}\end{equation}

\begin{equation}\begin{split}\label{app:bftuq}
{\left(4\pi\right)}^2\frac{d\bftuq_{ij}}{dt}=&\left[\sn^2\h+\cs^2\Hh\right]\bftuq_{ik}(\bdf_u^\dagger\bdf_u)_{kj}\\
&+\bftuq_{ik}\left[\frac{4}{9}\sbi\sul\bgtpur^*_{kl}\bgtpur^T_{lj}+\frac{4}{3}\sgl\sul\bgtsur^*_{kl}\bgtsur^T_{lj}+\sh\sql\bftuq^\dagger_{kl}\bftuq_{lj}\right]\\
&+\frac{3}{2}\sh\bftuq_{ij}\left[\suk\bftur^\dagger_{kl}\bftur_{lk}+\sql\bftuq^\dagger_{kl}\bftuq_{lk}\right]\\
&+\frac{1}{4}\sh\left[\sn^2\h+\cs^2\Hh\right]\bftuq_{ij}\left\{3\swi\mgthusq+\sbi\mgtphusq\right\}\\
&+\left[\sn^2\h+\cs^2\Hh\right]\left\{-3\swi\gthu\bgtq^*_{ik}+\frac{1}{3}\sbi\gtphu\bgtpq^*_{ik}\right\}(\bdf_u)_{kj}\\
&-\frac{4}{9}\sbi\sul\bgtpq^*_{ik}\bftur_{kl}\bgtpur^T_{lj}-\frac{16}{3}\sgl\sul\bgtsq^*_{ik}\bftur_{kl}\bgtsur^T_{lj}\\
&+\sql\left[\frac{3}{2}\swi\bgtq^*_{ik}\bgtq^T_{kl}+\frac{1}{18}\sbi\bgtpq^*_{ik}\bgtpq^T_{kl}+\frac{8}{3}\sgl\bgtsq^*_{ik}\bgtsq^T_{kl}\right]\bftuq_{lj}\\
&+\sh\sql\left[\bftuq_{ik}\bftuq^\dagger_{kl}+\bftdq_{ik}\bftdq^\dagger_{kl}\right]\bftuq_{lj}\\
&-\bftuq_{ij}\left[\frac{25}{12}g'^2+\frac{9}{4}g^2_2+4g^2_s\right]
\end{split}\end{equation}

\begin{equation}\begin{split}
{\left(4\pi\right)}^2\frac{d\bftdq_{ij}}{dt}=&\left[\cs^2\h+\sn^2\Hh\right]\bftdq_{ik}(\bdf_d^\dagger\bdf_d)_{kj}\\
&+\bftdq_{ik}\left[\frac{1}{9}\sbi\sdl\bgtpdr^*_{kl}\bgtpdr^T_{lj}+\frac{4}{3}\sgl\sdl\bgtsdr^*_{kl}\bgtsdr^T_{lj}+\sh\sql\bftdq^\dagger_{kl}\bftdq_{lj}\right]\\
&+\frac{1}{2}\sh\bftdq_{ij}\left[3\sdk\bftdr^\dagger_{kl}\bftdr_{lk}+\sek\bfter^\dagger_{kl}\bfter_{lk}\right.\\
&\left.\qquad\qquad\qquad+3\sql\bftdq^\dagger_{kl}\bftdq_{lk}+\sll\bftel^\dagger_{kl}\bftel_{lk}\right]\\
&+\frac{1}{4}\sh\left[\cs^2\h+\sn^2\Hh\right]\bftdq_{ij}\left\{3\swi\mgthdsq+\sbi\mgtphdsq\right\}\\
&+\left[\cs^2\h+\sn^2\Hh\right]\left\{-3\swi\gthd\bgtq^*_{ik}-\frac{1}{3}\sbi\gtphd\bgtpq^*_{ik}\right\}(\bdf_d)_{kj}\\
&+\frac{2}{9}\sbi\sdl\bgtpq^*_{ik}\bftdr_{kl}\bgtpdr^T_{lj}-\frac{16}{3}\sgl\sdl\bgtsq^*_{ik}\bftdr_{kl}\bgtsdr^T_{lj}\\
&+\sql\left[\frac{3}{2}\swi\bgtq^*_{ik}\bgtq^T_{kl}+\frac{1}{18}\sbi\bgtpq^*_{ik}\bgtpq^T_{kl}+\frac{8}{3}\sgl\bgtsq^*_{ik}\bgtsq^T_{kl}\right]\bftdq_{lj}\\
&+\sh\sql\left[\bftuq_{ik}\bftuq^\dagger_{kl}+\bftdq_{ik}\bftdq^\dagger_{kl}\right]\bftdq_{lj}\\
&-\bftdq_{ij}\left[\frac{13}{12}g'^2+\frac{9}{4}g^2_2+4g^2_s\right]
\end{split}\end{equation}

\begin{equation}\begin{split}
{\left(4\pi\right)}^2\frac{d\bftel_{ij}}{dt}=&\left[\cs^2\h+\sn^2\Hh\right]\bftel_{ik}(\bdf_e^\dagger\bdf_e)_{kj}\\
&+\bftel_{ik}\left[\sbi\sel\bgtper^*_{kl}\bgtper^T_{lj}+\sh\sll\bftel^\dagger_{kl}\bftel_{lj}\right]\\
&+\frac{1}{2}\sh\bftel_{ij}\left[3\sdk\bftdr^\dagger_{kl}\bftdr_{lk}+\sek\bfter^\dagger_{kl}\bfter_{lk}\right.\\
&\left.\qquad\qquad\qquad+3\sql\bftdq^\dagger_{kl}\bftdq_{lk}+\sll\bftel^\dagger_{kl}\bftel_{lk}\right]\\
&+\frac{1}{4}\sh\left[\cs^2\h+\sn^2\Hh\right]\bftel_{ij}\left\{3\swi\mgthdsq+\sbi\mgtphdsq\right\}\\
&+\left[\cs^2\h+\sn^2\Hh\right]\left\{-3\swi\gthd\bgtl^*_{ik}+\sbi\gtphd\bgtpl^*_{ik}\right\}(\bdf_e)_{kj}\\
&-2\sbi\sel\bgtpl^*_{ik}\bfter_{kl}\bgtper^T_{lj}\\
&+\frac{1}{2}\sll\left[3\swi\bgtl^*_{ik}\bgtl^T_{kl}+\sbi\bgtpl^*_{ik}\bgtpl^T_{kl}\right]\bftel_{lj}\\
&+\sh\sll\bftel_{ik}\bftel^\dagger_{kl}\bftel_{lj}-\bftel_{ij}\left[\frac{15}{4}g'^2+\frac{9}{4}g^2_2\right]
\end{split}\end{equation}

\begin{equation}\begin{split}\label{app:bftur}
{\left(4\pi\right)}^2\frac{d\bftur_{ij}}{dt}=&\frac{3}{2}\sh\left[\suk\bftur_{lk}\bftur^\dagger_{kl}+\sql\bftuq_{lk}\bftuq^\dagger_{kl}\right]\bftur_{ij}\\
&+\frac{1}{4}\sh\left[\sn^2\h+\cs^2\Hh\right]\bftur_{ij}\left\{3\swi\mgthusq+\sbi\mgtphusq\right\}\\
&+\frac{1}{2}\left\{\left[\sn^2\h+\cs^2\Hh\right](\bdf_u\bdf_u^\dagger)_{ik}+\left[\cs^2\h+\sn^2\Hh\right](\bdf_d\bdf_d^\dagger)_{ik}\right\}\bftur_{kj}\\
&+\frac{1}{2}\left[\sh\suk\bftur_{ik}\bftur^\dagger_{kl}+\sh\sdk\bftdr_{ik}\bftdr^\dagger_{kl}+\frac{3}{2}\swi\sqk\bgtq^T_{ik}\bgtq^*_{kl}\right.\\
&\left.\qquad\ +\frac{1}{18}\sbi\sqk\bgtpq^T_{ik}\bgtpq^*_{kl}+\frac{8}{3}\sgl\sqk\bgtsq^T_{ik}\bgtsq^*_{kl}\right]\bftur_{lj}\\
&-\frac{4}{9}\sbi\sqk\bgtpq^T_{ik}\bftuq_{kl}\bgtpur^*_{lj}-\frac{16}{3}\sgl\sqk\bgtsq^T_{ik}\bftuq_{kl}\bgtsur^*_{lj}\\
&-\frac{4}{3}\sbi\left[\sn^2\h+\cs^2\Hh\right]\gtphu(\bdf_u)_{ik}\bgtpur^*_{kj}+2\sh\suk\bftur_{ik}\bftur^\dagger_{kl}\bftur_{lj}\\
&+\suk\bftur_{ik}\left[\frac{8}{9}\sbi\bgtpur^T_{kl}\bgtpur^*_{lj}+\frac{8}{3}\sgl\bgtsur^T_{kl}\bgtsur^*_{lj}\right]\\
&-\bftur_{ij}\left[\frac{5}{6}g'^2+\frac{9}{2}g^2_2+4g^2_s\right]
\end{split}\end{equation}

\begin{equation}\begin{split}
{\left(4\pi\right)}^2\frac{d\bftdr_{ij}}{dt}=&\frac{1}{2}\sh\left[3\sdk\bftdr_{lk}\bftdr^\dagger_{kl}+3\sql\bftdq_{lk}\bftdq^\dagger_{kl}+\sek\bfter_{lk}\bfter^\dagger_{kl}\right.\\
&\left.\qquad+\sll\bftel_{lk}\bftel^\dagger_{kl}\right]\bftdr_{ij}\\
&+\frac{1}{4}\sh\left[\cs^2\h+\sn^2\Hh\right]\bftdr_{ij}\left\{3\swi\mgthdsq+\sbi\mgtphdsq\right\}\\
&+\frac{1}{2}\left\{\left[\sn^2\h+\cs^2\Hh\right](\bdf_u\bdf_u^\dagger)_{ik}+\left[\cs^2\h+\sn^2\Hh\right](\bdf_d\bdf_d^\dagger)_{ik}\right\}\bftdr_{kj}\\
&+\frac{1}{2}\left[\sh\suk\bftur_{ik}\bftur^\dagger_{kl}+\sh\sdk\bftdr_{ik}\bftdr^\dagger_{kl}+\frac{3}{2}\swi\sqk\bgtq^T_{ik}\bgtq^*_{kl}\right.\\
&\left.\qquad\ +\frac{1}{18}\sbi\sqk\bgtpq^T_{ik}\bgtpq^*_{kl}+\frac{8}{3}\sgl\sqk\bgtsq^T_{ik}\bgtsq^*_{kl}\right]\bftdr_{lj}\\
&+\frac{2}{9}\sbi\sqk\bgtpq^T_{ik}\bftdq_{kl}\bgtpdr^*_{lj}-\frac{16}{3}\sgl\sqk\bgtsq^T_{ik}\bftdq_{kl}\bgtsdr^*_{lj}\\
&-\frac{2}{3}\sbi\left[\cs^2\h+\sn^2\Hh\right]\gtphd(\bdf_d)_{ik}\bgtpdr^*_{kj}+2\sh\sdk\bftdr_{ik}\bftdr^\dagger_{kl}\bftdr_{lj}\\
&+\sdk\bftdr_{ik}\left[\frac{2}{9}\sbi\bgtpdr^T_{kl}\bgtpdr^*_{lj}+\frac{8}{3}\sgl\bgtsdr^T_{kl}\bgtsdr^*_{lj}\right]\\
&-\bftdr_{ij}\left[\frac{5}{6}g'^2+\frac{9}{2}g^2_2+4g^2_s\right]
\end{split}\end{equation}

\begin{equation}\begin{split}
{\left(4\pi\right)}^2\frac{d\bfter_{ij}}{dt}=&\frac{1}{2}\sh\left[3\sdk\bftdr_{lk}\bftdr^\dagger_{kl}+3\sql\bftdq_{lk}\bftdq^\dagger_{kl}+\sek\bfter_{lk}\bfter^\dagger_{kl}\right.\\
&\left.\qquad+\sll\bftel_{lk}\bftel^\dagger_{kl}\right]\bfter_{ij}\\
&+\frac{1}{4}\sh\left[\cs^2\h+\sn^2\Hh\right]\bfter_{ij}\left\{3\swi\mgthdsq+\sbi\mgtphdsq\right\}\\
&+\frac{1}{2}\left[\cs^2\h+\sn^2\Hh\right](\bdf_e\bdf_e^\dagger)_{ik}\bfter_{kj}\\
&+\frac{1}{2}\left[\sh\sek\bfter_{ik}\bfter^\dagger_{kl}+\frac{3}{2}\swi\slk\bgtl^T_{ik}\bgtl^*_{kl}\right.\\
&\left.\qquad\ +\frac{1}{2}\sbi\slk\bgtpl^T_{ik}\bgtpl^*_{kl}\right]\bfter_{lj}\\
&-2\sbi\slk\bgtpl^T_{ik}\bftel_{kl}\bgtper^*_{lj}-2\sbi\left[\cs^2\h+\sn^2\Hh\right]\gtphd(\bdf_e)_{ik}\bgtper^*_{kj}\\
&+2\sh\sek\bfter_{ik}\bfter^\dagger_{kl}\bfter_{lj}+2\sbi\sek\bfter_{ik}\bgtper^T_{kl}\bgtper^*_{lj}\\
&-\bfter_{ij}\left[\frac{3}{2}g'^2+\frac{9}{2}g^2_2\right]
\end{split}\end{equation}

\begin{equation}\begin{split}\label{app:bgtpq}
{\left(4\pi\right)}^2\frac{d\bgtpq_{ij}}{dt}=&\frac{1}{2}\sbi\left[\frac{1}{3}\sql\bgtpq_{lk}\bgtpq^\dagger_{kl}+\sll\bgtpl_{lk}\bgtpl^\dagger_{kl}+\frac{8}{3}\suk\bgtpur_{lk}\bgtpur^\dagger_{kl}\right.\\
&\left.\qquad\ +\frac{2}{3}\sdk\bgtpdr_{lk}\bgtpdr^\dagger_{kl}+2\sek\bgtper_{lk}\bgtper^\dagger_{kl}\right]\bgtpq_{ij}\\
&+\frac{1}{2}\sh\sbi\bgtpq_{ij}\left\{\left[\sn^2\h+\cs^2\Hh\right]\mgtphusq+\left[\cs^2\h+\sn^2\Hh\right]\mgtphdsq\right\}\\
&+\frac{1}{2}\bgtpq_{ik}\left\{\left[\sn^2\h+\cs^2\Hh\right](\bdf_u^*\bdf_u^T)_{kj}+\left[\cs^2\h+\sn^2\Hh\right](\bdf_d^*\bdf_d^T)_{kj}\right\}\\
&+\frac{1}{2}\bgtpq_{ik}\left[\sh\sul\bftur^*_{kl}\bftur^T_{lj}+\sh\sdl\bftdr^*_{kl}\bftdr^T_{lj}+\frac{3}{2}\swi\sql\bgtq^\dagger_{kl}\bgtq_{lj}\right.\\
&\left.\qquad\qquad\quad\ +\frac{1}{18}\sbi\sql\bgtpq^\dagger_{kl}\bgtpq_{lj}+\frac{8}{3}\sgl\sql\bgtsq^\dagger_{kl}\bgtsq_{lj}\right]\\
&+4\sh\left[-2\sul\bftuq^*_{ik}\bgtpur_{kl}\bftur^T_{lj}+\sdl\bftdq^*_{ik}\bgtpdr_{kl}\bftdr^T_{lj}\right]\\
&+6\sh\left\{\left[\sn^2\h+\cs^2\Hh\right]\gtphu\bftuq^*_{ik}(\bdf_u)^T_{kj}-\left[\cs^2\h+\sn^2\Hh\right]\gtphd\bftdq^*_{ik}(\bdf_d)^T_{kj}\right\}\\
&+\frac{1}{2}\sql\left[3\swi\bgtq_{ik}\bgtq^\dagger_{kl}+\frac{1}{9}\sbi\bgtpq_{ik}\bgtpq^\dagger_{kl}+\frac{16}{3}\sgl\bgtsq_{ik}\bgtsq^\dagger_{kl}\right]\bgtpq_{lj}\\
&+\sh\sql\left[\bftuq^*_{ik}\bftuq^T_{kl}+\bftdq^*_{ik}\bftdq^T_{kl}\right]\bgtpq_{lj}\\
&-\bgtpq_{ij}\left[\frac{1}{12}g'^2+\frac{9}{4}g^2_2+4g^2_s\right]
\end{split}\end{equation}

\begin{equation}\begin{split}
{\left(4\pi\right)}^2\frac{d\bgtpl_{ij}}{dt}=&\frac{1}{2}\sbi\left[\frac{1}{3}\sql\bgtpq_{lk}\bgtpq^\dagger_{kl}+\sll\bgtpl_{lk}\bgtpl^\dagger_{kl}+\frac{8}{3}\suk\bgtpur_{lk}\bgtpur^\dagger_{kl}\right.\\
&\left.\qquad\ +\frac{2}{3}\sdk\bgtpdr_{lk}\bgtpdr^\dagger_{kl}+2\sek\bgtper_{lk}\bgtper^\dagger_{kl}\right]\bgtpl_{ij}\\
&+\frac{1}{2}\sh\sbi\bgtpl_{ij}\left\{\left[\sn^2\h+\cs^2\Hh\right]\mgtphusq+\left[\cs^2\h+\sn^2\Hh\right]\mgtphdsq\right\}\\
&+\frac{1}{2}\bgtpl_{ik}\left[\cs^2\h+\sn^2\Hh\right](\bdf_e^*\bdf_e^T)_{kj}\\
&+\frac{1}{2}\bgtpl_{ik}\left[\sh\sel\bfter^*_{kl}\bfter^T_{lj}+\frac{3}{2}\swi\sll\bgtl^\dagger_{kl}\bgtl_{lj}+\frac{1}{2}\sbi\sll\bgtpl^\dagger_{kl}\bgtpl_{lj}\right]\\
&-4\sh\sel\bftel^*_{ik}\bgtper_{kl}\bfter^T_{lj}+2\sh\left[\cs^2\h+\sn^2\Hh\right]\gtphd\bftel^*_{ik}(\bdf_e)^T_{kj}\\
&+\frac{1}{2}\sll\left[3\swi\bgtl_{ik}\bgtl^\dagger_{kl}+\sbi\bgtpl_{ik}\bgtpl^\dagger_{kl}\right]\bgtpl_{lj}\\
&+\sh\sll\bftel^*_{ik}\bftel^T_{kl}\bgtpl_{lj}-\bgtpl_{ij}\left[\frac{3}{4}g'^2+\frac{9}{4}g^2_2\right]
\end{split}\end{equation}

\begin{equation}\begin{split}\label{app:bgtpur}
{\left(4\pi\right)}^2\frac{d\bgtpur_{ij}}{dt}=&\left[\sn^2\h+\cs^2\Hh\right](\bdf_u^T\bdf_u^*)_{ik}\bgtpur_{kj}\\
&+\left[\frac{4}{9}\sbi\suk\bgtpur_{ik}\bgtpur^\dagger_{kl}+\frac{4}{3}\sgl\suk\bgtsur_{ik}\bgtsur^\dagger_{kl}\right.\\
&\left.\quad\ +\sh\sqk\bftuq^T_{ik}\bftuq^*_{kl}\right]\bgtpur_{lj}\\
&+\frac{1}{2}\sbi\bgtpur_{ij}\left[\frac{1}{3}\sql\bgtpq^\dagger_{kl}\bgtpq_{lk}+\sll\bgtpl^\dagger_{kl}\bgtpl_{lk}+\frac{8}{3}\suk\bgtpur^\dagger_{kl}\bgtpur_{lk}\right.\\
&\left.\qquad\qquad\qquad\ +\frac{2}{3}\sdk\bgtpdr^\dagger_{kl}\bgtpdr_{lk}+2\sek\bgtper^\dagger_{kl}\bgtper_{lk}\right]\\
&+\frac{1}{2}\sbi\sh\bgtpur_{ij}\left\{\left[\sn^2\h+\cs^2\Hh\right]\mgtphusq+\left[\cs^2\h+\sn^2\Hh\right]\mgtphdsq\right\}\\
&-3\sbi\sh\left[\sn^2\h+\cs^2\Hh\right]\gtphu(\bdf_u)^T_{ik}\bftur^*_{kj}\\
&-\sh\sqk\bftuq^T_{ik}\bgtpq_{kl}\bftur^*_{lj}+2\sh\suk\bgtpur_{ik}\bftur^T_{kl}\bftur^*_{lj}\\
&+\suk\bgtpur_{ik}\left[\frac{8}{9}\sbi\bgtpur^\dagger_{kl}\bgtpur_{lj}+\frac{8}{3}\sgl\bgtsur^\dagger_{kl}\bgtsur_{lj}\right]\\
&-\bgtpur_{ij}\left[\frac{4}{3}g'^2+4g^2_s\right]
\end{split}\end{equation}

\begin{equation}\begin{split}
{\left(4\pi\right)}^2\frac{d\bgtpdr_{ij}}{dt}=&\left[\cs^2\h+\sn^2\Hh\right](\bdf_d^T\bdf_d^*)_{ik}\bgtpdr_{kj}\\
&+\left[\frac{1}{9}\sbi\sdk\bgtpdr_{ik}\bgtpdr^\dagger_{kl}+\frac{4}{3}\sgl\sdk\bgtsdr_{ik}\bgtsdr^\dagger_{kl}\right.\\
&\left.\quad\ +\sh\sqk\bftdq^T_{ik}\bftdq^*_{kl}\right]\bgtpdr_{lj}\\
&+\frac{1}{2}\sbi\bgtpdr_{ij}\left[\frac{1}{3}\sql\bgtpq^\dagger_{kl}\bgtpq_{lk}+\sll\bgtpl^\dagger_{kl}\bgtpl_{lk}+\frac{8}{3}\suk\bgtpur^\dagger_{kl}\bgtpur_{lk}\right.\\
&\left.\qquad\qquad\qquad\ +\frac{2}{3}\sdk\bgtpdr^\dagger_{kl}\bgtpdr_{lk}+2\sek\bgtper^\dagger_{kl}\bgtper_{lk}\right]\\
&+\frac{1}{2}\sbi\sh\bgtpdr_{ij}\left\{\left[\sn^2\h+\cs^2\Hh\right]\mgtphusq+\left[\cs^2\h+\sn^2\Hh\right]\mgtphdsq\right\}\\
&-6\sbi\sh\left[\cs^2\h+\sn^2\Hh\right]\gtphd(\bdf_d)^T_{ik}\bftdr^*_{kj}\\
&+2\sh\sqk\bftdq^T_{ik}\bgtpq_{kl}\bftdr^*_{lj}+2\sh\sdk\bgtpdr_{ik}\bftdr^T_{kl}\bftdr^*_{lj}\\
&+\sdk\bgtpdr_{ik}\left[\frac{2}{9}\sbi\bgtpdr^\dagger_{kl}\bgtpdr_{lj}+\frac{8}{3}\sgl\bgtsdr^\dagger_{kl}\bgtsdr_{lj}\right]\\
&-\bgtpdr_{ij}\left[\frac{1}{3}g'^2+4g^2_s\right]
\end{split}\end{equation}

\begin{equation}\begin{split}
{\left(4\pi\right)}^2\frac{d\bgtper_{ij}}{dt}=&\left[\cs^2\h+\sn^2\Hh\right](\bdf_e^T\bdf_e^*)_{ik}\bgtper_{kj}\\
&+\left[\sbi\sek\bgtper_{ik}\bgtper^\dagger_{kl}+\sh\slk\bftel^T_{ik}\bftel^*_{kl}\right]\bgtper_{lj}\\
&+\frac{1}{2}\sbi\bgtper_{ij}\left[\frac{1}{3}\sql\bgtpq^\dagger_{kl}\bgtpq_{lk}+\sll\bgtpl^\dagger_{kl}\bgtpl_{lk}+\frac{8}{3}\suk\bgtpur^\dagger_{kl}\bgtpur_{lk}\right.\\
&\left.\qquad\qquad\qquad+\frac{2}{3}\sdk\bgtpdr^\dagger_{kl}\bgtpdr_{lk}+2\sek\bgtper^\dagger_{kl}\bgtper_{lk}\right]\\
&+\frac{1}{2}\sbi\sh\bgtper_{ij}\left\{\left[\sn^2\h+\cs^2\Hh\right]\mgtphusq+\left[\cs^2\h+\sn^2\Hh\right]\mgtphdsq\right\}\\
&-2\sbi\sh\left[\cs^2\h+\sn^2\Hh\right]\gtphd(\bdf_e)^T_{ik}\bfter^*_{kj}\\
&-2\sh\slk\bftel^T_{ik}\bgtpl_{kl}\bfter^*_{lj}+2\sh\sek\bgtper_{ik}\bfter^T_{kl}\bfter^*_{lj}\\
&+2\sbi\sek\bgtper_{ik}\bgtper^\dagger_{kl}\bgtper_{lj}-3\bgtper_{ij}g'^2
\end{split}\end{equation}

\begin{equation}\begin{split}
{\left(4\pi\right)}^2\frac{d\left(\sn\gtphu\right)}{dt}=&\frac{2}{4}\sbi\left[\frac{1}{3}\sql\bgtpq_{lk}\bgtpq^\dagger_{kl}+\sll\bgtpl_{lk}\bgtpl^\dagger_{kl}+\frac{8}{3}\suk\bgtpur_{lk}\bgtpur^\dagger_{kl}\right.\\
&\left.\qquad\ +\frac{2}{3}\sdk\bgtpdr_{lk}\bgtpdr^\dagger_{kl}+2\sek\bgtper_{lk}\bgtper^\dagger_{kl}\right]\sn\gtphu\\
&+\frac{1}{2}\sh\sbi\left\{\left[\sn^2\h+\cs^2\Hh\right]\mgtphusq+\left[\cs^2\h+\sn^2\Hh\right]\mgtphdsq\right\}\sn\gtphu\\
&+\frac{1}{2}\sh\left[3\suk\sn\gtphu\bftur^*_{lk}\bftur^{T}_{kl}+3\sql\sn\gtphu\bftuq^*_{lk}\bftuq^T_{kl}\right]\\
&+\frac{1}{4}\sh\left[\sn^2\h+\cs^2\Hh\right]\sn\gtphu\left\{3\swi\mgthusq+\sbi\mgtphusq\right\}\\
&+2\sql\sn\bgtpq_{lk}(\bdf_u)^*_{km}\bftuq^T_{ml}-8\suk\sn\bftur^{T}_{km}(\bdf_u)^*_{ml}\bgtpur_{lk}\\
&+\sbi\sh\cs^2\left(\h-\Hh\right)\mgtphdsq\sn\gtphu+3\swi\sh\cs^2\left(\h-\Hh\right)\gtphd\gthds\sn\gthu\\
&+\sn\gtphu\left[\left(\sn^2\h+\cs^2\Hh\right)\mathrm{Tr}\{3\bdf_u^\dagger\bdf_u\}+\cs^2\left(\h-\Hh\right)\mathrm{Tr}\left\{3\bdf_d^\dagger\bdf_d+\bdf_e^\dagger\bdf_e\right\}\right]\\
&+\frac{1}{2}\sh\sn\gtphu\left\{3\swi\left[\left(\sn^2\h+\cs^2\Hh\right)\mgthusq+\cs^2\left(\h-\Hh\right)\mgthdsq\right]\right.\\
&\left.\qquad\qquad\qquad+\sbi\left[\left(\sn^2\h+\cs^2\Hh\right)\mgtphusq+\cs^2\left(\h-\Hh\right)\mgtphdsq\right]\right\}\\
&-\sn\gtphu\left[\frac{3}{4}g'^2+\frac{9}{4}g^2_2\right]
\end{split}\end{equation}

\begin{equation}\begin{split}
{\left(4\pi\right)}^2\frac{d\left(\cs\gtphd\right)}{dt}=&\frac{2}{4}\sbi\left[\frac{1}{3}\sql\bgtpq_{lk}\bgtpq^\dagger_{kl}+\sll\bgtpl_{lk}\bgtpl^\dagger_{kl}+\frac{8}{3}\suk\bgtpur_{lk}\bgtpur^\dagger_{kl}\right.\\
&\left.\qquad\ +\frac{2}{3}\sdk\bgtpdr_{lk}\bgtpdr^\dagger_{kl}+2\sek\bgtper_{lk}\bgtper^\dagger_{kl}\right]\cs\gtphd\\
&+\frac{1}{2}\sh\sbi\left\{\left[\sn^2\h+\cs^2\Hh\right]\mgtphusq+\left[\cs^2\h+\sn^2\Hh\right]\mgtphdsq\right\}\cs\gtphd\\
&+\frac{1}{2}\sh\left[3\sdk\cs\gtphd\bftdr^*_{lk}\bftdr^{T}_{kl}+\sek\cs\gtphd\bfter^*_{lk}\bfter^{T}_{kl}\right.\\
&\left.\qquad\qquad+3\sql\cs\gtphd\bftdq^*_{lk}\bftdq^T_{kl}+\sll\cs\gtphd\bftel^*_{lk}\bftel^T_{kl}\right]\\
&+\frac{1}{4}\sh\left[\cs^2\h+\sn^2\Hh\right]\cs\gtphd\left\{3\swi\mgthdsq+\sbi\mgtphdsq\right\}\\
&+2\left[-\sql\cs\bgtpq_{lk}(\bdf_d)^*_{km}\bftdq^T_{ml}+\sll\cs\bgtpl_{lk}(\bdf_e)^*_{km}\bftel^T_{ml}\right]\\
&-4\left[\sdk\cs\bftdr^{T}_{km}(\bdf_d)^*_{ml}\bgtpdr_{lk}+\sek\cs\bfter^T_{km}(\bdf_e)^*_{ml}\bgtper_{lk}\right]\\
&+\sbi\sh\sn^2\left(\h-\Hh\right)\mgtphusq\cs\gtphd+3\swi\sh\sn^2\left(\h-\Hh\right)\gtphu\gthus\cs\gthd\\
&+\cs\gtphd\left[\sn^2\left(\h-\Hh\right)\mathrm{Tr}\{3\bdf_u^\dagger\bdf_u\}+\left(\cs^2\h+\sn^2\Hh\right)\mathrm{Tr}\left\{3\bdf_d^\dagger\bdf_d+\bdf_e^{\dagger}\bdf_e\right\}\right]\\
&+\frac{1}{2}\sh\cs\gtphd\left\{3\swi\left[\sn^2\left(\h-\Hh\right)\mgthusq+\left(\cs^2\h+\sn^2\Hh\right)\mgthdsq\right]\right.\\
&\left.\qquad\qquad\qquad+\sbi\left[\sn^2\left(\h-\Hh\right)\mgtphusq+\left(\cs^2\h+\sn^2\Hh\right)\mgtphdsq\right]\right\}\\
&-\cs\gtphd\left[\frac{3}{4}g'^2+\frac{9}{4}g^2_2\right]
\end{split}\end{equation}

\begin{equation}\begin{split}
{\left(4\pi\right)}^2\frac{d\bgtq_{ij}}{dt}=&\frac{1}{2}\swi\left[3\sql\bgtq_{lk}\bgtq^\dagger_{kl}+\sll\bgtl_{lk}\bgtl^\dagger_{kl}\right]\bgtq_{ij}\\
&+\frac{1}{2}\sh\swi\bgtq_{ij}\left\{\left[\sn^2\h+\cs^2\Hh\right]\mgthusq+\left[\cs^2\h+\sn^2\Hh\right]\mgthdsq\right\}\\
&+\frac{1}{2}\bgtq_{ik}\left\{\left[\sn^2\h+\cs^2\Hh\right](\bdf_u^*\bdf_u^T)_{kj}+\left[\cs^2\h+\sn^2\Hh\right](\bdf_d^*\bdf_d^T)_{kj}\right\}\\
&+\frac{1}{2}\bgtq_{ik}\left[\sh\sul\bftur^*_{kl}\bftur^T_{lj}+\sh\sdl\bftdr^*_{kl}\bftdr^T_{lj}+\frac{3}{2}\swi\sql\bgtq^\dagger_{kl}\bgtq_{lj}\right.\\
&\left.\qquad\qquad\quad+\frac{1}{18}\sbi\sql\bgtpq^\dagger_{kl}\bgtpq_{lj}+\frac{8}{3}\sgl\sql\bgtsq^\dagger_{kl}\bgtsq_{lj}\right]\\
&-2\sh\left\{\left[\sn^2\h+\cs^2\Hh\right]\gthu\bftuq^*_{ik}(\bdf_u)^T_{kj}+\left[\cs^2\h+\sn^2\Hh\right]\gthd\bftdq^*_{ik}(\bdf_d)^T_{kj}\right\}\\
&+\frac{1}{2}\sql\left[3\swi\bgtq_{ik}\bgtq^\dagger_{kl}+\frac{1}{9}\sbi\bgtpq_{ik}\bgtpq^\dagger_{kl}+\frac{16}{3}\sgl\bgtsq_{ik}\bgtsq^\dagger_{kl}\right]\bgtq_{lj}\\
&+\sh\sql\left[\bftuq^*_{ik}\bftuq^T_{kl}+\bftdq^*_{ik}\bftdq^T_{kl}\right]\bgtq_{lj}\\
&-\bgtq_{ij}\left[\frac{1}{12}g'^2+\frac{33}{4}g^2_2+4g^2_s\right]
\end{split}\end{equation}

\begin{equation}\begin{split}
{\left(4\pi\right)}^2\frac{d\bgtl_{ij}}{dt}=&\frac{1}{2}\swi\left[3\sql\bgtq_{lk}\bgtq^\dagger_{kl}+\sll\bgtl_{lk}\bgtl^\dagger_{kl}\right]\bgtl_{ij}\\
&+\frac{1}{2}\sh\swi\bgtl_{ij}\left\{\left[\sn^2\h+\cs^2\Hh\right]\mgthusq+\left[\cs^2\h+\sn^2\Hh\right]\mgthdsq\right\}\\
&+\frac{1}{2}\left[\cs^2\h+\sn^2\Hh\right]\bgtl_{ik}(\bdf_e^*\bdf_e^T)_{kj}\\
&+\frac{1}{2}\bgtl_{ik}\left[\sh\sel\bfter^*_{kl}\bfter^T_{lj}+\frac{3}{2}\swi\sll\bgtl^\dagger_{kl}\bgtl_{lj}+\frac{1}{2}\sbi\sll\bgtpl^\dagger_{kl}\bgtpl_{lj}\right]\\
&-2\sh\left[\cs^2\h+\sn^2\Hh\right]\gthd\bftel^*_{ik}(\bdf_e)^T_{kj}\\
&+\frac{1}{2}\sll\left[3\swi\bgtl_{ik}\bgtl^\dagger_{kl}+\sbi\bgtpl_{ik}\bgtpl^\dagger_{kl}\right]\bgtl_{lj}\\
&+\sh\sll\bftel^*_{ik}\bftel^T_{kl}\bgtl_{lj}-\bgtl_{ij}\left[\frac{3}{4}g'^2+\frac{33}{4}g^2_2\right]
\end{split}\end{equation}

\begin{equation}\begin{split}
{\left(4\pi\right)}^2\frac{d\left(\sn\gthu\right)}{dt}=&\frac{1}{2}\swi\left[3\sql\bgtq_{lk}\bgtq^\dagger_{kl}+\sll\bgtl_{lk}\bgtl^\dagger_{kl}\right]\sn\gthu\\
&+\frac{1}{2}\sh\swi\left\{\left[\sn^2\h+\cs^2\Hh\right]\mgthusq+\left[\cs^2\h+\sn^2\Hh\right]\mgthdsq\right\}\sn\gthu\\
&+\frac{1}{2}\sh\left[3\suk\sn\gthu\bftur^*_{lk}\bftur^T_{kl}+3\sql\sn\gthu\bftuq^*_{lk}\bftuq^T_{kl}\right]\\
&+\frac{1}{4}\sh\left[\sn^2\h+\cs^2\Hh\right]\sn\gthu\left\{3\swi\mgthusq+\sbi\mgtphusq\right\}\\
&-6\sql\sn\bgtq_{lk}(\bdf_u)^*_{km}\bftuq^T_{ml}\\
&+\sbi\sh\cs^2\left(\h-\Hh\right)\gthd\gtphds\sn\gtphu-\swi\sh\cs^2\left(\h-\Hh\right)\mgthdsq\sn\gthu\\
&+\sn\gthu\left[\left(\sn^2\h+\cs^2\Hh\right)\mathrm{Tr}\{3\bdf_u^\dagger\bdf_u\}+\cs^2\left(\h-\Hh\right)\mathrm{Tr}\left\{3\bdf_d^\dagger\bdf_d+\bdf_e^\dagger\bdf_e\right\}\right]\\
&+\frac{1}{2}\sh\sn\gthu\left\{3\swi\left[\left(\sn^2\h+\cs^2\Hh\right)\mgthusq+\cs^2\left(\h-\Hh\right)\mgthdsq\right]\right.\\
&\left.\qquad\qquad\qquad+\sbi\left[\left(\sn^2\h+\cs^2\Hh\right)\mgtphusq+\cs^2\left(\h-\Hh\right)\mgtphdsq\right]\right\}\\
&-\sn\gthu\left[\frac{3}{4}g'^2+\frac{33}{4}g^2_2\right]
\end{split}\end{equation}

\begin{equation}\begin{split}
{\left(4\pi\right)}^2\frac{d\left(\cs\gthd\right)}{dt}=&\frac{1}{2}\swi\left[3\sql\bgtq_{lk}\bgtq^\dagger_{kl}+\sll\bgtl_{lk}\bgtl^\dagger_{kl}\right]\cs\gthd\\
&+\frac{1}{2}\sh\swi\left\{\left[\sn^2\h+\cs^2\Hh\right]\mgthusq+\left[\cs^2\h+\sn^2\Hh\right]\mgthdsq\right\}\cs\gthd\\
&+\frac{1}{2}\sh\left[3\sdk\cs\gthd\bftdr^*_{lk}\bftdr^{T}_{kl}+\sek\cs\gthd\bfter^*_{lk}\bfter^{T}_{kl}\right.\\
&\left.\qquad\qquad+3\sql\cs\gthd\bftdq^*_{lk}\bftdq^T_{kl}+\sll\cs\gthd\bftel^*_{lk}\bftel^T_{kl}\right]\\
&+\frac{1}{4}\sh\left[\cs^2\h+\sn^2\Hh\right]\cs\gthd\left\{3\swi\mgthdsq+\sbi\mgtphdsq\right\}\\
&-2\left[3\sql\cs\bgtq_{lk}(\bdf_d)^*_{km}\bftdq^T_{ml}+\sll\cs\bgtl_{lk}(\bdf_e)^*_{km}\bftel^T_{ml}\right]\\
&+\sbi\sh\sn^2\left(\h-\Hh\right)\gthu\gtphus\cs\gtphd-\swi\sh\sn^2\left(\h-\Hh\right)\mgthusq\cs\gthd\\
&+\cs\gthd\left[\sn^2\left(\h-\Hh\right)\mathrm{Tr}\{3\bdf_u^\dagger\bdf_u\}+\left(\cs^2\h+\sn^2\Hh\right)\mathrm{Tr}\left\{3\bdf_d^\dagger\bdf_d+\bdf_e^\dagger\bdf_e\right\}\right]\\
&+\frac{1}{2}\sh\cs\gthd\left\{3\swi\left[\sn^2\left(\h-\Hh\right)\mgthusq+\left(\cs^2\h+\sn^2\Hh\right)\mgthdsq\right]\right.\\
&\left.\qquad\qquad\qquad+\sbi\left[\sn^2\left(\h-\Hh\right)\mgtphusq+\left(\cs^2\h+\sn^2\Hh\right)\mgtphdsq\right]\right\}\\
&-\cs\gthd\left[\frac{3}{4}g'^2+\frac{33}{4}g^2_2\right]
\end{split}\end{equation}

\begin{equation}\begin{split}
{\left(4\pi\right)}^2\frac{d\bgtsq_{ij}}{dt}=&\frac{1}{2}\sgl\left[2\sql\bgtsq_{lk}\bgtsq^\dagger_{kl}+\suk\bgtsur_{lk}\bgtsur^\dagger_{kl}+\sdk\bgtsdr_{lk}\bgtsdr^\dagger_{kl}\right]\bgtsq_{ij}\\
&+\frac{1}{2}\bgtsq_{ik}\left\{\left[\sn^2\h+\cs^2\Hh\right](\bdf_u)^*_{kl}(\bdf_u)^T_{lj}+\left[\cs^2\h+\sn^2\Hh\right](\bdf_d)^*_{kl}(\bdf_d)^T_{lj}\right\}\\
&+\frac{1}{2}\bgtsq_{ik}\left[\sh\sul\bftur^*_{kl}\bftur^T_{lj}+\sh\sdl\bftdr^*_{kl}\bftdr^T_{lj}+\frac{3}{2}\swi\sql\bgtq^\dagger_{kl}\bgtq_{lj}\right.\\
&\left.\qquad\qquad\quad+\frac{1}{18}\sbi\sql\bgtpq^\dagger_{kl}\bgtpq_{lj}+\frac{8}{3}\sgl\sql\bgtsq^\dagger_{kl}\bgtsq_{lj}\right]\\
&-2\sh\left[\sul\bftuq^*_{ik}\bgtsur_{kl}\bftur^T_{lj}+\sdl\bftdq^*_{ik}\bgtsdr_{kl}\bftdr^T_{lj}\right]\\
&+\frac{1}{2}\sql\left[3\swi\bgtq_{ik}\bgtq^\dagger_{kl}+\frac{1}{9}\sbi\bgtpq_{ik}\bgtpq^\dagger_{kl}+\frac{16}{3}\sgl\bgtsq_{ik}\bgtsq^\dagger_{kl}\right]\bgtsq_{lj}\\
&+\sh\sql\left[\bftuq^*_{ik}\bftuq^T_{kl}+\bftdq^*_{ik}\bftdq^T_{kl}\right]\bgtsq_{lj}\\
&-\bgtsq_{ij}\left[\frac{1}{12}g'^2+\frac{9}{4}g^2_2+13g^2_s\right]
\end{split}\end{equation}

\begin{equation}\begin{split}
{\left(4\pi\right)}^2\frac{d\bgtsur_{ij}}{dt}=&\left[\sn^2\h+\cs^2\Hh\right](\bdf_u^T\bdf_u^*)_{ik}\bgtsur_{kj}\\
&+\left[\frac{4}{9}\sbi\suk\bgtpur_{ik}\bgtpur^\dagger_{kl}+\frac{4}{3}\sgl\suk\bgtsur_{ik}\bgtsur^\dagger_{kl}\right.\\
&\left.\quad\ +\sh\sqk\bftuq^T_{ik}\bftuq^*_{kl}\right]\bgtsur_{lj}\\
&+\sgl\sql\bgtsur_{ij}\bgtsq^\dagger_{kl}\bgtsq_{lk}\\
&+\frac{1}{2}\sgl\bgtsur_{ij}\left[\suk\bgtsur^\dagger_{kl}\bgtsur_{lk}+\sdk\bgtsdr^\dagger_{kl}\bgtsdr_{lk}\right]\\
&-4\sh\sqk\bftuq^T_{ik}\bgtsq_{kl}\bftur^*_{lj}+2\sh\suk\bgtsur_{ik}\bftur^T_{kl}\bftur^*_{lj}\\
&+\suk\bgtsur_{ik}\left[\frac{8}{9}\sbi\bgtpur^\dagger_{kl}\bgtpur_{lj}+\frac{8}{3}\sgl\bgtsur^\dagger_{kl}\bgtsur_{lj}\right]\\
&-\bgtsur_{ij}\left[\frac{4}{3}g'^2+13g^2_s\right]
\end{split}\end{equation}

\begin{equation}\begin{split}
{\left(4\pi\right)}^2\frac{d\bgtsdr_{ij}}{dt}=&\left[\cs^2\h+\sn^2\Hh\right](\bdf_d^T\bdf_d^*)_{ik}\bgtsdr_{kj}\\
&+\left[\frac{1}{9}\sbi\sdk\bgtpdr_{ik}\bgtpdr^\dagger_{kl}+\frac{4}{3}\sgl\sdk\bgtsdr_{ik}\bgtsdr^\dagger_{kl}\right.\\
&\left.\quad\ +\sh\sqk\bftdq^T_{ik}\bftdq^*_{kl}\right]\bgtsdr_{lj}\\
&+\sgl\sql\bgtsdr_{ij}\bgtsq^\dagger_{kl}\bgtsq_{lk}\\
&+\frac{1}{2}\sgl\bgtsdr_{ij}\left[\suk\bgtsur^\dagger_{kl}\bgtsur_{lk}+\sdk\bgtsdr^\dagger_{kl}\bgtsdr_{lk}\right]\\
&-4\sh\sqk\bftdq^T_{ik}\bgtsq_{kl}\bftdr^*_{lj}+2\sh\sdk\bgtsdr_{ik}\bftdr^T_{kl}\bftdr^*_{lj}\\
&+\sdk\bgtsdr_{ik}\left[\frac{2}{9}\sbi\bgtpdr^\dagger_{kl}\bgtpdr_{lj}+\frac{8}{3}\sgl\bgtsdr^\dagger_{kl}\bgtsdr_{lj}\right]\\
&-\bgtsdr_{ij}\left[\frac{1}{3}g'^2+13g^2_s\right]
\end{split}\end{equation}


\end{document}